\documentclass[11pt]{article}
\pdfoutput=1

\usepackage{formatting}

\usepackage{shortcuts}

\usepackage{braket}

\begin{document} 
\begin{titlepage}
\begin{center}
\phantom{ }
\vspace{1.5cm}

{\bf \Large{Cosmology from random entanglement}}
\vskip 0.7cm
{Stefano Antonini$^{\dagger}$, Martin Sasieta$^{\mathsection}$, Brian Swingle$^{\ddagger}$}
\vskip 0.3in
\small{$^{\dagger}$\textit{University of Maryland, College Park, MD, 20742, USA}}
\vskip 0.03in
\small{ $^{\mathsection}$$^{\ddagger}$\textit{Martin Fisher School of Physics, Brandeis University,}}
\vskip -.4cm
\small{\textit{Waltham, Massachusetts 02453, USA}}
\vskip .8in

\begin{abstract}
We construct entangled microstates of a pair of holographic CFTs whose dual semiclassical description includes big bang-big crunch AdS cosmologies in spaces without boundaries. The cosmology is supported by inhomogeneous heavy matter and it partially purifies the bulk entanglement of two disconnected auxiliary AdS spacetimes. We show that the island formula for the fine grained entropy of one of the CFTs follows from a standard gravitational replica trick calculation. In generic settings, the cosmology is contained in the entanglement wedge of one of the two CFTs. We then investigate properties of the cosmology-to-boundary encoding map, and in particular, its non-isometric character. Restricting our attention to a specific class of states on the cosmology, we provide an explicit, and state-dependent, boundary representation of operators acting on the cosmology. Finally, under genericity assumptions, we argue for a non-isometric to approximately-isometric transition of the cosmology-to-boundary map for ``simple'' states on the cosmology as a function of the bulk entanglement, with tensor network toy models of our setup as a guide.

\end{abstract}
\end{center}

\vfill 

\small{\, \\
${}^{\dagger}$ \href{mailto:santonin@terpmail.umd.edu}{santonin@terpmail.umd.edu} \\
${}^{\mathsection}$ \href{mailto:martinsasieta@brandeis.edu}{martinsasieta@brandeis.edu}\\
${}^{\ddagger}$ \href{mailto:headrick@brandeis.edu}{bswingle@brandeis.edu} \\
}
\end{titlepage}

\setcounter{tocdepth}{2}

\hrule 
{\parskip = .4\baselineskip \tableofcontents}

\vskip 1cm
\hrule

\section{Introduction}
\label{sec:1}

Any theory of quantum gravity aiming at capturing physics in our universe must be able to describe cosmology. Holography \cite{tHooft:1993dmi,Susskind:1994vu}, and its realization in the Anti-de Sitter/Conformal Field Theory (AdS/CFT) correspondence \cite{Maldacena:1997re}, offers a promising framework for the formulation of theories of quantum gravity, one which has advanced our understanding of several fundamental aspects of quantum gravity in recent years, see e.g. \cite{Ryu2006a,Swingle:2009bg,VanRaamsdonk:2010pw,Lewkowycz:2013nqa,Maldacena:2013xja,Faulkner:2013ana,Stanford:2014jda,Almheiri:2014lwa,Maldacena:2015waa,Pastawski:2015qua,Brown:2015bva,Hayden:2016cfa,Penington:2019npb,Almheiri:2019psf,Penington:2019kki,Almheiri:2019qdq}. However, describing realistic cosmologies within the holographic paradigm remains a central open problem. 

An intuitive way to visualize the difficulties arising when trying to attain a holographic model of cosmology is to recall that in AdS/CFT the bulk theory lives on an asymptotically AdS spacetime and the holographic CFT lives on the conformal boundary of this manifold. In contrast, most cosmological spacetimes, and in particular the homogeneous and isotropic Friedmann-Lema\^itre-Robertson-Walker (FLRW) spacetimes relevant for the description of our universe, do not possess timelike asymptotic boundaries or weakly gravitating regions where a holographic description could be located. It is therefore reasonable to expect that some new ingredient is needed in order to describe cosmology within holography, or at least that a holographic cosmology would need to be encoded in the dual theory in a non-standard fashion. 

Moreover, seeking realistic models of cosmology which can be described using holography introduces additional difficulties. For example, cosmological observations show that the expansion of our universe is currently accelerating. This feature is explained in the standard model of cosmology by the presence of a positive cosmological constant. In contrast, the familiar version of holography realized by AdS/CFT describes quantum gravity in the presence of a negative cosmological constant, $\Lambda<0$, seemingly in tension with observation. While new ingredients seem to thus be \textit{a priori} needed, there do exist alternative explanations of the observed accelerated expansion which include a negative cosmological constant, see e.g. \cite{Hartle:2012qb,Peebles:1987ek,Ratra:1987rm,Tsujikawa:2013fta,Antonini:2022ptt,Antonini:2022fna,VanRaamsdonk:2023ion}. 

In the last two decades, progress has been made in several different approaches to holographic cosmology \cite{Strominger:2001pn,Alishahiha:2004md,dsds,Gorbenko:2018oov,Coleman:2021nor,Susskind:2021dfc,Susskind:2021omt,Susskind:2021esx,Shaghoulian:2022fop,McFadden:2009fg,Afshordi:2017ihr,Nastase:2018cbf,Nastase:2019rsn,Maldacena:2004rf,McInnes:2004nx,McInnes:2005sa,Freivogel:2005qh,Engelhardt:2014mea,Cooper:2018cmb,Antonini:2019qkt,Chen:2020tes,VanRaamsdonk:2020tlr,VanRaamsdonk:2021qgv,Antonini:2022blk,Antonini:2022fna,Fallows:2022ioc,Antonini:2022ptt,Ross:2022pde,VanRaamsdonk:2023ion,Sahu:2023fbx,Akers:2022max}. These include proposals for a holographic description of de Sitter space \cite{Strominger:2001pn,Alishahiha:2004md,dsds,Gorbenko:2018oov,Coleman:2021nor,Susskind:2021dfc,Susskind:2021omt,Susskind:2021esx,Shaghoulian:2022fop}, phenomenologically-oriented descriptions of $\Lambda>0$ cosmological spacetimes \cite{McFadden:2009fg,Afshordi:2017ihr,Nastase:2018cbf,Nastase:2019rsn}, and, more recently, holographic descriptions of $\Lambda<0$ big bang-big crunch cosmologies \cite{Cooper:2018cmb,Antonini:2019qkt,Chen:2020tes,VanRaamsdonk:2020tlr,VanRaamsdonk:2021qgv,Antonini:2022blk,Antonini:2022fna,Fallows:2022ioc,Antonini:2022ptt,Ross:2022pde,VanRaamsdonk:2023ion,Sahu:2023fbx}. The first two approaches are based on novel---and not yet fully developed---ingredients, i.e. a new de Sitter holography framework and a dual microscopic theory with complex parameters, respectively. The last approach makes use of more standard AdS/CFT tools, but nonetheless the cosmology is expected to be encoded in the state of the dual microscopic theory in a highly non-trivial way.

The cosmological spacetimes described in refs. \cite{Cooper:2018cmb,Antonini:2019qkt,VanRaamsdonk:2020tlr,Chen:2020tes,VanRaamsdonk:2021qgv,Antonini:2022blk,Antonini:2022fna,Fallows:2022ioc,Antonini:2022ptt,Ross:2022pde,VanRaamsdonk:2023ion,Sahu:2023fbx} have a well-defined analytic continuation to asymptotically AdS Euclidean spaces. These are saddle point geometries of the gravitational path integral dual to the Euclidean path integral preparing a specific state of the microscopic boundary theory. In the constructions of refs. \cite{VanRaamsdonk:2020tlr,VanRaamsdonk:2021qgv,Antonini:2022blk,Antonini:2022fna,Antonini:2022ptt}, the cosmology is an ``entanglement island'' for the dual microscopic system, similar to the setups studied in refs. \cite{Hartman:2020khs,Balasubramanian:2020xqf,Balasubramanian:2020coy,Balasubramanian:2021wgd,Bousso:2022gth}. In the braneworld cosmology approach studied in refs. \cite{Cooper:2018cmb,Antonini:2019qkt}, which can be understood as a a doubly-holographic description of the setup of refs. \cite{VanRaamsdonk:2020tlr,VanRaamsdonk:2021qgv,Antonini:2022blk,Antonini:2022fna,Antonini:2022ptt}, the cosmology lives instead behind the horizon of a higher-dimensional AdS black hole. In both cases, the dictionary relating cosmological observables to microscopic CFT observables is expected to be highly non-trivial, as is the case for the black hole interior  \cite{Papadodimas:2012aq,Brown:2019rox,Engelhardt:2021mue,Engelhardt:2021qjs,Akers:2022qdl,Cao:2023gkw}.

In this paper, we build a new class of cosmological microstates in AdS/CFT and study how cosmological physics is encoded in the dual CFT description. The Lorentzian bulk description of the state includes a $\Lambda<0$, big bang-big crunch, inhomogeneous closed cosmology, along with two additional disconnected AdS spacetimes. The cosmology is supported by heavy matter and contains thermal radiation which partially purifies the bulk entanglement of thermal radiation present in the two AdS spacetimes. The global bulk state is time reversal-symmetric, and it is prepared on the time reflection-symmetric slice by a Euclidean gravitational path integral, evaluated in the saddle point approximation. In the boundary description, the dual Euclidean path integral prepares a low-temperature \textit{partially entangled thermal state} (PETS) of two copies of a holographic CFT.

The present setup is similar in spirit to the constructions of refs. \cite{VanRaamsdonk:2020tlr,VanRaamsdonk:2021qgv,Antonini:2022blk,Antonini:2022fna,Antonini:2022ptt}, in the sense that it describes $\Lambda<0$ cosmologies with an associated bulk Euclidean saddle point geometry, and the microscopic state encoding the cosmology is prepared via a well-defined dual boundary Euclidean path integral. However, the simplicity of our construction, which we will briefly review below, will allow us to have more microscopic control over the system and study the holographic encoding of the cosmology in more detail. The price to pay is a cosmology which is not phenomenologically realistic: in its present realization, its evolution is not driven by a perfect fluid, it is inhomogeneous, and it does not undergo any phase of accelerated expansion. 

Although the final goal of any holographic cosmology program is certainly to describe our own universe using the holographic framework, understanding the properties of the holographic encoding of any cosmological spacetime---even one which is not phenomenologically relevant---is already a complicated and worthwhile endeavor. Hopefully, this will teach us general lessons that could guide us towards a holographic description of more realistic cosmologies.

We will now provide a summary of the main constructions and results of the present paper, which will be explored in detail in Secs. \ref{sec:2}-\ref{sec:5}.

\subsection{Overview and summary of results}
\subsection*{Cosmological states}

\begin{figure}[h]
\centering
\includegraphics[width = .77\textwidth]{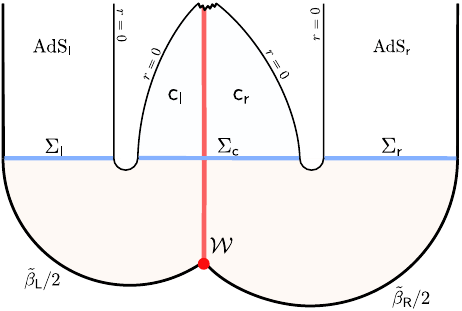}
\caption{Semiclassical state prepared by the gravitational path integral at low preparation temperatures. The geometry of the Euclidean section $X$ consists of two thermal AdS spaces. Each space has topology $\mathbf{D}^d\times \mathbf{S}^1$, where $\mathbf{D}^d$ is a $d$-dimensional ball, and $\mathbf{S}^1$ is the additional non-contractible thermal circle. The two solutions are glued together along the trajectory of the thin shell (red line). The Euclidean manifold $X$ is homotopic to a pair of pants. The semiclassical state is prepared on the time reflection-symmetric slice $\Sigma$ (blue slice), which contains the initial data of two global AdS timeslices $\Sigma_\lef$ and $\Sigma_\ri$, and a third closed universe $\Sigma_\Co$ supported by the thin shell. In the Loretzian section, obtained from analytic continuation along $\Sigma$, the universe crunches towards the future.}
\label{fig:cosm}
\end{figure}

The bulk Euclidean saddle point geometry $X$ preparing the cosmological state of interest is built by gluing two patches of thermal AdS spaces (generically at different temperatures) along the worldvolume of a thin shell of pressureless matter, as illustrated in Fig. \ref{fig:cosm}. In this regard, our construction is reminiscent of the construction in ref. \cite{Balasubramanian:2022gmo} of two-sided black hole microstates with long Einstein-Rosen bridges. In order for $X$ to be dominant over the competing black hole solution, we require the inverse temperatures $\tilde{\beta}_\Le,\tilde{\beta}_\Ri$\footnote{The inverse temperatures $\tilde{\beta}_\Le,\tilde{\beta}_\Ri$ represent the Euclidean preparation time of the CFT microstate. The corresponding physical inverse temperatures $\beta_\Le,\beta_\Ri$ of the left and right Euclidean AdS spaces are given by the sum of the preparation temperatures and the Euclidean time elapsed by the shell's trajectory, see eq. (\ref{eq:phystemps}).} to be large in units of the AdS radius $\ell$. In other words, we are interested in studying the gravitational system below its characteristic Hawking-Page temperature. 

In this regime, the time reflection-symmetric slice $\Sigma$ of $X$, on which the the bulk semiclassical state is prepared, contains three connected components, $\Sigma = \Sigma_\lef \cup \Sigma_\Co \cup \Sigma_\ri$. Each of the two components $\Sigma_\lef$ and $\Sigma_\ri$ corresponds to a hyperbolic disk, providing the initial data of two independent AdS spacetimes (AdS${}_\lef$ and AdS${}_\ri$). The last component, $\Sigma_\Co$, is a closed universe of spherical topology. It is formed by two hyperbolic disks $\Sigma_{\Co_\lef},\Sigma_{\Co_\ri}$ with a radial cut off at $r =R_*$ (which is equal for the two components), glued together along the position of the thin shell. The causal development of the initial data on $\Sigma_\Co$, denoted by $\Co=\Co_\lef\cup\Co_\ri$, consists of two patches of global AdS, glued along the spherical worldvolume of the thin shell. The shell collapses towards the past and the future, generating the big bang and big crunch singularities of the cosmological spacetime $\Co$.

More concretely, the metric on each one of the two AdS patches forming the cosmology can be locally characterized by the open FLRW metric
\be 
\text{d}s^2 =  - \text{d}t_c^2 + \ell^2\cos^2(t_c/\ell)\, \text{d}\mathbf{H}_{d}^2\;,
\label{eq:flrwintro}
\ee 
where $\text{d}\mathbf{H}_{d}^2 = \text{d}\rho^2 + \sinh^2 \rho \,\text{d}\Omega_{d-1}^2$ is the hyperbolic metric of a FLRW slice. In this form, the metric is homogeneous and isotropic, with the scale factor given by $a(t_c)=\ell \cos(t_c/\ell)$. The singularities arising at $t_c=\pm \pi\ell/2$ are only coordinate singularities. In fact, this metric covers a hyperbolic slicing of a Wheeler-DeWitt (WDW) patch of a pure AdS spacetime. 

However, $\Co_\lef$ and $\Co_\ri$ are strictly contained within the WDW patch covered by the metric (\ref{eq:flrwintro}) and they become truly singular due to the presence of the collapsing shell. We will show in Sec. \ref{sec:2} that each FLRW time slice of $\Co_\lef$ and $\Co_\ri$ is cut off at the location of the shell, implicitly given by $\rho=\rho_{\max}(t_c)$, where $\rho_{\max}(t_c)$ parametrizes the radial trajectory of the shell in these coordinates (see Fig. \ref{fig:WDWpatch}). In particular, the size of the spatial slices shrinks to zero at $t_c=\pm t_f$, with $t_f<\pi\ell/2$, corresponding to the endpoints of the shell's collapse. The spacetime curvature is singular at these points, which we identify with the big bang and the big crunch singularities of our cosmological spacetime.\footnote{The geometry of $\Co_\lef$ and $\Co_\ri$ is similar to the one arising inside a collapsing shell of matter during black hole formation. The future singularity of our cosmological spacetime can then be identified with the black hole singularity. Unlike usual black hole formation setups, our construction is time reversal-symmetric and there is a past singularity as well.}

$\Co_\lef$ and $\Co_\ri$ can then be understood to be two patches of a closed,\footnote{Although the spatial FLRW slices of each patch $\Co_\lef$ and $\Co_\ri$ are hyperbolic, each spatial slice is cut off at the location of the shell and glued to another identical portion of a spatial slice, giving rise to a closed cosmology.} $\Lambda<0$, big bang-big crunch cosmology. Although this cosmology looks homogeneous and isotropic locally (i.e. within the $\Co_\lef$ and $\Co_\ri$ patches), it is actually inhomogeneous due to the presence of the shell. In particular, the cosmological singularities are inhomogeneous: they do not arises from the scale factor vanishing in each patch, but rather from the size of the spatial slices shrinking to zero due to the collapse of the shell of matter.

The classical solutions built this way provide sensible (macroscopic) cosmologies if the shell is heavy, of rest mass $m\ell \gg 1$. As we will explain, this regime can be approached microscopically, by taking the large-$N$ limit of the holographic CFT, and considering that the dual operator to the thin shell is heavy, with a conformal dimension scaling with $N^2$. Additionally, we will be alluding to the case of a very heavy shell throughout this paper, treated formally in the $m\ell \rightarrow \infty$ limit. In this parametric regime, many of the calculations that we will encounter simplify considerably. Of course, we expect that this limit provides a good approximation to very large but finite $m\ell$ as well. 

A clear example where such simplification occurs is at the level of the entanglement structure of the state of the quantum fields on $\Sigma$, in the bulk effective field theory (EFT). Specifically, the path integral of the bulk fields on $X$ prepares a specific bulk EFT state, via a bulk time-dependent Hamiltonian coupling all of the three components of $\Sigma$. At intermediate values of $m\ell$, the bulk state prepared this way generally shares entanglement between the three connected components of $\Sigma$. In the limit of large mass for the shell, however, the shell's worldvolume $\Ws$ effectively pinches off in Fig. \ref{fig:cosm}. This drastically reduces the coupling between the two sides of the cosmology $\Sigma_{\Co_\lef}, \Sigma_{\Co_\ri} \subset \Sigma_\Co$, in the Euclidean preparation. As a result, the bulk state prepared by $X$ will contain mostly bipartite entanglement between each side of the cosmology ($\Sigma_{\Co_\lef}, \Sigma_{\Co_\ri}$), and the respective AdS region ($\Sigma_\lef,\Sigma_\ri$). Moreover, in this case, the bulk state prepared by $X$ roughly correspond to two bulk thermofield doubles, at temperatures $\beta_{\Le}$ (for AdS${}_{\lef}$ and $\Co_\lef$) and $\beta_{\Ri}$ (for AdS${}_{\ri}$ and $\Co_\ri$), with fields in the cosmology purifying fields in the AdS regions.

Given the construction of the Euclidean geometry $X$ preparing the bulk state, it is natural (and convenient for the reasons described in Sec. \ref{sec:2}) to consider more general bulk states on $\Sigma$, with additional entanglement between the AdS regions and the cosmology. Such states can be easily prepared with the addition of local operator insertions along the Euclidean preparation manifold $X$ (see Sec. \ref{sec:2.3}). For similar reasons, in the large mass limit for the thin shell, all of the bulk states prepared in this way contain mostly bipartite entanglement.

\subsection*{Partially entangled thermal states}

The gravitational path integral preparing the cosmological state holographically corresponds to the Euclidean CFT path integral on flat cylinders of lengths $\tbeta_{\Le}/2$ and $\tbeta_{\Ri}/2$, with the insertion of a heavy operator $\Op$ dual to the shell in between these Euclidean evolutions (see Fig. \ref{fig:PETS}). This path integral prepares a so-called \textit{partially entangled thermal state} (PETS) of two independent copies of the holographic CFT,
\be\label{eq:PETSintro}
\ket{\Psi_{\Op}} = \ket{\rho_{\tbeta_{\Le}/2}\Op\rho_{\tbeta_{\Ri}/2}}\,= \dfrac{1}{\sqrt{Z_1}}\sum_{n,m} e^{-\frac{1}{2}(\tbeta_{\Le} E_n + \tbeta_{\Ri} E_m)} \Op_{nm}\ket{E_n}^*_{\Le}\otimes \ket{E_m}_{\Ri}\;,
\ee 
where each CFT lives on a spatial $\mathbf{S}^{d-1}$ of radius $\ell$. The two-point function of the operator $Z_1 = \text{Tr}(e^{-\tbeta_{\Le} H}\mathcal{O}^\dagger e^{-\tbeta_{\Ri} H}\mathcal{O})$ normalizes the state. 

The insertion of the heavy CFT operator $\mathcal{O}$ in the path integral provides the boundary condition for the worldvolume of the thin shell at the asymptotic boundary of the Euclidean geometry $X$. The specific operator corresponding to the thin shell is $\mathcal{O}\sim \prod_i\mathcal{O}_\Delta(\Omega_i)$, formed by a large number of single trace operator insertions $\mathcal{O}_{\Delta}(\Omega_i)$, of low conformal dimension $\Delta \ll N^2$, distributed in an approximately homogeneous way along the sphere, $\Omega_i\in \mathbf{S}^{d-1}$. In what follows, we consider that the number of operator insertions is $O(N^2)$, which implies that the associated bulk backreaction cannot be treated perturbatively in the $1/N$ expansion. Still, the matrix elements of the shell operator $\Op$ admit an effective gaussian description in the energy basis, at least at the level of the two-point function, given that the shell effectively behaves as a spherical heavy particle (see \cite{Sasieta:2022ksu}).

In the absence of any operator insertion (i.e. when $\mathcal{O}=\mathbb{1}$), the PETS becomes a low-temperature TFD state of two independent CFTs in the confined phase, dual to a pair of AdS spacetimes with bulk quantum fields entangled between the two components to form a bulk thermofield-double state. The insertion of the heavy thin shell operator causes the non-perturbative modification of the semiclassical state, leading to the presence of the closed cosmology in the way described above. Similarly, in the high temperature regime, i.e. above the associated confinement/deconfinement phase transition temperature for the state, the semiclassical description of the PETS becomes that one of the two-sided black hole microstates constructed in ref. \cite{Balasubramanian:2022gmo}, where the black hole contains a large interior geometry, supported by the presence of the shell.

The construction of these cosmological states admitting a holographic dual description should be regarded as the first main result of the present paper. 

\subsection*{Cosmology as an entanglement island}

 Given the construction of the cosmological microstates described above, in Sec. \ref{sec:3} we derive, through a standard gravitational replica trick calculation, that the entanglement entropy of the reduced state $\rho_{\Le}$ to the CFT${}_\Le$ follows a version of the quantum-corrected RT formula. We denote by $\rho_i$, $i=\lef,\ri,\Co$, the density matrices for the bulk quantum fields restricted on the time symmetric slices of the two AdS regions $\Sigma_\lef$, $\Sigma_\ri$ and of the cosmology $\Sigma_\Co$, respectively. The global state on the bulk Cauchy slice $\Sigma = \Sigma_\lef \cup \Sigma_\Co \cup \Sigma_\ri$ is pure, i.e. $S(\rho_{\lef  \Co\ri})=0$. 
 
 In the present setup, the entropy of the bulk fields is $O(G^0)$ and therefore subleading with respect to any ${O}(G^{-1})$ area term. This implies that all possible geometrically non-empty extremal surfaces are disfavored and that the RT surface is indeed empty, $\gamma_\Le=\emptyset$, for the CFT${}_\Le$.\footnote{An interesting classical extremal surface is given by the initial sphere associated to the position of the shell, at $r=R_*$, in the time reflection-symmetric slice $\Sigma$. This surface has index 1 at least, which means that the surface is maximal in at least one direction of deformation space, as opposed to the candidate RT surfaces, which are always minimal surfaces. In the regime where the quantum-corrected RT prescription is valid, the whole cosmology resembles a python's lunch, encoded in either CFT${}_\Le$ or CFT${}_\Ri$.}  However, due to the presence of the closed cosmology, it is still necessary to specify the homology hypersurface $\Sigma_{\Le}$, with $\partial \Sigma_{\Le} = \Le$. The two possibilities are:
\begin{itemize}
    \item The homology hypersurface is $\Sigma_{\Le} = \Sigma_\lef$, and it does not include the cosmology. The bulk entanglement entropy of $\rho_{\Sigma_{\Le}}$ given by this candidate is $S(\rho_{\lef})$.
    \item The homology hypersurface is $\Sigma_{\Le} = \Sigma_\lef \cup \Sigma_\Co$, which includes the full closed cosmology. The bulk entanglement entropy of $\rho_{\Sigma_{\Le}}$ given by this candidate is $S(\rho_{\lef \Co}) = S(\rho_{\ri}) $.
\end{itemize}

\begin{figure}[h]
\centering
\includegraphics[width = .8\textwidth]{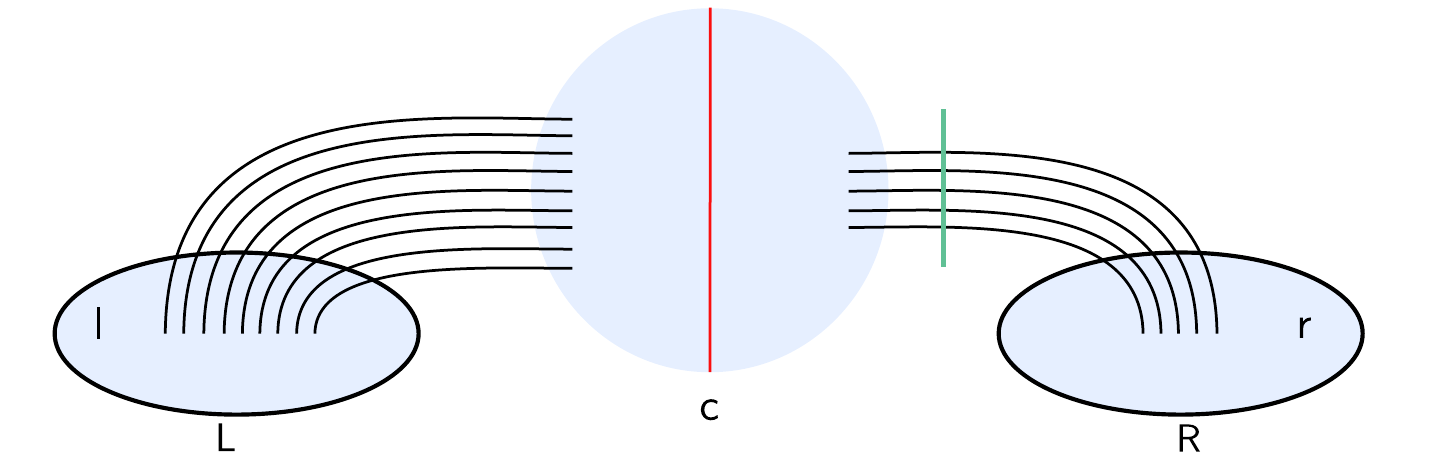}
\caption{Schematic representation of the semiclassical state. The lines connecting the cosmology to the left and right AdS spacetimes represent entanglement between bulk fields. The green line cutting the entanglement lines represents the dominant RT surface for the whole CFT${}_\Le$, which is geometrically empty. The cosmology $\Co$ is contained within the entanglement wedge of CFT${}_\Le$.}
\label{fig:gen}
\end{figure}

From these considerations, the naive application of the quantum-corrected RT prescription leads to the island formula 
\be\label{eq:RTisland}
S(\rho_{\Le})=\min \left\{S(\rho_{\lef}),S(\rho_{\lef \Co})\right\}\;.
\ee
This results suggests that whenever $S(\rho_{\lef \Co})<S(\rho_\lef)$---i.e. whenever the entanglement between AdS${}_\lef$ and $\Co_\lef$ is larger than the entanglement between AdS${}_\ri$ and $\Co_\ri$, given that $S(\rho_{\lef \Co})=S(\rho_\ri)$---the entanglement wedge of CFT${}_\Le$ will contain the cosmology $\Co$. If, on the contrary, $S(\rho_\ri)>S(\rho_\lef)$, the cosmology will be contained in the entanglement wedge of CFT${}_\Ri$. See Fig. \ref{fig:gen} for a visual representation of the former case.

We would like to remark that the situation is more subtle. The derivation of the minimization prescription of eq. \eqref{eq:RTisland} requires that one of the two terms dominates over the other, with corrections which are negligible when $\exp(|S(\rho_{\lef})- S(\rho_{\ri})|)\gg 1$. This condition is generally satisfied by the excited cosmological states we will be interested in even if the bulk entropies $S(\rho_\ri)$ and $S(\rho_\lef)$ are in general $O(N^0)$ in the large-$N$ expansion, as we will explain in Sec. \ref{sec:2.3}. If, however, the two terms become comparable or if both terms are small, then we cannot consider the cosmology to be encoded in any one of the two CFTs, but only in the union of both CFTs. For example, this is the case for the simple PETS state below the Hawking-Page temperature \eqref{eq:PETSintro}, for which $S(\rho_\lef)$ and $S(\rho_\ri)$ are generically small in string theory constructions of AdS/CFT (see Appendix \ref{app:B}), as well as in any bottom-up model, unless a parametrically large number of bulk species is added by hand.

In Sec. \ref{sec:3}, we derive the island formula \eqref{eq:RTisland} from Euclidean gravity, adopting the FLM replica trick prescription of refs. \cite{Lewkowycz:2013nqa,Faulkner:2013ana} for these particular cosmological microstates. In this case, different saddle point geometries arise from different thin shell contractions between replicated CFT contours. These provide different contributions to the R\'{e}nyi entropies of $\rho_{\Le}$, which in the end leads to \eqref{eq:RTisland} upon analytic continuation. In particular, the island term, $S(\rho_{\lef \Co})$, originates from a ``replica wormhole'' saddle point geometry with thin shell contractions between different boundary replicas. This phenomenon was first noticed in a different context in the ``west coast model'' of black hole evaporation \cite{Penington:2019kki}.  An important aspect of the present replica trick calculation is that the classical contribution of every saddle point geometry contributing to the R\'{e}nyi entropies factors out, and cancels with the semiclassical normalization of the state $\rho_{\Le}$. The remaining non-trivial contributions arise from suitable one-loop partition functions of the bulk fields on top of each geometry, leading to eq. \eqref{eq:RTisland}.

\subsection*{Cosmology-to-boundary map}

The construction of the cosmological microstates sets the stage for an interesting question: how is the cosmological EFT encoded holographically in the CFTs? We address this question in Sec. \ref{sec:4} by studying the bulk-to-boundary map between the bulk EFT and the microscopic dual CFTs. The encoding of the bulk EFT on the AdS${}_\lef$ and AdS${}_\ri$ spacetimes into the CFT${}_\Le$ and CFT${}_\Ri$, respectively, presents no difficulty and is given in terms of the standard AdS/CFT dictionary. Therefore, we are particularly interested in understanding the trickier question of how different bulk states in the cosmology $\Co$ are mapped to CFT states, and how to reconstruct the action of bulk operators on $\Co$ from the dual boundary theory.

For this reason, we focus our attention on the bulk Hilbert space $\mathcal{H}_{\psi}\subset \mathcal{H}_\Sigma$ obtained by acting on a reference state $\ket{\psi}\in \mathcal{H}_{\Sigma}$ with local unitary operators supported only on the time-symmetric slice of the cosmology $\Sigma_\Co$, where $\mathcal{H}_\Sigma$ is the full bulk EFT Hilbert space defined on the time reflection-symmetric slice $\Sigma=\Sigma_\lef\cup\Sigma_\Co\cup\Sigma_\ri$. In other words, we fix the state $\rho_{\lef\ri}$ and thus the bulk entanglement entropy between the AdS regions and the cosmology, and consider different bulk states by acting on the cosmological degrees of freedom (DOF). In this paper we will refer to the bulk theory restricted to this class of states as the ``cosmological EFT''. We call the resulting bulk-to-boundary map $V_\psi:\mathcal{H}_{\psi}\rightarrow \mathcal{H}_{\Le}\otimes \mathcal{H}_\Ri$ (with $\mathcal{H}_{\Le,\Ri}$ Hilbert spaces of the left and right CFTs) the \textit{cosmology-to-boundary} map.\footnote{Note that we will discuss the global cosmology-to-boundary map. Some of our results apply to the single-sided cosmology-to-boundary map given by entanglement wedge reconstruction, when the cosmology is encoded in only one of the two CFTs, i.e. when the island formula \eqref{eq:RTisland} holds.}

In order to simplify our analysis, we restrict our attention to operator insertions $\phi_i$ at the time reflection-symmetric slice in the cosmology that create localized particles on top of some reference bulk state $\ket{\psi}$. These particles are heavy (i.e. $m_i\ell\gg 1$), but light enough that their backreaction on the geometry can be neglected. The resulting bulk states have the form
\be\label{eq:bulkstatecintro} 
\ket{\psi_I} = \dfrac{1}{\sqrt{z_I}}\prod_i \phi_i(y_i)\ket{\psi}\;,
\ee
where $z_I$ is a normalization factor and $y_i\in\Sigma_\Co$. The condition $m_i\ell\gg 1$ guarantees that we can use the geodesic approximation in the Euclidean geometry to relate the operator insertions in the cosmology to corresponding local CFT operator insertions---which we will collectively represent by $\mathbb{O}_I(\mathbf{x}_I)$---in the Euclidean path integral preparing the reference state $\ket{\Psi}=V_\psi\ket{\psi}$ (see Fig. \ref{fig:cosmomapads}):
\be\label{eq:bdystateexcintro}
\ket{\Psi_I} = V_\psi\ket{\psi_I} =  \ket{\dfrac{1}{\sqrt{Z_I}}\mathcal{T}\lbrace \Op(\tbeta_{\Ri}/2) \mathcal{O}_\Psi(\mathbf{x}_\Psi) \mathbb{O}_I(\mathbf{x}_I) \rbrace}  \;,
\ee 
where $\mathcal{T}$ denotes Euclidean time ordering, the operator $\Op(\tbeta_{\Ri}/2)$ inserts the shell, and we collectively represented by $\mathcal{O}_\Psi(\mathbf{x}_\Psi)$ the operator insertions in the Euclidean path integral preparing the reference state $\ket{\Psi}$.

Given this simple class of bulk states and their dual boundary states, we can investigate the properties of the cosmology-to-boundary map $V_\psi$ relating the two. The first result that we derive in Sec. \ref{sec:4.1} is that $V_\psi$ is non-isometric. To show this, we consider two orthogonal bulk states $\ket{\psi_{I_i}}$, $\ket{\psi_{I_j}}$ with different configurations of particles in the cosmology. In the bulk EFT, $\bra{\psi_{I_i}}\ket{\psi_{I_j}}=0$ by assumption.\footnote{We consider states with particles in the cosmology carrying different internal DOF, so that the bulk overlap vanishes up to $1/N$-suppressed corrections due to gravitational interactions between the particles.} The CFT overlap of the dual states, $G_{ij}=\bra{\Psi_{I_i}}\ket{\Psi_{I_j}}=\bra{\psi_{I_i}}V_\psi^\dag V_\psi\ket{\psi_{I_j}}$, is however non-vanishing. Computing it semiclassically requires to use the bulk Euclidean gravitational path integral. The connected CFT overlap vanishes, $\overline{G_{ij}}=0$, where the overline indicates that we evaluated the quantity within the semiclassical gravity approximation. The square of the overlap, $\overline{|G_{ij}|^2}$, is however non-vanishing in this approximation, due to the contribution coming from an Euclidean wormhole saddle point geometry (see Fig. \ref{fig:overlapsq}). In the large mass limit, the contribution of this overlap becomes universal, independent of the details of $\ket{\psi_{I_i}}$ and $\ket{\psi_{I_j}}$. It is given by
\be
    \overline{|G_{ij}|^2}=e^{-\left(S_2(\rho_\lef)+S_2(\rho_\ri)\right)}
    \label{eq:nonvanoverlap}
\ee
where $S_2(\rho_{\lef,\ri})$ is the second R\'enyi entropy of the bulk state $\rho_{\lef,\ri}$ of the AdS${}_{\lef,\ri}$ region, respectively. Note that, as we have discussed above, $S_2(\rho_{\lef,\ri})\sim O(N^0)$ and therefore the square of the overlap is not parametrically suppressed in the large-$N$ limit. However, for general excited reference states $\ket{\psi}$ of the bulk EFT, with a large amount of entanglement between the cosmology and the AdS regions, the overlap is small.

This calculation signals that in general the CFT overlap $G_{ij}$ is non-vanishing. The bulk EFT only reproduces the average value of the CFT overlap, whose structure is pseudo-random and depends on microscopic details of the states. It however cannot capture deviations from it, even if these contributions are not parametrically suppressed in $\exp(-N^2)$, see Sec. \ref{sec:4.1} for additional details. Remarkably, the bulk Euclidean path integral is nonetheless able to reproduce some structure of these deviations, once Euclidean wormhole saddles are allowed when computing the moments. From this perspective, we can conclude that the map $V_\psi$ is non-isometric (i.e. $V^\dagger_\psi V_\psi \neq \mathbb{1}$) since it does not preserve the inner product structure of the cosmological EFT.

We want to emphasize that the non-isometricity of the cosmology-to-boundary map does not imply that such a map is non-invertible. This would only be the case in finite-dimensional toy models of $V_\psi$, where a clear ``bag of gold paradox'' arises \cite{Wheeler:1964qna,Fu:2019oyc,Penington:2019kki}. In AdS/CFT, however, the boundary description has an infinite dimensional Hilbert space, and we generally expect that orthogonal states in the bulk EFT are mapped to non-orthogonal but still linearly-independent states in the dual boundary theory, so that $V_\psi^{-1}$ exists.\footnote{We thank Mark Van Raamsdonk for discussions on this point.} The information distinguishing all of the bulk states is in principle stored on thermal tails of the CFT wavefunction of these states (see Sec. \ref{sec:4.2} for additional details on this point). In this way, all of the bulk EFT states can at least in principle be distinguished by a boundary observer, provided that this observer can probe the CFT states to an arbitrary accuracy.

On the other hand, for CFT observers with finite resolution, such as those only sensitive to a specific microcanonical band, a bag of gold paradox does arise. Under this restriction, we show in Sec. \ref{sec:4.2} that the universal overlap \eqref{eq:nonvanoverlap} and its higher moments are consistent with the fact that the cosmology-to-boundary map, restricted to a microcanonical window, has a non-trivial kernel and is therefore non-invertible. Intuitively, ignoring the information contained in thermal tails of the microscopic states, and restricting to a CFT microcanonical window, null states appear and many of the cosmological states look linearly dependent as an artifact of this truncation. We further show that the overlap \eqref{eq:nonvanoverlap} and its higher moments are consistent with the low-energy microcanonical density of states of the CFTs. By choosing a number of bulk EFT orthogonal states $K>e^{\textbf{S}}$ (where $\textbf{S}=S_\lef+S_\ri$ with $S_{\lef,\ri}$ bulk microcanonical entropies in AdS${}_{\lef,\ri}$) the dual CFT microstates span the microcanonical window.

Finally, in Sec. \ref{sec:4.3}, we turn our attention to the reconstruction of bulk operators acting on the cosmology. Consider a bulk operator $\phi(y)$ acting on a bulk state $\ket{\psi_I}$ and creating a particle at a point $y\in \Sigma_\Co$ in the cosmology, and the corresponding boundary operator $\Phi=V_\psi \phi V_\psi^{-1}$ acting on the CFT state $\ket{\Psi}$. In standard AdS/CFT, the bulk operator $\phi$ is dual to a given CFT generalized free field $\mathcal{O}_\phi$ which is identified by means of the extrapolate dictionary. One then writes down a smeared version of the generalized free field, with an appropriate kernel, which provides the HKLL representation of the local bulk operator in the CFT \cite{Hamilton:2005ju,Hamilton:2006az}. In the case of the cosmology, any causal reconstruction is doomed to fail, since the cosmology $\Co$ is geometrically disconnected from the asymptotic boundaries where the dual theory is defined.

In order to obtain an explicit reconstruction of the bulk operator, we can use the fact that the state $\Phi\ket{\Psi_I}=V_\psi \phi(y)\ket{\psi_I}$ can be prepared by a Euclidean path integral of the form \eqref{eq:bdystateexcintro} with the addition of an extra operator insertion $\mathcal{O}_\phi$ at a specific Euclidean time $\tau_\phi$.\footnote{For a generic bulk operator $\phi$ with small mass, the corresponding CFT operator insertion in the Euclidean path integral is non-local. We will again restrict for simplicity to the case of operators that can be treated under the geodesic worldline approximation, for which a local insertion of $\phi$ in the bulk corresponds to a local insertion of $\mathcal{O}_\phi$ in the Euclidean boundary.} The matrix elements of the CFT operator $\Phi$ then take the form
\be
\Phi_{IJ} = \bra{\Psi_I}V_\psi \phi(y) V_\psi^{-1}\ket{\Psi_J} = \dfrac{1}{\sqrt{Z_IZ_J}}\text{Tr}(\mathcal{T}\lbrace \Op(\tbeta_{\Ri}/2)\mathcal{O}_\Psi(\mathbf{x}_\Psi) \mathbb{O}_J(\mathbf{x}_J) \mathcal{O}_\phi(\mathbf{x}_\phi) \Op^\dagger(-\tbeta_{\Ri}/2)\mathcal{O}^\dagger_\Psi(\mathbf{x}_\Psi')\mathbb{O}_I^\dagger(\mathbf{x}'_I)\rbrace)\;.
\label{eq:matrixintro}
\ee 
In this expression, the Euclidean time ordering $\mathcal{T}$ of the CFT operator insertions is crucial for the correct definition of the matrix elements. Since different bulk states correspond to insertions of different operators $\mathcal{O}_\Psi,\mathbb{O}_I$ at different Euclidean times $\tau_\Psi,\tau_I$, the right-hand-side of eq. \eqref{eq:matrixintro} depends on the states $\ket{\Psi_I}$ and $\ket{\Psi_J}$, and the matrix elements of the CFT operator $\Phi$ need to be defined one by one. In this sense, the explicit Euclidean reconstruction \eqref{eq:matrixintro} of the bulk operators acting on the cosmology is manifestly state-dependent.

The properties of the cosmology-to-boundary map $V_\psi$ outlined above and the state-dependent reconstruction of the cosmological EFT operators described above should be regarded as the second main result of the present paper.

\subsection*{Tensor network toy model}

\begin{figure}[h]
\centering
\includegraphics[width = .8\textwidth]{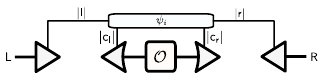}
\caption{TN model of our setup defining the cosmology-to-boundary map $V_\psi$.}
\label{fig:TNintro}
\end{figure}

In order to study the properties of the cosmology-to-boundary map more explicitly, we introduce a tensor network (TN) toy model for the cosmological states that qualitatively captures all of the properties of the AdS/CFT setup described above, see Fig. \ref{fig:TNintro}.

In this TN, described in detail in Sec. \ref{sec:5}, the triangles on the left and right represent MERA TNs modeling the initial slices of the AdS${}_{\lef,\ri}$ spacetimes; the triangles with a curved edge are MERAs cut off at $r=R_*$ which represent the $\Co_{\lef,\ri}$ initial slices; the projection labeled by $\mathcal{O}$ glues the two cosmological MERAs together and has the interpreation of the shell operator $\mathcal{O}$; the lines labeled by $|\lef|=2^{S_\lef},|\ri|=2^{S_\ri}$ represent the EPR pairs between the AdS$_{\lef,\ri}$ regions and the cosmology; the lines labeled by $|\Co_{\lef,\ri}|$ represent bulk DOF on the cosmology, including those entangled with AdS${}_{\lef,\ri}$ and additional, DOF which are not entangled; finally, the box labeled by $\psi_i$ represents a bulk state $\ket{\psi_i}$ on $\Sigma=\Sigma_\lef\cup\Sigma_\Co\cup\Sigma_\ri$. We can tune the entanglement between the AdS${}_{\lef,\ri}$ spaces and the cosmological components $\Co_{\lef,\ri}$, by increasing the number of EPR pairs by hand, given by the parameters $S_{\lef,\ri}$ of the TN. Note that the state $\ket{\psi}$ is a state defined on \textit{all} of the bulk DOF, and in particular also on the AdS${}_{\lef,\ri}$ DOF which are not entangled with the cosmology (which we did not represent here for simplicity, given that they are irrelevant for our discussion).

Moreover, we choose the tensor $\mathcal{O}$ to be Gaussian random, with zero mean and unit variance.\footnote{This assumption is motivated by the coarse-grained semiclassical description of the thin shell operator $\mathcal{O}$ in terms of a Gaussian random matrix in the energy basis \cite{Sasieta:2022ksu}.} With this choice, a TN replica trick calculation shows that this random TN approximately saturates the RT formula on the TN, which is the analog of the island formula \eqref{eq:RTisland} described above for the AdS/CFT setup. Interestingly, the island term in the TN replica trick arises from a replica-connected contribution present when taking averages over the $\mathcal{O}$ tensor, in analogy with the gravitational replica wormhole contribution in AdS/CFT.

Given the TN of Fig. \ref{fig:TNintro}, we consider general states $\ket{\psi_i}$ of the cosmological EFT, that is, those states generated by acting on a reference state $\ket{\psi}$ with local unitaries $U_{\Co}^{(i)}$ supported on the DOF of the cosmology. The TN in Fig. \ref{fig:TNintro} defines a cosmology-to-boundary map
\be\label{eq:linearctbmapintro}
V_\psi\ket{\psi_i} = I_{\lef}I_{\ri}(\Pi_{\text{MAX}}^\lef \Pi_{\text{MAX}}^\ri\bra{\Pi_{\Op}}_{\Co})\ket{\psi_i}\;,
\ee
where $\bra{\Pi_{\Op}}_{\Co} = \bra{\Op}I_{\Co_\lef}I_{\Co_\ri}$ is the ``cosmological projector'' associated with the random shell tensor and the $I_{\Co_{\lef,\ri}}$ cutoff MERA isometries, $I_{\lef,\ri}$ are the AdS MERA isometries, and $\Pi^i_{\text{MAX}} = \ket{\text{MAX}_i}\bra{\text{MAX}_i}$ is the orthogonal projector to the maximally entangled state of $S_{\lef,\ri}$ pairs between AdS${}_{\lef,\ri}$ and $\Co_{\lef,\ri}$, respectively. 

In general, the dimension of the Hilbert space of cosmological DOF can be larger than the entanglement between the cosmology and the AdS regions, i.e. $|\Co_\lef||\Co_\ri|>|\lef||\ri|$. Since the TN defines a linear map between finite-dimensional Hilbert spaces, this implies that, in this regime, there are null states in the cosmological EFT, i.e. $\text{dim}\{\text{Ker}(V_\psi)\}=|\Co_\lef||\Co_\ri|-|\lef||\ri|>0$, and the map $V_\psi$ is non-isometric. Thus, unlike the previous discussion in AdS/CFT, the TN model of the cosmology-to-boundary map in this case has a non-trivial kernel. This is a consequence of the finite dimensional nature of the TN, in which high energy states have been projected out and thus the information contained on thermal tails (which in principle enables to distinguish different cosmological states in the CFT) is lost. The TN should be regarded as a model of what a CFT observer with finite resolution can distinguish, in analogy with the situation described above in which we truncate the CFT Hilbert space by projecting onto a microcanonical band of the CFT.

By exploiting the gaussian isoperimetric inequalities for the random $\mathcal{O}$ tensor, we can estimate under which condition the map $V_\psi$ is approximately isometric for all states that a computationally limited bulk observer can prepare (both the notion of approximate isometry and the definition of these ``simple'' states are made precise in Sec. \ref{sec:5.2}) with extremely high probability. We find that this is the case if
\be\label{eq:introtrans}
    |\lef||\ri|\gtrsim (\log |\Co|)^q
\ee
where $|\Co|=|\Co_\lef||\Co_\ri|$ and $q$ is a constant given in eq. \eqref{eq:complexityobserver}. This result shows that in order to have an approximately isometric encoding of all ``simple'' cosmological states, it is enough that the number of entangled pairs $S_{\lef}+S_{\ri}$ between cosmology and AdS regions scales logarithmically with the number of DOF in the cosmology.  

We expect that these considerations to also hold under genericity assumptions for the full-fledged cosmology-to-boundary map in AdS/CFT. A simple estimate of the cosmological dimension in this case is $\log_2 |\Co| \sim 2N^p$, for $p<2$, associated to the microcanonical entropy of two gases of particles in AdS, one on each of the patches $\Co_\lef$ and $\Co_\ri$. The regime \eqref{eq:introtrans} is still within the microcanonical stability of the gas of particles in AdS, and it can therefore be attained in our construction.

The non-isometric to approximately isometric transition for the cosmology-to-boundary map, when the entanglement to the cosmology exceeds the bound \eqref{eq:introtrans}, should be regarded as the third main result of the present paper.

\subsection*{Organization of the paper}

The rest of the paper is organized as follows. In Sec. \ref{sec:2} we present a detailed construction of the cosmological states in the bulk and boundary theories and describe the properties of the cosmological spacetime. In Sec. \ref{sec:3} we derive the island formula from the FLM replica trick calculation for these states. In Sec. \ref{sec:4} we analyze the properties of the cosmology-to-boundary map restricted to specific states prepared via the insertion of localized matter operators in the CFT Euclidean path integral, and provide an explicit state-dependent reconstruction of operators acting on the cosmology. In Sec. \ref{sec:5} we present a TN toy model of our setup and analyze generic properties of the cosmology-to-boundary map using this model. We end with a discussion of our results in Sec. \ref{sec:6}. Technical details related to our results can be found in Appendices \ref{app:hawkingpage}-\ref{app:E}.

\section{Cosmological states}
\label{sec:2}

To construct CFT states with closed cosmological duals, we will be guided by the construction of semiclassical states of a two-sided black hole with long Einstein-Rosen bridges. It is well-known that such states can be created by adding heavy matter with substantial backreaction in the black hole interior. In JT gravity coupled to matter particles, this class of states were dubbed \textit{partially entangled thermal states} (PETS) in \cite{Goel:2018ubv}. More recently, analogous microstates have been considered in a variety of different contexts, including extremal supersymmetric black holes \cite{Lin:2022rzw,Lin:2022zxd} and higher dimensional neutral black holes \cite{Sasieta:2022ksu,Balasubramanian:2022gmo, Chandra:2023rhx} (see also \cite{Kourkoulou:2017zaj,Chandra:2022fwi,Balasubramanian:2022lnw} for one-sided analogs and B-states).

\begin{figure}[h]
\centering
\includegraphics[width = .5\textwidth]{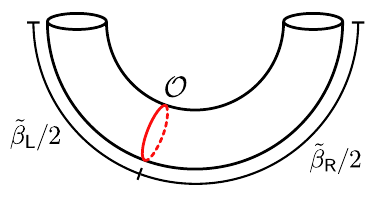}
\caption{Euclidean path integral on the flat cylinder preparing $\ket{\Psi_{\Op}}$.}
\label{fig:PETS}
\end{figure}

A PETS is a state on the Hilbert space of two independent copies of the CFT, $\ket{\Psi_{\Op}} \in \mathcal{H}_{\Le} \otimes \mathcal{H}_{\Ri}$, where each CFT lives on a spatial $\mathbf{S}^{d-1}$ of radius $\ell$. The state is prepared by an Euclidean CFT path integral on a flat cylinder of finite length (see Fig. \ref{fig:PETS}). The preparation includes two portions of Euclidean time-evolutions of lengths $\tbeta_{\Le}/2$ and $\tbeta_{\Ri}/2$, driven by the CFT Hamiltonian $H = H_{\Le} = H_{\Ri}$, and the additional insertion of, at least, a matter operator $\mathcal{O}$ in between. The wavefunction of the state in the energy basis reads 
\be\label{eq:PETS}
\ket{\Psi_{\Op}} = \ket{\rho_{\tbeta_{\Le}/2}\Op\rho_{\tbeta_{\Ri}/2}}\,= \dfrac{1}{\sqrt{Z_1}}\sum_{n,m} e^{-\frac{1}{2}(\tbeta_{\Le} E_n + \tbeta_{\Ri} E_m)} \Op_{nm}\ket{E_n}^*_{\Le}\otimes \ket{E_m}_{\Ri}\;,
\ee 
where the state has been normalized by a suitable thermal two-point function, $Z_1 = \text{Tr}(e^{-\tbeta_{\Le} H}\mathcal{O}^\dagger e^{-\tbeta_{\Ri} H}\mathcal{O})$.

In this paper, we consider an operator $\mathcal{O} \sim \prod_{i} \mathcal{O}_{\Delta}(\Omega_i)$ comprised of a large number of single-trace operator insertions $\mathcal{O}_{\Delta}$ along the sphere $\Omega_i \in \mathbf{S}^{d-1}$, distributed in an approximately homogeneous way.\footnote{We can also include different conformal primaries, see Appendix \ref{app:hawkingpage}.} In the limit of a large number of insertions, these operators create classical spherically symmetric domain walls of dust particles in the bulk (see \cite{Anous:2016kss}). We will work with an effective hydrodynamic description of these shells of dust, characterizing them as localized pressureless perfect fluids with mass density $\sigma$. The bulk EFT that we will consider to describe this system is
\be \label{eq:actionEFT}
I[X]=-\frac{1}{16\pi G}\int_X (\, R -2\Lambda ) + \dfrac{1}{8\pi G} \int_{\partial X} K + I_{\text{matter}}[X] + \int_{\mathcal{W}} \sigma + I_{\text{ct}}[\partial X] \;,
\ee
where $I_{\text{matter}}[X]$ is the action of the bulk quantum fields, $\Ws$ is the codimension-one worldvolume of the thin shell and $I_{\text{ct}}[\partial X]$ are local counterterms at asymptotic infinity that implement holographic renormalization and render the value of the action finite \cite{deHaro:2000vlm,Skenderis:2002wp}.

The semiclassical state dual to $\ket{\Psi_{\Op}}$ is prepared, according to the rules of the holographic dictionary, by the dominant bulk Euclidean saddle point geometry $X$ satisfying the boundary conditions imposed by the CFT path integral which computes the two-point function $Z_1$. The time reflection-symmetric slice $\Sigma$ of the Euclidean solution $X$, provides the initial data of the semiclassical state and singles out a particular state of the bulk fields.

\begin{figure}[h]
\centering
\includegraphics[width = .75\textwidth]{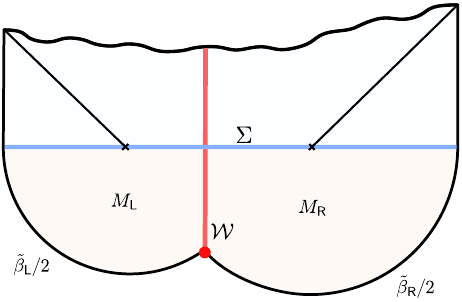}
\caption{Semiclassical state prepared by the gravitational path integral at high preparation temperatures. The Euclidean section $X$ consists of two Euclidean black holes of masses $M_{\Le}$ and $M_{\Ri}$. Each solution is topologically $\mathbf{D}^2 \times \mathbf{S}^{d-1}$, where $\mathbf{D}^2$ is a two dimensional disk (each point in the figure corresponds to a transverse $\mathbf{S}^{d-1}$). The two solutions are  glued together along the worldvolume of the thin shell $\Ws$, in red. The semiclassical bulk state is prepared at the time reflection-symmetric slice $\Sigma$ (blue slice), which contains a long Einstein-Rosen bridge supported by the shell. In the Lorentzian section, the shell hides behind the horizon of the two-sided black hole.}
\label{fig:bh}
\end{figure}

At high preparation temperatures, $\tbeta_{\Le},\tbeta_{\Ri} \lesssim \ell$, the dominant Euclidean geometry $X$ is constructed out of the Euclidean Schwarzschild AdS black hole solution, of topology $\mathbf{D}^2 \times \mathbf{S}^{d-1}$, where $\mathbf{D}^2$ is a two dimensional disk, transverse to the spatial spheres. $X$ is formed out of two black hole solutions, with different masses $M_{\Le},M_{\Ri}$. They are glued together along the worldvolume of the thin shell $\Ws$, which backreacts in this localized way (see Fig. \ref{fig:bh}).\footnote{As shown in \cite{Balasubramanian:2022gmo}, the preparation moduli of the cylinder $\tbeta_{\Le},\tbeta_{\Ri}$ can always be chosen such that the thin shell is prepared in the black hole interior, irrespective of its rest mass $m$, for given fixed physical temperatures $\beta_{\Le},\beta_{\Ri}$ of the two black holes. To be more precise, there is a lower bound for the mass of the shell set by the difference in ADM masses, $|M_{\Le}-M_{\Ri}|\leq m$. There is no upper bound for $m$ in these solutions, which allows to explore the Hilbert space of the black hole by varying the value of $m$ for fixed $M_{\Le}, M_{\Ri}$ (cf. \cite{Balasubramanian:2022gmo}).}. The semiclassical state on $\Sigma$ corresponds to the initial data of two AdS-Schwarzschild black holes with a shared Einstein-Rosen bridge, which is supported by the presence of the spherical heavy shell of matter. The CFT wavefunction \eqref{eq:PETS} in this regime is dominated by high-energy microcanonical bands, with energies associated to the ADM masses $M_{\Le},M_{\Ri}$ of the two large AdS black holes.

Our initial motivation is the observation in \cite{Sasieta:2022ksu} that AdS solutions with the heavy shell operator $\mathcal{O}$ inserted also exist and dominate below the analog of the Hawking-Page preparation temperature. A detailed analysis of the dominance of the different phases in our setup can be found in Appendix \ref{app:hawkingpage} together with the discussion of other possible competing configurations. At low preparation temperatures, $\tbeta_{\Le}, \tbeta_{\Ri}\gtrsim \ell$, the support of the wavefunction \eqref{eq:PETS} is dominated by AdS microcanonical bands, dual to low-energy states in the confined phase of the holographic CFT. The dominant $X$ now consists of two Euclidean thermal AdS spaces, each of topology $\mathbf{D}^d\times \mathbf{S}^1$ (where $\mathbf{D}^d$ is a $d$-dimensional ball), which are again glued along $\Ws$ (see Fig. \ref{fig:cosm}). The time reflection-symmetric slice $\Sigma$ of this saddle is geometrically disconnected. It contains three connected components: two initial slices of global AdS ($\Sigma_{\lef},\Sigma_{\ri}$), and a third closed initial slice, $\Sigma_{\Co}$, supported by the presence of the heavy shell. Under Lorentzian evolution, $\Sigma_{\Co}$ develops a big-crunch singularity towards the future, and arises from a big-bang singularity in the past. In this way, the geometrically disconnected space corresponds to the configuration of a semiclassical closed AdS cosmology, microscopically described by the low-temperature PETS \eqref{eq:PETS} of two holographic CFTs. We now move onto the detailed construction of these states.

\subsection{Semiclassical state}

In the EFT \eqref{eq:actionEFT}, the shell propagates freely at tree level, following a codimension-one worldvolume $\mathcal{W}$ which bisects the Euclidean geometry $X$ into two components $X_\pm$.  To leading order, we shall neglect the backreaction exerted by the state of the quantum fields on $X$. At this level, the metric on $X_\pm$ will be that of thermal AdS
\be\label{eq:metricpm}
\text{d}s_\pm^2 =  f(r) \text{d}\tau_\pm^2 + \dfrac{\text{d}r^2}{f(r)}+ r^2 \text{d}\Omega_{d-1}^2\;,
\ee
where
\be 
f(r) = \frac{r^2}{\ell^2} +1\;.
\ee 
Each Euclidean time coordinate is periodic with periodicity given by the physical inverse temperature of the corresponding AdS space, $\tau_- \sim \tau_- + \beta_{\Le}$ and $\tau_+ \sim \tau_+ + \beta_{\Ri}$. The radial coordinate must remain continuous across the gluing of these two solutions across $\Ws$, given that it measures the proper areas of transverse spheres. It therefore has the same label for both components in \eqref{eq:metricpm}.

From spherical symmetry of $\mathcal{W}$, the embedding functions of $\Ws$ in $X_\pm$ reduce to $ \tau_\pm = \tau_\pm(T)$ and $ r = R(T) $, where $T$ is the thin shell's proper Euclidean time, and $R(T)$ is its radial trajectory. The rest of the angular coordinates over $\mathbf{S}^{d-1}$ are inherited from the metrics \eqref{eq:metricpm}. The induced metric on $\Ws$ is simply
\be\label{eq:inducedmetric}
\text{d}s^2_{\Ws} = \text{d}T^2 + R(T)^2 \text{d}\Omega_{d-1}^2\;.
\ee 

The gluing of $X_\pm$ is performed solving Einstein's field equations locally in a neighbourhood of $\mathcal{W}$. This leads to the so-called \textit{Israel-Darmois junction conditions} \cite{Darmois:1927,Israel}. The first condition states that the metric should remain continuous along $\Ws$, which, from \eqref{eq:metricpm} and \eqref{eq:inducedmetric}, translates to
\be\label{eq:firstcondition}
1 = f(R) \dot{\tau}_\pm^2 + \dfrac{\dot{R}^2}{f(R)}\;,
\ee 
where the dot represents $\frac{\text{d}}{\text{d}T}$. The second junction condition quantifies how the normal derivative of the induced metric jumps across $\mathcal{W}$. The total jump is given by the energy-momentum tensor of the thin shell, which from \eqref{eq:action} has dust form, $T_{\mu\nu}|_{\Ws} = \sigma u_{\mu} u_{\nu}$, for $u_\mu = (\text{d}T)_\mu$ the proper velocity of each particle forming the shell. Combined with \eqref{eq:firstcondition} this condition yields (see e.g. \cite{Sasieta:2021pzj})
\be\label{eq:secondcondition}
\left(\sign(\dot{\tau}_+) - \sign(\dot{\tau}_-)\right)\sqrt{-\dot{R}^2 + f(R)} = \dfrac{8\pi G m}{(d-1)V_\Omega R^{d-2}}\;.
\ee 
In this expression we have introduced the total rest mass of the shell
\be
m = V_\Omega R^{d-1} \sigma\;,
\ee 
where $V_\Omega = \text{Vol}(\mathbf{S}^{d-1})$. The total rest mass $m$ is conserved by virtue of the angular components of the second junction condition. It correspond to an intrinsic property which characterizes the shell operator used in the construction of the state.\footnote{In the microscopic description of the shell, given by the CFT operator $\Op$, the parameter $m$ is related to the total conformal dimension $m \sim \Delta_{\Op} \sim n \Delta$, where $n\sim N^2$ is the number of local operator insertions that form $\Op$, each of conformal dimension $\Delta$.}

Squaring \eqref{eq:secondcondition} produces the equation of motion of a non-relativistic particle with zero total energy, moving in one dimension, 
\be\label{eq:effectiveeom}
\dot{R}^2 + V_{\text{eff}}(R) = 0\;,
\ee 
which describes the radial motion of the thin shell. The effective potential is
\be \label{eq:veffeuc}
V_{\text{eff}}(R) = -f(R) + \left( \dfrac{4\pi G m}{(d-1)V_\Omega R^{d-2}}\right)^2\;.
\ee 
The second term of the effective potential corresponds to the gravitational self-energy of the shell, which creates a repulsive force in the Euclidean section. This makes the shell bounce at a given radius $R_*>0$, satisfying $V_{\text{eff}}(R_*) =0$.\footnote{Setups similar to those described in this paper cannot be achieved using holographic BCFT constructions dual to spacetimes containing end-of-the-world (ETW) branes like in \cite{Takayanagi:2011zk,Fujita:2011fp,Cooper:2018cmb,Antonini:2019qkt,Fallows:2022ioc,Ross:2022pde}. In fact, at least in these minimal models, the brane's equation of state is that of constant proper energy density (its stress-energy tensor is proportional to the induced metric), leading to a brane's trajectory without turning points when the background is pure AdS: the brane simply shrinks from $r=\infty$ to $r=0$ in Euclidean signature. Therefore, the insertion of two branes at symmetric Euclidean times leads to two disconnected ETW branes (see \cite{Cooper:2018cmb} for an explicit construction), and the corresponding Lorentzian manifold does not contain a disconnected component (the cosmology) like in the present setup.} The shell's trajectory then starts from the Euclidean asymptotic boundary at $R = \infty$, dives into the bulk until it turns around at $R=R_*$, and then goes back to $R = \infty$ again (see Fig. \ref{fig:cosm}).\footnote{For $d=2$, these solutions only exist for $R_*>0$ if $2Gm >1$. In higher dimensions, they always exist.}

The turning radius $R_*$ is a function of the shell's mass, $R_* = R_*(m)$. For large mass in AdS units, $m \ell  \gg 1$, the turning radius scales as
\be\label{eq:rstar}
\dfrac{R_*}{\ell_P} \sim (m\ell)^{\frac{1}{d-1}} \;,
\ee 
measured in units of the $(d+1)$-dimensional Planck length $\ell_P = G^{\frac{1}{d-1}}$. The classical solution can thus be trusted if the shell is parametrically heavy in AdS units. In this case, its mass density $\sigma$ remains parametrically smaller than the Planck density along $\Ws$.

The total Euclidean time elapsed by the shell from the point of view of each component $X_\pm$ can be obtained by plugging \eqref{eq:effectiveeom} into \eqref{eq:firstcondition}. It is given by
\be 
\label{eq:shifttime}
\Delta \tau_{\pm}  = 2 \int_{R_*}^{\infty}\dfrac{\text{d}R}{f(R)} \, \sqrt{\dfrac{f(R) + V_{\text{eff}}(R)}{- V_{\text{eff}}(R)}}\;.
\ee 
Note that the value of \eqref{eq:shifttime} only depends on $m$. 

Moreover, we squared \eqref{eq:secondcondition} to reduce it to the form \eqref{eq:effectiveeom}. Plugging \eqref{eq:effectiveeom} back into \eqref{eq:secondcondition} imposes the additional requirement that the trajectory of the shell satisfies
\be
\begin{split}
\sign(\dot{\tau}_+) = +1 \;, \\
\sign(\dot{\tau}_-) = -1 \;.
\end{split}
\label{eq:timeeuclid}
\ee
These gluing conditions can only be satisfied if the gluing is done as indicated in Fig. \ref{fig:cosm}. Namely, the Euclidean time has to be running in the opposite directions in both components, with the correct overall sign imposed by \eqref{eq:timeeuclid}. This enforces  $-\frac{1}{4}\beta_{\Le} <\tau_- (T) < \frac{1}{4}\beta_{\Le}$ and $\frac{1}{4}\beta_{\Ri} <\tau_+ (T) < \frac{3}{4}\beta_{\Ri}$, where the Euclidean time runs clockwise in both components.

The Euclidean asymptotic boundary $\partial X$ of the solution constructed this way includes two portions of a thermal circle, of lengths $\tbeta_{\Le}$ and $\tbeta_{\Ri}$, corresponding to (twice) the inverse preparation temperatures of the dual microscopic state. These two portions are glued along the two spheres where $\Ws$ meets the boundary, which correspond to the heavy operator insertions in the Euclidean CFT path integral. In the bulk, the physical temperatures of each component is fully determined by the preparation temperatures, and by the additional Euclidean time elapsed by the shell on each component,
\be
\begin{split}
\beta_{\Ri} = \tbeta_{\Ri} + \Delta \tau_+\;,\\
\beta_{\Le} = \tbeta_{\Le} + \Delta \tau_-\;.
\end{split}
\label{eq:phystemps}
\ee

The saddle point geometry $X$ prepares a bulk semiclassical state at the Euclidean time reflection-symmetric slice
\be 
\Sigma = \Sigma_{\lef}\, \cup \,\Sigma_{\Co} \, \cup\,\Sigma_{\ri} \;.
\ee
The slice $\Sigma$ is now disconnected and contains $\Sigma_{\lef} = \lbrace \tau_- =0 \rbrace$ and $\Sigma_{\ri} = \lbrace \tau_+ =0\rbrace$, which consist of two hyperbolic disks $\mathbf{H}^{d}$, providing the initial data of two disconnected AdS spaces. The third connected component is
\be 
\Sigma_\Co = \lbrace \tau_- =-\beta_{\Le}/2\rbrace \cup \lbrace \tau_+ = \beta_{\Ri}/2\rbrace \;,
\ee 
and corresponds to the initial slice of a closed universe, of topology $\mathbf{S}^{d}$, comprised of two hyperbolic disks, $\Sigma_{\Co_{\lef}} = \lbrace \tau_- =-\beta_{\Le}/2\rbrace$ and $\Sigma_{\Co_{\ri}} = \lbrace \tau_+ =\beta_{\Ri}/2\rbrace$. The disks are regulated and glued together at a finite radial distance $r = R_*$---which is the same for both disks---precisely associated to the position of the thin shell.

The total proper volume of $\Sigma_\Co$ is controlled by the mass of the thin shell
\be 
\dfrac{\text{Vol}(\Sigma_{\Co})}{G\ell} = \dfrac{2 V_\Omega}{G\ell} \int_0^{R_*} \dfrac{r^{d-1}\text{d}r}{\sqrt{f(r)}}  \sim \dfrac{8\pi}{(d-1)^2} m\ell  \;.
\label{eq:spatialvol}
\ee 
Again, we see that the symmetric slice $\Sigma_{\Co}$ of the closed universe is macroscopic and the Euclidean manifold $X$ is under semiclassical control if the shell is parametrically heavy in AdS units.

\subsection*{Limit of large mass}

\begin{figure}[h]
\centering
\includegraphics[width = .9\textwidth]{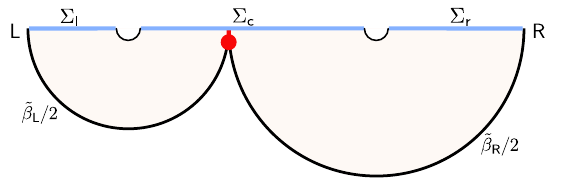}
\caption{Half of the Euclidean manifold $X$ in the limit of large mass for the thin shell operator. In this illustration, the trajectory of the shell pinches off to the asymptotic region, $\Delta \tau_\pm \rightarrow 0$, and the physical preparation time becomes $\beta_{\Le,\Ri} = \tbeta_{\Le,\Ri}$. The slice $\Sigma_\Co \subset \Sigma$ corresponds to the initial data of a very large closed universe.}
\label{fig:largemass}
\end{figure}

A regime which allows us to simplify technical details considerably, and that we shall adopt at different points throughout this paper, is the case of a very heavy shell, treated in the limit $m\ell \rightarrow \infty$.  In this limit, the Euclidean time traveled by the shell vanishes, $\Delta \tau_\pm \rightarrow 0$ by virtue of \eqref{eq:shifttime}, given that the turning radius $R_*$ in \eqref{eq:rstar} becomes infinite. The physical temperatures solving \eqref{eq:phystemps} are $\beta_{\Le} = \tbeta_{\Le}$ and $\beta_{\Ri}=\tbeta_{\Ri}$ in this regime.  The worldvolume $\Ws$ then effectively ``pinches off'' and the Euclidean solution $X$ in Fig. \ref{fig:cosm} becomes effectively disconnected, consisting of two complete Euclidean thermal AdS spaces glued at a point of the diagram (see Fig. \ref{fig:largemass}).\footnote{The large mass limit is not a pinching limit for $X$ in the physical sense, since the proper distance travelled by $\Ws$ remains infinite.} The volume \eqref{eq:spatialvol} of the reflection-symmetric slice of the closed universe $\Sigma_{\Co} \subset \Sigma$ becomes also infinite in this limit.

\subsection*{State of the bulk fields}

The Euclidean gravitational path integral, in the saddle point approximation provided by $X$, prepares a specific state for bulk quantum fields on $\Sigma$, which we shall denote by $\ket{\psi_0}$. Since $X$ is connected, the bulk preparation is driven by a (time-dependent) Hamiltonian which couples the quantum fields on $\Sigma_\Co$ and on $\Sigma_\lef$ and $\Sigma_\ri$. The outcome of this coupling is to generally produce an entangled state $\ket{\psi_0}$ between the three components.

In the large mass limit, the pinching of $X$ makes the two components $X_\pm$ approximately decouple, since the time-dependent Hamiltonian coupling them acts for a very small Euclidean time $\Delta \tau_\pm \rightarrow 0$. Therefore, generically, we can assume that the state of the bulk fields $\ket{\psi_0}$ manifestly factorizes into the product
\be\label{eq:statebulkfact}
\ket{\psi_0} \approx \ket{\Phi_-}_{\lef\Co_{\lef}}  \ket{\Phi_+}_{\Co_{\ri}\ri}\;,
\ee 
where $\ket{\Phi_\pm}$ correspond to two entangled states between the respective side of the closed universe and the AdS region. The approximate factorization of the bulk state \eqref{eq:statebulkfact} implicitly assumes that the entanglement entropy of any such states $\ket{\Phi_\pm}$ is much greater than the very small residual entanglement entropy between both sides of the closed universe $\Sigma_\Co$. As we will see in Sec. \ref{sec:2.3}, depending on the bulk theory, this might not be the case for the semiclassical bulk state prepared by $X$, given that the preparation temperature is below the gap of the system. In these cases, the bulk state $\ket{\psi_0}$ could retain a more complicated entanglement structure, even in the large mass limit. However, in this paper, we will generally consider excited states of the bulk fields $\ket{\psi}$ on $\Sigma$, prepared by the addition of bulk local operator insertions in the same Euclidean preparation manifold $X$ (see Fig. \ref{fig:cosmogeneral}). Generally such states contain additional entanglement and will satisfy \eqref{eq:statebulkfact} in the large mass limit.

\subsection*{Lorentzian geometry and inhomogeneous cosmology}

The Lorentzian manifold $X^L$ corresponding to the initial data prepared on $\Sigma$ is obtained from analytic continuation $\tau_\pm \rightarrow i t_\pm$ of the Euclidean manifold $X$. This leads to a real time reversal-symmetric geometry $X^L$. $X^L$ contains two disconnected global AdS components AdS${}_\lef$, AdS${}_\ri\subset X^L$, with metric 
\be\label{eq:metricpmL}
\text{d}s_{\lef,\ri}^2 =  -f(r) \text{d}t_\pm^2 + \dfrac{\text{d}r^2}{f(r)}+ r^2 \text{d}\Omega_{d-1}^2\;,
\ee
corresponding to the evolution of the initial data on $\Sigma_\lef$ and $\Sigma_\ri$, respectively.

Additionally, there is a third connected component $\Co \subset X^L$ given by the causal development of the initial data on $\Sigma_{\Co}$. This development is comprised of two patches of pure AdS space, $\Co_\lef$ and $\Co_\ri$, which are glued together along the worldvolume $\Ws^L$ of the thin shell in the Lorentzian section, as illustrated in Fig. \ref{fig:cosm}. The metric of each one of these patches is locally given by a pure AdS metric of the form \eqref{eq:metricpmL}, with a time-dependent IR cutoff at the trajectory of the shell $r=R(t_i)$, for $t_i$ the corresponding global time variable defined on each patch $i=\Co_\lef,\Co_\ri$. The two patches $\Co_\lef$ and $\Co_\ri$ are geometrically identical, but the state of the bulk fields living on them is different (in particular, the temperatures $\beta_\Le$ and $\beta_\Ri$ of the thermal gas in the two patches is different).

The shell's equation of motion is obtained from the analytic continuation $T\rightarrow iT$ of its proper Euclidean time in \eqref{eq:effectiveeom}. The corresponding equation for the trajectory of the shell 
\be\label{eq:effectiveeomL}
\dot{R}^2 + V^L_{\text{eff}}(R) = 0\;
\ee 
is indeed equivalent to the Lorentzian junction condition, given the effective potential
\be 
V^L_{\text{eff}}(R) = f(R) - \left( \dfrac{4\pi G m}{(d-1)V_\Omega R^{d-2}}\right)^2\;.
\ee 
Note that this potential has a minus sign with respect to the Euclidean potential \eqref{eq:veffeuc}. In particular, the gravitational self-energy is attractive in the Lorentzian section, as expected. Together with the AdS potential, this makes the shell collapse towards the future and towards the past. Namely, in the Lorentzian section the shell starts with zero velocity at $r= R_*$ at $T=0$ on $\Sigma_{\Co}$, and collapses to $r=0$ in a finite proper time $T=\pm T_f$ both towards the future and the past. The value of this time is given by
\be 
T_f = \int_{0}^{R_*} \dfrac{\text{d}R}{\sqrt{-V^L_{\text{eff}}(R)}}\;.
\ee 
The classical solution stops being sensible and develops curvature singularities slightly before this time, when the shell's mass density $\sigma$ becomes Planckian. For $d=2$, $T_f = \frac{\pi}{2}\ell$, independently of the rest mass of the shell. For $d>2$, the proper time is $T_f\approx \frac{\ell}{\sqrt{2d}}$ for $m\ell$ parametrically large. These moments in time thus define the big bang and big crunch of the closed universe $\Co$ as illustrated in Fig. \ref{fig:cosm}.

\begin{figure}[h]
    \centering    \includegraphics[width=0.65\textwidth]{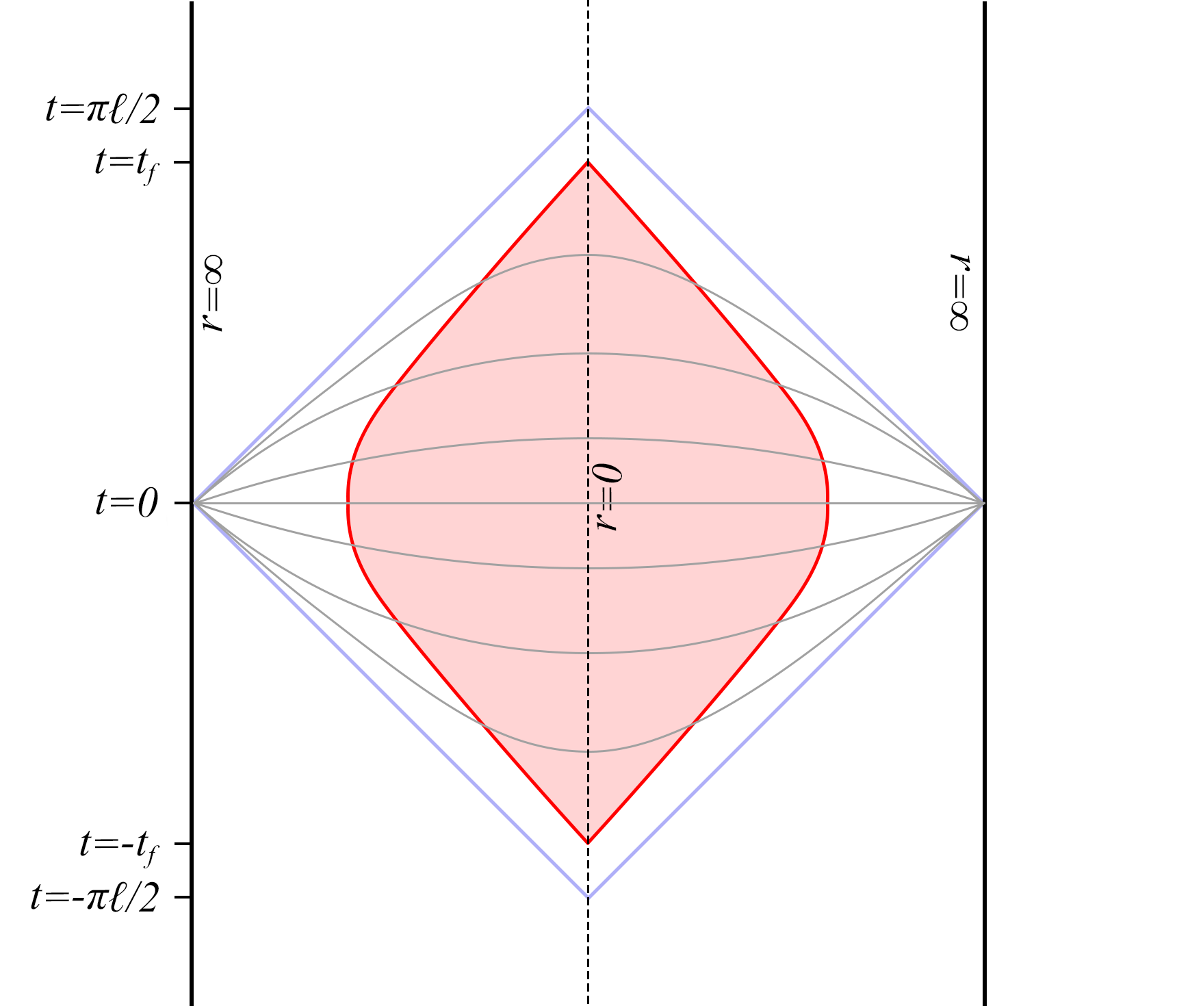}
    \caption{Each one of the locally AdS patches $\Co_\lef$ and $\Co_\ri$ (shaded in red) is bounded by the collapsing shell (red solid line) and contained within the $t=0$ Wheeler-DeWitt patch (bounded by the solid blue line) in the corresponding pure AdS spacetime without cutoff. The gray lines represent constant $t_c$ slices in the hyperbolic foliation described by the metric (\ref{eq:flrw}). In our setup, each slice is cut off when it intersects the shell and glued to an analogous slice of the other cosmological component. A curvature singularity arises at the tips of the shell's trajectory. In this figure we represent the full Lorentzian geometry obtained by evolving the initial state for $\Co_\lef$ or $\Co_\ri$ at $t=0$ backwards and forward in time.}
    \label{fig:WDWpatch}
\end{figure}

\begin{figure}
    \centering
    \includegraphics[width=0.6\textwidth]{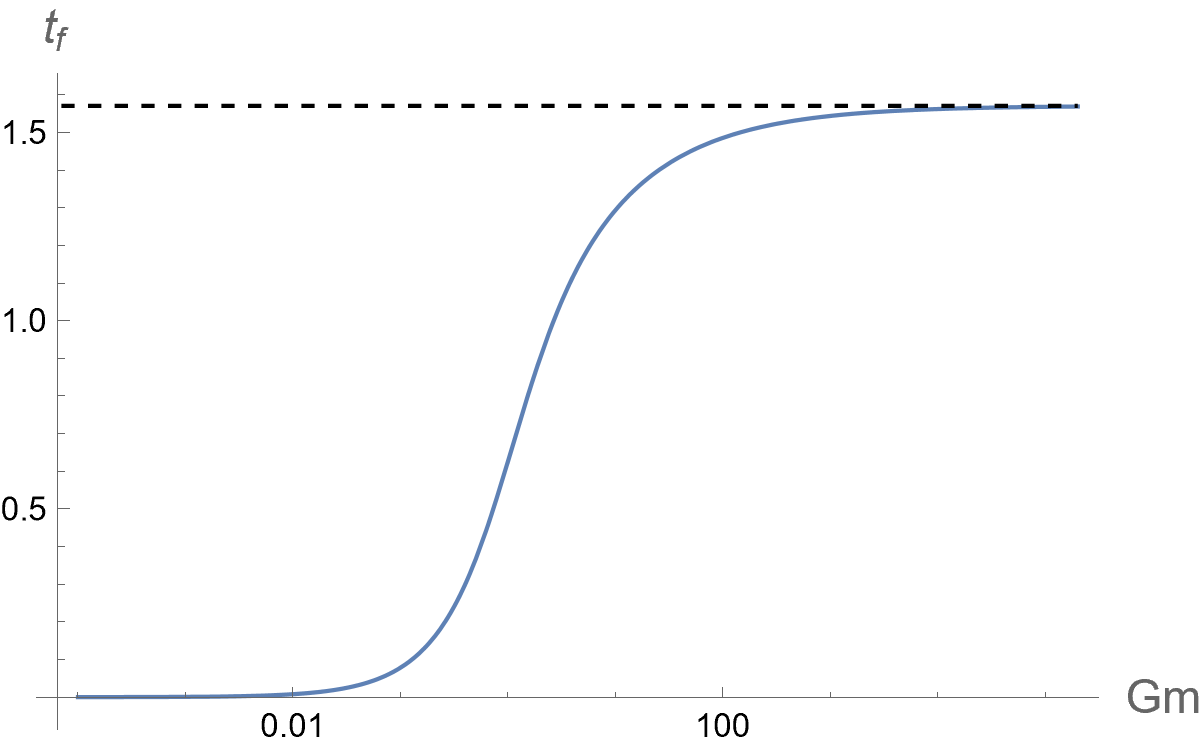}
    \caption{Global AdS time elapsed during the shell's trajectory from $r=R_*$ to $r=0$ as a function of the shell's mass $m$ for $d=3$. We set here $\ell=1$, so the $x$ axis is measured in units of the AdS radius. $t_f<\pi \ell /2$ for all values of $m$, and $t_f\to \pi\ell /2$ in the large $m$ limit: the shell is always contained within the $t=0$ Wheeler-DeWitt patch. }
    \label{fig:tf}
\end{figure}

It is interesting to note that $\Co_\lef$ and $\Co_\ri$ are contained within the Wheeler-DeWitt patches associated with boundary times $t_\pm =0$ in the corresponding pure AdS spacetimes without IR cutoff (see Fig. \ref{fig:WDWpatch}). This can be readily understood by computing the total coordinate global time $t_\pm = t_f$ needed for the shell to collapse from its maximum radius $r=R_*$ to $r=0$. This time is 
\be 
\label{eq:shifttimeL}
 t_f  =  \int_{0}^{R_*}\dfrac{\text{d}R}{f(R)} \, \sqrt{\dfrac{f(R) - V^L_{\text{eff}}(R)}{- V^L_{\text{eff}}(R)}}\;.
\ee 
For $d=2$ we again obtain $t_f=\frac{\pi\ell}{2}$ independently of $m$. For $d>2$, a numerical evaluation shows that $t_f<\pi \ell/2$ for any $m$, with $t_f\to \pi \ell/2$ in the large mass limit, as illustrated in Fig. \ref{fig:tf}.

The (null) boundary of the $t=0$ Wheeler-DeWitt patch of a pure AdS spacetime has infinite radius at $t=0$ and shrinks to zero radius at $t=\pm \pi \ell/2$, independently of the dimension $d$. The full (timelike) trajectory of the shell, which starts at $r=R_*<\infty$ for $t=0$ and ends at $r=0$ for $t=t_f<\pi\ell/2$ is contained within such Wheeler-DeWitt patch. 

As a consequence of this result, the entirety of each one of the two (identical) patches $\Co_\lef$ and $\Co_\ri$ of the cosmology is separately covered by the FLRW coordinates describing a hyperbolic slicing of the Wheeler-DeWitt patch (see Fig. \ref{fig:WDWpatch}). The change of coordinates from global AdS to FLRW coordinates is given by $\tan(t/\ell)=\tan (t_c/\ell)/\cosh \rho$, $r=\ell \cos(t_c/\ell)\sinh \rho$, in which case the AdS metric takes the form
\be 
\text{d}s_\pm^2 =  - \text{d}t_c^2 + \ell^2\cos^2(t_c/\ell)\, \text{d}\mathbf{H}_{d}^2\;,
\label{eq:flrw}
\ee 
where $\text{d}\mathbf{H}_{d}^2 = \text{d}\rho^2 + \sinh^2 \rho \,\text{d}\Omega_{d-1}^2$ is the hyperbolic metric of a FLRW time slice. 

In the case of empty AdS, this metric describes an open universe and is a solution of the Friedmann equation with negative spatial curvature and a negative cosmological constant, with no additional matter. Since this metric describes a Wheeler-DeWitt patch of pure AdS, it is clearly non-singular. The ``big bang'' and ``big crunch'' at $t_c=\pm\pi\ell/2$ are simply coordinate singularities corresponding to the tips of the Wheeler-DeWitt patch. 

On the other hand, in the case of $\Co$, each FLRW time slice $t=t_c$ in each one of the two patches $\Co_\lef$ and $\Co_\ri$ is cut off at the location of the shell, implicitly given by $\rho_{\text{max}}(t_c) = \sinh^{-1} \left[R(t_c)/\left(\ell\cos(t_c/\ell)\right)\right]$, where $R(t_c)\equiv R(t(t_c))$, and glued to another identical cutoff FLRW slice. Therefore, the cosmology $\Co=\Co_\lef\cup \Ws^L \cup \Co_\ri$ is actually closed (i.e. each time slice is compact). Although $\Co$ is locally homogeneous and isotropic inside each one of the two patches $\Co_\lef$ and $\Co_\ri$, the presence of the shell introduces an inhomogeneity. In particular, there are two contributions to the expansion and contraction of the universe:\footnote{Here we are considering the full Lorentzian geometry $\Co$ obtained by evolving backwards and forward the initial state prepared by the Euclidean path integral at global time $t_+=t_-\equiv t=0$. If we want to restrict to the contracting phase, it is sufficient to restrict our analysis to $t>0$, corresponding to $t_c>0$.} the first one is homogeneous and driven by the evolution of the scale factor $a(t_c)=\ell\cos(t_c/\ell)$, and the second one is inhomogeneous and determined by the contraction of the shell. Unlike the hyperbolic slicing of pure AdS, our cosmology is singular, with the big bang and big crunch singularities approximately corresponding to the initial and final loci of the shell's worldvolume $\Ws^L$ at $r=0$ and $t=\mp t_f$, respectively. Note that these are not the usual homogeneous singularities arising in FLRW universes, but rather inhomogeneous singularities associated with the collapse of the shell, similar to what happens in the geometry enclosed by a collapsing shell during black hole formation. By inverting the coordinate transformation between global and FLRW coordinates:
\be
    \tanh \rho = \frac{r}{\ell\sqrt{1+\frac{r^2}{\ell^2}}\cos\left(\frac{t}{\ell}\right)}; \hspace{1cm} \tan\left(\frac{t_c}{\ell}\right)=\frac{\tan\left(\frac{t}{\ell}\right)}{\sqrt{1-\frac{r^2}{\ell^2\left(1+r^2/\ell^2\right)\cos^2(t/\ell)}}}\;,
\ee
we obtain that the big bang and big crunch singularities sit at FLRW coordinates $\rho=0$ and $t_c=\mp t_f$ in each one of the two cosmological patches $\Co_\lef$ and $\Co_\ri$.

\subsection{Excited bulk states} 
\label{sec:2.3}

\begin{figure}[h]
    \centering
    \includegraphics[width=0.9\textwidth]{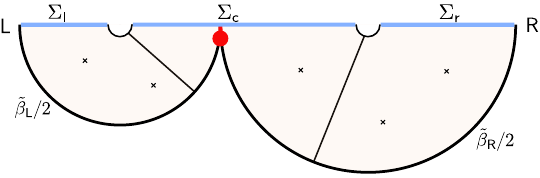}
    \caption{Preparation of the excited state $\ket{\psi}$ from local operator insertions along $X$. The radial slices correspond to constant Euclidean time slices, evolved with the action of the time-independent Hamiltonians $H^{\lef,\ri}_b$. Note that here and in all the following figures, we add probe operator insertions in the bulk manifold $X$ that do not backreact on the geometry. These local operators are inserted at different points on the $\mathbf{S}^{d-1}$ (which is suppressed in the figures).}
    \label{fig:cosmogeneral}
\end{figure}

Given the construction of the Euclidean section $X$ supported by the heavy shell, which prepares the state $\ket{\psi_0}$ of the quantum fields on the EFT Hilbert space $\mathcal{H}_\Sigma$ defined on the initial slice $\Sigma = \Sigma_\lef \cup \Sigma_\Co \cup \Sigma_\ri$, it is natural to consider more general bulk states, $\ket{\psi} \in \mathcal{H}_\Sigma$, prepared by additional light operator insertions $\phi(x)$ at different points $x\in X$, as illustrated in Fig. \ref{fig:cosmogeneral}.\footnote{Such a naive notion of EFT of ``free gravitons and matter'' $\mathcal{H}_\Sigma$, at $G=0$, should be a reasonable approximation to the true EFT, when $G\neq 0$. At a first glance, this seems problematic, given that the cosmology $\Sigma_\Co \subset \Sigma$ is a closed space, and that it carries all of the well-know problems of implementing the gravitational Gauss' law non-trivially in closed spaces. The way in which this setup avoids this issue is to note that all of the modes of the quantum fields in the state $\ket{\psi_0}$ contain some entanglement with the corresponding modes in the AdS region, given that the preparation of the state is thermal-like. Entanglement allows to dress all of the modes on the cosmology, in a state-dependent way, to the physical asymptotic boundaries of the AdS regions.  In Sec. \ref{sec:5} we comment on this issue, which affects all of the discussions of entanglement islands, in a simplified toy model.} Each bulk state prepared this way defines a corresponding CFT state $\ket{\Psi} \in \mathcal{H}_{\Le}\otimes \mathcal{H}_{\Ri}$ prepared by inserting an appropriate (and in general non-local) CFT operator in the CFT Euclidean path integral. The associated bulk-to-boundary map will be the object of study in Secs. \ref{sec:4} and \ref{sec:5} of this paper.

In the limit of large mass $m\ell \rightarrow \infty $ for the shell, the bulk Euclidean preparation via $X$ effectively decouples both sides of the cosmology (see Fig. \ref{fig:largemass}) and, as a consequence, the bulk state approximately factorizes, $\ket{\psi} \approx \ket{\Phi_-}_{\lef\Co_{\lef}}  \ket{\Phi_+}_{\Co_{\ri}\ri}$. In this regime, the reduced density matrices prepared by the bulk Euclidean path integral become simply
\begin{gather}
\rho_{\lef} \approx \dfrac{\exp(-\beta_{\Le}H^{\lef}_b)}{Z_{\lef}(\beta_{\Le})} \mathcal{T}\lbrace \phi^\lef_I(x_I)\rbrace \;,\label{eq:stateredl}\\
\rho_{\ri} \approx \dfrac{\exp(-\beta_{\Ri} H^{\ri}_b)}{Z_{\ri}(\beta_{\Ri})} \mathcal{T}\lbrace \phi^\ri_I(x_I)\rbrace \;,\label{eq:stateredr}
\end{gather}
where $H^{\lef,\ri}_{b}$ is the time-independent bulk Hamiltonian, associated with the $U(1)$ Euclidean time translation symmetry of each component of $X$. Each $U(1)$ is weakly broken in this limit by the presence of the thin shell in the near-boundary region. The corresponding Hamiltonians are given by local integral on constant Euclidean time slices illustrated in Fig. \ref{fig:cosmogeneral}.  Additionally, the light fields $\phi^{\lef,\ri}_I(x_I)$ represent the corresponding operator insertions at the points $x_I \in X$, where $\mathcal{T}$ denotes the Euclidean time-ordering in the bulk preparation.\footnote{Note that with this time-ordering prescription the Euclidean time runs clockwise on the right component $X_+$, and counter-clockwise in the left component $X_-$.} The factors $Z_{\lef,\ri}(\beta) = \text{Tr}(\mathcal{T}\lbrace e^{-\beta H^{\lef,\ri}_b}\phi^{\lef,\ri}_I\rbrace)$ serve as normalization factors for the states, and correspond to suitable bulk partition functions on thermal AdS spaces.

The bulk state $\ket{\psi}$ constructed this way contains mostly bipartite entanglement between each of the AdS regions and the corresponding side of the cosmology. The reduced bulk density matrices $\rho_{\lef}, \rho_{\ri}$ and $\rho_{\Co}$ will have entanglement entropies given by $S(\rho_{\lef}), S(\rho_{\ri})$ and $S(\rho_{\Co}) \approx S(\rho_{\Co_{\lef}}\otimes \rho_{\Co_{\ri}}) \approx S(\rho_{\lef}) + S(\rho_{\ri})$, respectively. Therefore, the cosmology effectively purifies the bulk entanglement of the quantum fields in the AdS regions.

For most of our analysis, we will be interested in excited states $\ket{\psi}$ in which the entanglement $S(\rho_{\lef,\ri})$ and the difference $|S(\rho_{\lef}) - S(\rho_{\ri})|$ can be taken to be large. This can be achieved with the addition of suitable operator insertions along $X$, since these can create light entangled particles between the cosmology and the AdS regions. The effect of adding such insertions is to change the ADM energy of the state. Microcanonically, a gas of particles in AdS is stable, up to energies of order $E \ell  \lesssim G^{-1}\sim N^2$, before it collapses into a small black hole. Therefore, by adding light entangled pairs between the cosmology and the AdS regions, it is possible to attain a large value of $S(\rho_{\lef,\ri})$ and $|S(\rho_{\lef}) - S(\rho_{\ri})|$. As long as the number of entangled pairs, and thus the total energy, is parametrically smaller than $O(N^2)$, the semiclassical wavefunction of the state will remain within the cosmological branch.

The reason why we are interested in considering more general excited states rather than the original state $\ket{\psi_0}$ described above, dual to the PETS, is that, in string theory realizations of AdS/CFT, the bulk (thermal) entanglement between the cosmology and the AdS regions is very small in the original state $\ket{\psi_0}$. This is the case because the preparation temperature is of the order of the gap of the system, and the number of species in these theories is not extremely large. For instance, in type IIB supergravity in AdS$_5 \times \mathbf{S}^5$, which contains $256$ local DOF, we provide a simple estimate of this entanglement in Appendix \ref{app:B}, showing that indeed $S(\beta)_{\text{IIB}} \lesssim 0.29$. As we will discuss in detail in Secs. \ref{sec:4} and \ref{sec:5}, this small amount of entanglement between the AdS regions and the cosmology implies that the cosmology-to-boundary map encoding the cosmology in the dual theory is highly non-isometric, and that the reconstruction of the cosmological physics by a CFT observer requires exceptionally precise control over the CFT. In the presence of a large amount of entanglement between the cosmology and the AdS regions, the situation greatly simplifies. For this reason, we will consider more general excited states $\ket{\psi}$ with larger amounts of bulk entanglement between the AdS regions and the cosmology.\footnote{An alternative avenue would be to work with the original PETS in bottom-up models of AdS/CFT, in which the number of bulk species can be chosen to be much larger.}

\section{Cosmology as an island}
\label{sec:3}

In this section we derive a version of the island formula for the fine-grained entanglement entropy between the CFT$_\Le$ and the CFT$_\Ri$ in the cosmological microstates built in the previous section. To do this, we will use a standard CFT replica trick, complemented by the rules of the semiclassical path integral of gravity, following FLM \cite{Lewkowycz:2013nqa,Faulkner:2013ana} closely.

Recall that the von Neumann entropy of $\rho_{\Le}$ can be computed via the replica trick
\be\label{eq:replicatrick}
S(\rho_{\Le}) = \lim_{n\rightarrow 1^+} \dfrac{1}{1-n} \log \text{Tr}\rho_{\Le}^n = - \partial_n \log \text{Tr} (\rho_{\Le}^n) |_{n=1}\;.
\ee 
To do this, one starts from $\text{Tr} (\rho_{\Le}^n)$ for integer $n>1$, which is related to the $n$-th R\'{e}nyi entropy,\footnote{In the following we will refer for brevity to $\text{Tr} (\rho_{\Le}^n)$ as the $n$-th R\'enyi entropy, although the latter is related to the former by $S_n(\rho_\Le)=\log \text{Tr} (\rho_{\Le}^n)/(1-n)$.} to then analytically continue in $n$, and evaluate it close to $n=1^+$, in order to take the derivative in \eqref{eq:replicatrick}.

\begin{figure}[h]
\centering
\includegraphics[width = .6\textwidth]{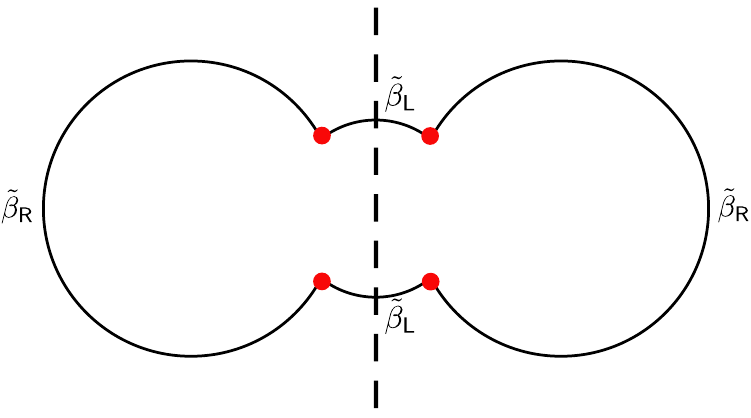}
\caption{Euclidean CFT path integral computing the unnormalized purity $\text{Tr}(\rhohat_{\Le}^2)$. Each point represents a spatial sphere $\mathbf{S}^{d-1}$. The two replicas are glued together across the dashed axis of $\mathbf{Z}_2$ symmetry, which cuts the Hilbert space of the $\Le$ subsystems. }
\label{fig:replica}
\end{figure}

In our case we can prepare the unnormalized state $\rhohat_{\Le}$ using the Euclidean CFT path integral. The $n$-th R\'{e}nyi entropy of the unnormalized state $\text{Tr} (\rhohat_{\Le}^n)$ is given by a CFT path integral on a manifold consisting of $n$ replicas of the original manifold. The contour of this path integral is shown in Fig. \ref{fig:replica} for the case $n=2$, which analogously extends to higher R\'{e}nyi entropies.

For the unnormalized $n$-th R\'{e}nyi entropy, there will be $n$ creation operator insertions $\Op_{i}$ and $n$ annihilation insertions $\Op^\dagger_{i}$ along the CFT contour, where $i=1,...,n$ is the replica label. Within the EFT \eqref{eq:actionEFT} with $n$ shells, the shells can only propagate freely at tree level.\footnote{A putative tadpole or three-point interaction between shells is expected to be exponentially suppressed in $m\ell$ if the mass of the shell is large, like in the case of classical matter. Moreover the operator $\mathcal{O}$ has to be thought of as an s-wave channel, and thus each shell is an irreducible object.} There will thus be $n!$ saddle point geometries, $X_g$, each of which corresponds to a symmetric group element $g \in \text{Sym}(n)$ representing that $O^\dagger_i$ is connected to $O_{g(i)}$ in this saddle. 

To compute the normalized $n$-th R\'enyi entropy, the normalization of the state by the norm $Z_1 = \text{Tr} (\rhohat_{\Le})$ also needs to be taken into account.\footnote{As customary in the gravitational Euclidean path integral computations of R\'enyi entropies (see for instance \cite{Lewkowycz:2013nqa,Faulkner:2013ana}), we assume that computing the trace of the unnormalized state $\text{Tr} (\rhohat_{\Le}^n)$ and the normalization factor separately using the gravitational Euclidean path integral and then taking the ratio is a good approximation to computing the trace of the normalized state.}

\subsection{Purity}

\begin{figure}[h]
\centering
\includegraphics[width = .7\textwidth]{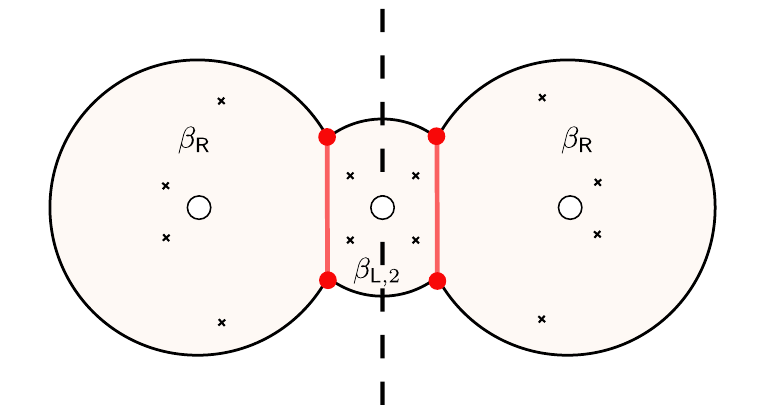}
\caption{The saddle point geometry $X_e$ consist of thin shell contractions between each replica. The center of replica symmetry corresponds to an AdS space of physical inverse temperature $\beta_{\Le,2} = 2\tbeta_{\Le} + 2 \Delta \tau_+$. Here and in all the following figures we consider generic saddles in which additional operator insertions $\phi_I$ are present (represented by the crosses), see Sec. \ref{sec:2.3}.}
\label{fig:filling1}
\end{figure}

\begin{figure}[h]
\centering
\includegraphics[width = \textwidth]{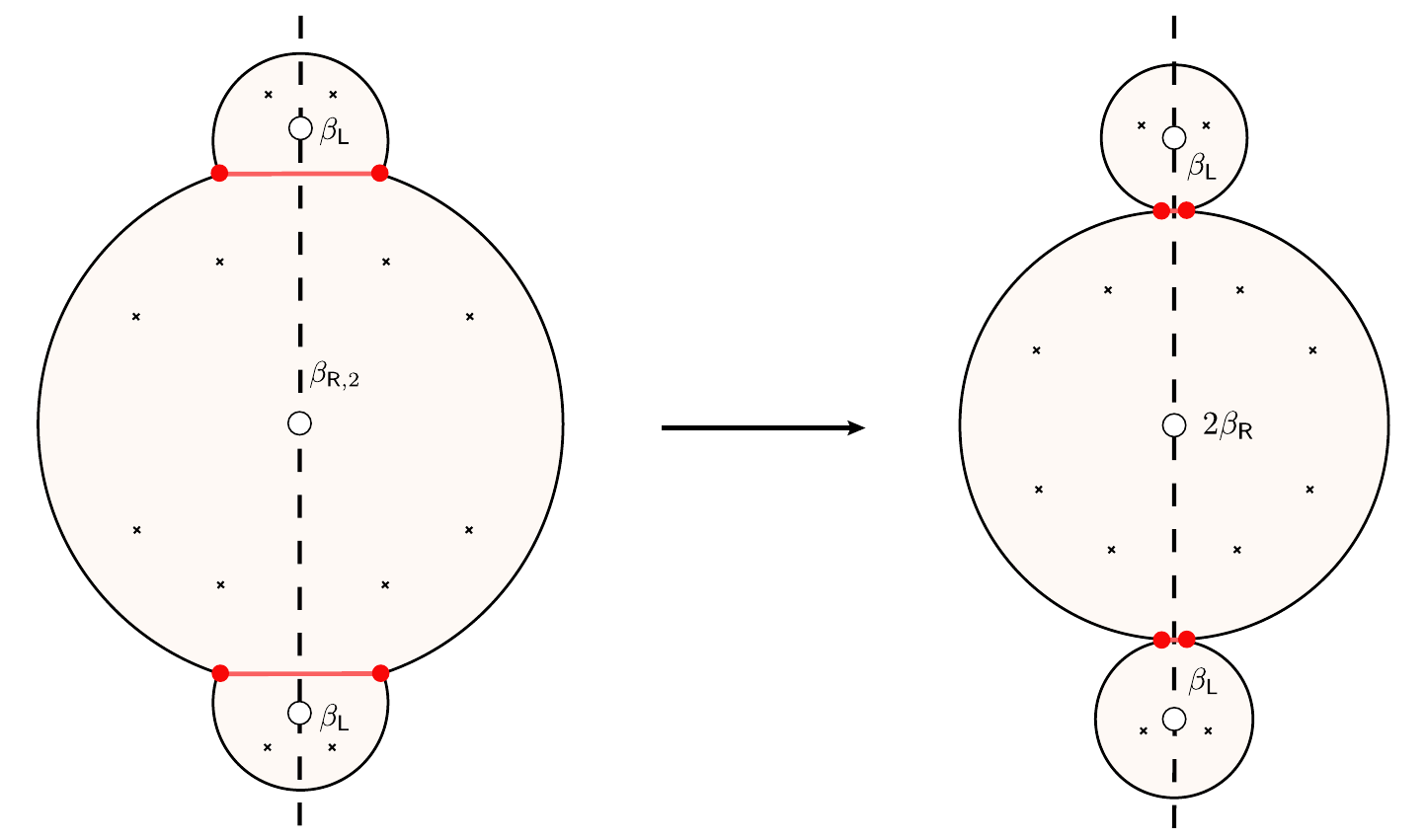}
\caption{The replica wormhole $X_\eta$ consist of thin shell contractions between replicas. The center of replica symmetry corresponds to an AdS space of temperature $\beta_{\Ri,2} = 2\tbeta_{\Ri} + 2 \Delta \tau_+$. On the right, the large mass limit of the shell, where the geometry becomes effectively composed of disconnected Euclidean AdS spaces, with $\beta_{\Ri,2} = 2\beta_{\Ri}$. Each of these spaces contributes with a factor of $Z_{\lef}(k\beta_{\Le})$ or $Z_{\ri}(k\beta_{\Ri})$.}
\label{fig:filling2}
\end{figure}

Consider for concreteness the case $n=2$. There will be two saddle point geometries $X_e$, $X_{\eta}$, where $e$ is the symmetric group identity element and $\eta =(12)$. These solutions are illustrated in Fig. \ref{fig:filling1} and Fig. \ref{fig:filling2}, respectively. The unnormalized purity is simply the sum of both contributions
\be\label{eq:purity1}
\overline{\text{Tr}(\rhohat_{\Le}^2)} \sim Z[X_e,\phi_I] + Z[X_\eta,\phi_I]\;,
\ee
where $Z[X_i,\phi_I] = e^{-I[X_i]} Z_{\text{bulk}}[X_i,\phi^{(1)}_I,\phi^{(2)}_I]$ is the semiclassical contribution of $X_i$, with $I[X_i] = O(G^{-1})$ the classical gravitational action. The replicated bulk operators $\phi^{(1)}_I,\phi^{(2)}_I$ are inserted at the corresponding Euclidean times, and the term $Z_{\text{bulk}}[X_i,\phi^{(1)}_I,\phi^{(2)}_I]$ corresponds to the correlation function of these fields on each manifold. The overline is to denote that this quantity is computed by the gravitational path integral, in the semiclassical $G\rightarrow 0$ asymptotic expansion.

\begin{figure}[h]
\centering
\includegraphics[width = .7\textwidth]{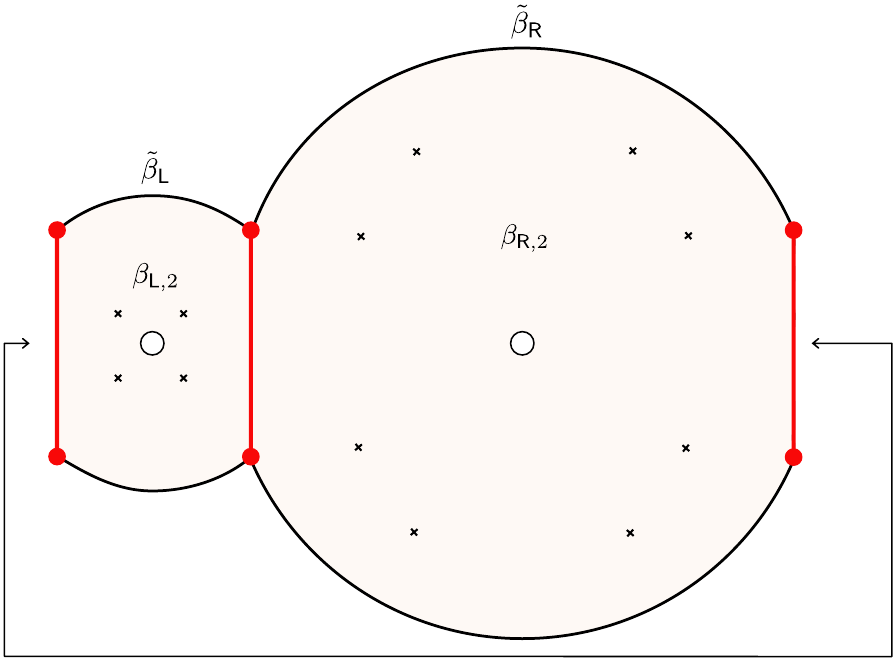}
\caption{The Euclidean wormhole $X_2$ contributing to the square of the norm of the state. In the space transverse to the spheres, $X_2$ is homotopic to a sphere with four punctures. The asymptotic boundary $\partial X_2$ contains two disconnected components, corresponding to Euclidean CFT path integrals computing the normalization $Z_1$. The wormhole is deconstructed in a pair of thermal AdS spaces at inverse temperatures $\beta_{\Ri,2} = 2\beta_{\Ri} + 2 \Delta \tau_+$ and $\beta_{\Le,2} = 2\beta_{\Le}+ 2 \Delta \tau_+$, glued along the trajectories of two thin shells. }
\label{fig:wh2}
\end{figure}

To compute the purity, the normalization of the state by the norm $Z_1 = \text{Tr} (\rhohat_{\Le})$ needs to be taken into account.\footnote{As customary in the gravitational Euclidean path integral computations of R\'enyi entropies (see for instance \cite{Lewkowycz:2013nqa,Faulkner:2013ana}), we assume that computing the unnormalized purity and the normalization factor separately using the gravitational Euclidean path integral and then taking the ratio is a good approximation to computing the normalized purity directly.} In particular, we need to consider the square of this norm, which enters in the normalization of \eqref{eq:purity1}. As it was shown in \cite{Sasieta:2022ksu,Balasubramanian:2022gmo} for the high-temperature PETS, moments of the overlap between PETS states, in this case the norm of one of them, admit connected contributions coming from Euclidean wormholes in the semiclassical path integral computing these quantities. These wormholes also exist at low temperatures,
\be 
\overline{(Z_1)^2} = Z[X,\phi_I]^2 + Z[X_2,\phi_I]\;.
\ee 
The corresponding Euclidean wormhole $X_2$ was constructed in \cite{Sasieta:2022ksu} (see Fig. \ref{fig:wh2}). It consists of a pair of AdS spaces, glued along the trajectories of the thin shells in such a way that $\partial X_2$ contains two disconnected components, each of which prepares a single copy of $Z_1$.

The normalized purity is then given by
\be\label{eq:purity2}
\overline{\text{Tr}(\rho_{\Le}^2)} \sim \dfrac{Z[X_e,\phi_I] + Z[X_\eta,\phi_I]}{Z[X,\phi_I]^2 + Z[X_2,\phi_I]}\;.
\ee
Since, at this level, the classical on-shell gravitational action of Euclidean AdS is linear in the boundary Euclidean time, the classical gravitational contributions to \eqref{eq:purity2} cancel between numerator and denominator, given that the boundary is always of length $2\tbeta_{\Le} + 2\tbeta_{\Ri}$ for all of these saddles, and that the contribution from the shell does not depend on the physical temperature of the solutions. 

All that is left to do to evaluate the purity \eqref{eq:purity2} is to compute the corresponding correlation function of the bulk fields on each manifold, $Z_{\text{bulk}}[X_i,\phi_I]$. The computation of these quantities is involved, given the irregular spectrum of the second order differential operator $\nabla^2$ in the space of normalizable functions $L_g^2(X_i)$ (for any of the above manifolds). A remarkable simplification occurs again in the limit of large mass of the thin shell, $m\ell \rightarrow \infty$. As explained in Sec. \ref{sec:2}, in this limit the manifolds effectively factorize and become a collection of Euclidean AdS spaces at different temperatures. The bulk partition functions factorize accordingly\footnote{A more detailed argument to show the factorization of the one-loop partition function will appear in \cite{Climent:2023}.}
\begin{gather}
Z_{\text{bulk}}[X_e,\phi_I] \approx Z_{\lef}(2\beta_{\Le}) Z_{\ri}(\beta_{\Ri})^2\;,\label{eq:gpur1}\\ 
Z_{\text{bulk}}[X_\eta,\phi_I] \approx Z_{\lef}(\beta_{\Le})^2 Z_{\ri}(2\beta_{\Ri})\;,\label{eq:gpur2}\\
Z_{\text{bulk}}[X,\phi_I] \approx Z_{\lef}(\beta_{\Le}) Z_{\ri}(\beta_{\Ri})\;, \label{eq:gnorm1}\\
Z_{\text{bulk}}[X_2,\phi_I] \approx Z_{\lef}(2\beta_{\Le}) Z_{\ri}(2\beta_{\Ri})\;,\label{eq:gnorm2}
\end{gather}
where 
\be\label{eq:bulkpf}
Z_{\lef,\ri}(n\beta_{\Le,\Ri}) = \text{Tr}\left( \mathcal{T}\lbrace e^{-n\beta_{\Le,\Ri} H^{\lef,\ri}_b}(\phi_I^{\lef,\ri})^{(1)}...(\phi_I^{\lef,\ri})^{(n)} \rbrace \right) 
\ee 
is the correlation function of the bulk operator insertions in the replicated Euclidean AdS spaces. These factors precisely correspond to traces of the replicated unnormalized bulk states $\hat{\rho}_{\lef}^n$ and $\hat{\rho}_{\ri}^n$, prepared by the bulk replicated manifolds at $\tau =0$, as in \eqref{eq:stateredl} and \eqref{eq:stateredr} for $n=1$.

Plugging these expressions back into \eqref{eq:purity2}, we get that the purity can be written as
\be\label{eq:purity4}
\overline{\text{Tr}(\rho_{\Le}^2)} \sim \dfrac{d_{\ri} + d_{\lef}}{1 + d_{\lef}d_{\ri}}\;,
\ee 
where $\log d_{\lef,\ri} = S_2(\rho_{\lef,\ri})$ is the second R\'{e}nyi entropy of the respective bulk states \eqref{eq:stateredl} and \eqref{eq:stateredr}. Assuming $d_{\lef,\ri}\gg 1$,\footnote{This condition is satisfied as long as the AdS regions and the cosmology share enough entanglement, see Sec. \ref{sec:2.3} for a discussion of the size of such entanglement.} the first term in the numerator, which comes from the replica disconnected saddle $X_e$, provides a contribution of approximately $d^{-1}_{\lef} = \text{Tr}(\rho_\lef^2)$, which corresponds to the naive purity of the state $\rho_\lef$ in the AdS${}_\lef$ space of Fig. \ref{fig:cosm}. The second term, on the other hand, comes from the replica-connected (also known as `replica wormhole') saddle $X_\eta$, and provides a contribution of approximately $d^{-1}_{\ri}= \text{Tr}(\rho_\ri^2)$, which corresponds to the purity of the state $\rho_\ri$ in the AdS${}_\ri$ region. Since the global bulk state is pure, $d^{-1}_{\ri}$ also corresponds to the purity of the state $\rho_{\lef \Co}$ of the AdS${}_\lef$ region \textit{and} the cosmology $\Co$ in Fig. \ref{fig:cosm}. 

Interestingly, when $d_{\lef} \gg d_{\ri}$, i.e. if the bulk entanglement of fields in the AdS${}_\lef$ space is sufficiently larger than the bulk entanglement of fields in the AdS${}_\ri$ space, the purity of the CFT${}_\Le$ is controlled by the purity of the bulk state $\rho_\ri$ in the AdS${}_\ri$ region (or, equivalently, by the purity of the state $\rho_{\lef\Co}$ on the AdS${}_\ri$ region and the cosmology $\Co$). This result is already suggestive of an island formula: if the entanglement of bulk fields in AdS${}_\lef$ is large, in order to compute the purity of the CFT${}_\Le$ using a bulk calculation we must include the contribution of an entanglement island given by the cosmology $\Co$. In Sec. \ref{sec:3.2} we will make this statement precise by computing the von Neumann entropy of the CFT${}_\Le$.

The expression \eqref{eq:purity4} has suggestive similarities with the purity of a random state (cf. \cite{Page:1993df}). As we explain in Appendix \ref{app:D}, this is no coincidence, and one can identify the cosmological state in the large mass limit with a \textit{random purification} of the bulk states $\rho_{\lef,\ri}$, at least at the level of the R\'{e}nyi entropies.  

\subsection{Entanglement spectrum}

The generalization to higher R\'{e}nyi entropies is a matter of combinatorics. In the regime of large mass $m\ell \rightarrow \infty$, evaluating all of the contributions becomes straightforward. First of all, the trajectories of the shells pinch off and the preparation temperatures become equal to the physical temperatures. All that is left to do is to evaluate the number of closed $\beta_{\Le}$ and $\beta_{\Ri}$ loops on each diagram, and assign a factor of the bulk partition function \eqref{eq:bulkpf} for each loop. For a permutation $g=\sigma_1 \sigma_2 ... \sigma_{s_g}\in \text{Sym}(n)$, with cycles of lengths $k_i= | \sigma_i| $, such that $\sum_{i=1}^{s_g}k_i = n$, the number of $\beta_{\Ri}$-loops in $X_g$ is simply $s_g$, and each of them has length $k_i\beta_{\Ri}$. Moreover let us define the left conjugate permutation $g' = \eta g$, with $\eta = (123...n)$, where  $g'=\sigma_1' \sigma_2' ... \sigma_{s_{g'}}'$ are cycles of lengths $k_i'= | \sigma'_i| $. The number of $\beta_{\Le}$ loops in $X_g$ is simply $s_{g'}$, and each of them has length $k_i'\beta_{\Le}$.  The formula for the unnormalized $n$-th R\'{e}nyi is then
\be\label{eq:renyisunn}
\overline{\text{Tr}(\rhohat_{\Le}^n)} \sim \,  \sum_{g \in \text{Sym}(n) }Z_{\ri}(k_1\beta_{\Ri}) ... Z_{\ri}(k_{s_g}\beta_{\Ri})Z_{\lef}(k'_1\beta_{\Le}) ... Z_{\lef}(k'_{s_{g'}}\beta_{\Le})\;,
\ee 
were, again, each factor of $Z_{\lef,\ri}(n\beta_{\Le,\Ri})$ corresponds to the trace of the unnormalized state of the bulk fields, prepared by the bulk EFT path integral, on $n$ copies of thermal AdS.

For the normalization of the state, $Z_1^n$, there will be $n$ creation operator insertions $\Op_{i}$ and $n$ annihilation insertions $\Op^\dagger_{i}$ along the disconnected CFT contours, where $i=1,...,n$ labels each connected copy of $Z_1$. Thus, there will be $n!$ saddle points $X'_g$ also labeled by a symmetric group element $g\in \text{Sym}(n)$. For $g\neq e$, $X'_g$ will contain at least an Euclidean wormhole solution, connecting different components of the boundary together. In the limit of large mass $m\ell \rightarrow \infty$, evaluating all the actions becomes also straightforward. Again, one needs to evaluate the number of closed $\beta_{\Le}$ and $\beta_{\Ri}$ loops on each diagram. It turns out that, in this case, both of them are given by $s_g$, and each of the loops has length $k_i\beta_{\Ri}$ and $k_i\beta_{\Le}$ respectively. With this in mind, the normalization becomes
\be\label{eq:normn}
\overline{\text{Tr}(\rhohat_{\Le})^n} \sim \,  \sum_{g \in \text{Sym}(n) }Z_{\lef}(k_1\beta_{\Le}) ... Z_{\lef}(k_{s_g}\beta_{\Le})Z_{\ri}(k_1\beta_{\Ri}) ... Z_{\ri}(k_{s_{g}}\beta_{\Ri})\,.
\ee 

Finally, we arrive to the $n$-th R\'{e}nyi entropy computed by semiclassical gravity\footnote{A consistency check of this expression is that when $\beta_{\Ri}\rightarrow \infty$, the state $\rho_{\Le}$ becomes exactly pure.}
\be\label{eq:renyis}
\overline{\text{Tr}(\rho_{\Le}^n)} \sim \,  \dfrac{\sum_{g \in \text{Sym}(n) }Z_{\ri}(k_1\beta_{\Ri}) ... Z_{\ri}(k_{s_g}\beta_{\Ri})Z_{\lef}(k'_1\beta_{\Le}) ... Z_{\lef}(k'_{s_{g'}}\beta_{\Le})}{\sum_{g \in \text{Sym}(n) }Z_{\lef}(k_1\beta_{\Le}) ... Z_{\lef}(k_{s_g}\beta_{\Le})Z_{\ri}(k_1\beta_{\Ri}) ... Z_{\ri}(k_{s_{g}}\beta_{\Ri})}\,.
\ee 

Again, this expression has similarities with the average R\'{e}nyi entropies of a \textit{random purification} of the bulk density matrices $\rho_{\lef,\ri}$, as we explain in Appendix \ref{app:D}.  

\subsection{Entanglement entropy}
\label{sec:3.2}

In order to compute the von Neumann entropy \eqref{eq:replicatrick}, we need to analytically continue eq. \eqref{eq:renyis}. This analytic continuation is subtle already from the CFT perspective, given that random states with thermal tails do not have self-averaging entanglement entropy in systems with infinite Hilbert space dimension. The reason is that it is possible to store an arbitrary amount of entanglement entropy by slightly modifying the high-energy tail of the wavefunction of such states.

In the gravity computation, the problem comes from replica symmetry breaking saddle points, such as the two-cycles
\be\label{eq:twocyclerenyis}
\overline{\text{Tr}(\rhohat_{\Le})^n} \supset {n\choose 2}  Z_\ri(2\beta_{\Ri})Z_\ri(\beta_{\Ri})^{n-2} Z_\lef(\beta_{\Le})Z_\lef((n-1)\beta_{\Le})\;,
\ee 
which give rise to terms proportional to $Z_\lef(0) = \infty$, when the derivative is taken in \eqref{eq:replicatrick}. 

A way to evade this problem is to analytically continue the gravitational saddles off-shell, which gives a more refined characterization of the cosmological microstate. In our setup the problematic saddle points cannot be physically continued near $n\rightarrow 1^+$. Some of them, like \eqref{eq:twocyclerenyis}, become perturbatively unstable due to the Jeans instability of the gas of particles in AdS at high temperatures \cite{HawkingPage}. Others directly fail to satisfy the Witten-Kontsevich-Segal criterium for allowable saddle point geometries \cite{Witten:2021nzp,Kontsevich:2021dmb}, since the local temperature of the solution is negative as $n\rightarrow 1^+$.  This motivates to only consider replica-symmetric saddle points, and in particular those which remain stable close to $n\rightarrow 1^+$. Strictly, we continue the quotient manifold $M_g = X_g/\mathbf{Z}_n$ off-shell in $n$, for any replica-symmetric $X_g$ (i.e. for $g$ satisfying $g = \eta g \eta^{-1}$), in the opening angle $2\pi/n$, with the corresponding conical defect at the center of replica symmetry. The two contributions which satisfy these criteria are
\be\label{eq:nicecontinuation}
\overline{\text{Tr}(\rho_{\Le})^n} = \dfrac{Z_\ri(n\beta_{\Ri})Z_\lef(\beta_{\Le})^n+Z_\lef(n\beta_{\Le})Z_\ri(\beta_{\Ri})^n}{Z_\ri(\beta_{\Ri})^n Z_\lef(\beta_{\Le})^n + Z_\ri(n\beta_{\Ri}) Z_\lef(n\beta_{\Le})} \hspace{1cm} 0<n-1 \ll 1\;,
\ee
corresponding to $g=e,\eta$ saddle points in the numerator and the denominator. The quotient manifold $X_\eta/\mathbf{Z}_n$ corresponding to the second term in the numerator of \eqref{eq:nicecontinuation} is the vacuum AdS analog of the `pinwheel' geometry (cf. \cite{Penington:2019kki}). 

We can now analytically continue eq. \eqref{eq:nicecontinuation} and compute the von Neumann entropy using eq. \eqref{eq:replicatrick}, which leads to the quantum corrected RT formula for the entanglement entropy
\be\label{eq:islandFLM}
S(\rho_{\Le}) \approx \text{min}\lbrace S(\rho_{\lef}), S(\rho_{\ri})\rbrace\;.
\ee 
In the derivation of eq. \eqref{eq:islandFLM} we have assumed that the state is such that the exponential in the entropy difference $\exp(|S(\rho_{\lef})-S(\rho_{\ri})|)$ is large enough, so that one of the two terms in eq. \eqref{eq:nicecontinuation} dominates. The bulk entropies are computed using the bulk replica trick
\be 
S(\rho_{\lef,\ri}) = -\partial_n (Z_{\lef,\ri}(n\beta_{\Le,\Ri}) - nZ_{\lef,\ri}(\beta_{\Le,\Ri}))|_{n=1} = -\text{Tr}(\rho_{\lef,\ri} \log \rho_{\lef,\ri})\;,
\ee 
where the reduced states are given by eqs. \eqref{eq:stateredl} and \eqref{eq:stateredr}, respectively. 

Given the fact that the global bulk state is pure, eq. \eqref{eq:islandFLM} can be rewritten as the island formula 
\be\label{eq:islandFLM2}
S(\rho_{\Le}) \approx \text{min}\lbrace S(\rho_{\lef}), S(\rho_{\lef \Co})\rbrace\;.
\ee 
The area term in both cases vanishes, since the RT is empty $\gamma_{\text{RT}} =\emptyset$, but one needs to still minimize over choices of entanglement entropy of the bulk fields on different homology hypersurfaces $\Sigma_{\Le}$ with $\partial \Sigma_{\Le} = \Le$ as the single boundary. Because of the presence of the closed cosmology, one needs to decide whether $\Sigma_{\Le}$ contains the cosmology $\Co$, or not. Thus, the candidates are $\Sigma_{\Le} = \Sigma_{\lef}$ and $\Sigma_{\Le}' =  \Sigma_{\lef} \cup \Sigma_{\Co}$. In the case where the latter dominates, the CFT${}_\Le$ contains the closed cosmology in its entanglement wedge, see Fig. \ref{fig:gen} for a schematic representation.\footnote{Note that the necessary conditions for the formation of cosmological islands in e.g. \cite{Hartman:2020khs,Bousso:2021sji} are trivially satisfied by the empty extremal surfaces of this paper, similar to the closed cosmological islands studied in \cite{Bousso:2022gth}.}

The derivation of eq. \eqref{eq:islandFLM} presented in this section supports the lesson drawn from the west coast model of black hole evaporation, namely that the island formula follows from the replica trick by considering Euclidean saddle points with connectivity between different replicas (i.e. replica wormholes) \cite{Penington:2019kki}. The rules of our calculation have simply been those in the standard FLM replica trick for this particular entangled microstate of two holographic CFTs, and the island formula is simply the quantum-corrected RT formula in this context. An important feature of these cosmological microstates is that the cosmology $\Co$ is \textit{always} in an entanglement island, for either the $\Le$ or the $\Ri$ holographic systems. In fact, the full CFT state is pure, implying $S(\rho_\Le)=S(\rho_\Ri)$. Therefore, when the bulk entropies satisfy $S(\rho_\lef)<S(\rho_\ri)$, eq. \eqref{eq:islandFLM2} gives $S(\rho_\Ri)=S(\rho_\Le)\approx S(\rho_\lef)=S(\rho_{\ri\Co})$ and the cosmology is now an island for the CFT${}_\Ri$.

It is important to remark that the microscopic description of the cosmology is non-perturbatively well defined: it corresponds to a state in the Hilbert space of two holographic CFTs. On the one hand, this setup explicitly shows that potential issues with islands and the mass of the graviton can be avoided (cf. \cite{Geng:2021hlu,Geng:2020qvw}), since in this construction the CFTs do not interact, and the bulk AdS graviton propagates $(d+1)(d-2)/2$ polarizations.\footnote{In Sec. \ref{sec:5}, we address the issue of dressing operators in the cosmology.} On the other hand, it is worth emphasizing the generality of the results in this section, which are expected to apply to any higher dimensional holographic system with Einstein gravity coupled to matter and negative cosmological constant in their low energy bulk effective description.

\section{Cosmology-to-boundary map}
\label{sec:4}

A central question given our construction is how the EFT on the cosmology is encoded in the boundary description. Let us denote by $\mathcal{H}_\Sigma$ the Hilbert space of the bulk EFT living on the two AdS components and the cosmology, by $\mathcal{H}_{\psi}\subset \mathcal{H}_\Sigma$ the Hilbert space obtained by acting on a reference state $\ket{\psi}\in\mathcal{H}_\Sigma$ with an arbitrary number of unitary operators supported only on $\Co$, and by $\mathcal{H}_{\Le}\otimes \mathcal{H}_{\Ri}$ the CFT Hilbert space. A complete answer to the question above would require to elucidate the nature of a \textit{cosmology-to-boundary map}\footnote{As we have discussed in Sec. \ref{sec:1}, this is more precisely a bulk-to-boundary map from the bulk Hilbert space of the cosmology and the two AdS components to the CFT Hilbert space, with the bulk Hilbert space restricted to states for which the reduced density matrix on $\Sigma_\lef\cup\Sigma_\ri$ is fixed, whereas the state of bulk fields in the cosmology can vary. In particular, for all these states the entanglement between the AdS regions and the cosmology is fixed.} of the form
\be\label{eq:cosmotobdyfull}
V_\psi: \mathcal{H}_{\psi} \rightarrow \mathcal{H}_{\Le}\otimes \mathcal{H}_{\Ri}\;.
\ee 
As we will show in this section, constructing concrete realizations of the cosmology-to-boundary map $V_\psi$, valid for certain classes of cosmological states $\ket{\psi}$, is straightforward because the Euclidean CFT preparation of the PETS is known. Our goal in this section is to study aspects of this restricted version of $V_\psi$. In particular we will show that $V_\psi$ does not preserve the inner product structure (i.e. it is non-isometric) in a way which is consistent with the finite microcanonical entropy of the holographic CFTs. Moreover, we will propose a way to explicitly reconstruct operators acting on the cosmology, in a state-dependent way.

Recall that the low-temperature PETS $\ket{\Psi_\Op}$ is prepared with the Euclidean CFT path integral on the cylinder illustrated in Fig. \ref{fig:PETS}. Given this preparation, we can consider additional insertions of local primaries, $\mathcal{O}_i(\mathbf{x}_i)$, where $\mathbf{x}_i = (\tau_i, \Omega_i)$ labels a point on the Euclidean cylinder. We will choose these new insertions to be light, $\Delta_i \ll N^2$. The microscopic CFT state prepared this way is 
\be\label{eq:bdystateexc}
\ket{\Psi_I} =  \ket{\dfrac{1}{\sqrt{Z_I}}\mathcal{T}\lbrace \Op(\tbeta_{\Ri}/2) \mathbb{O}_I(\mathbf{x}_I) \rbrace}  \;,
\ee 
where $\mathcal{T}$ corresponds to the Euclidean time ordering, where in this prescription the Euclidean time lies in the range $0\leq \tau \leq (\tbeta_{\Le} + \tbeta_{\Ri})/2$. The symbol $\mathbb{O}_I(\mathbf{x}_I)$ denotes the product of local operators
\be 
\mathbb{O}_I(\mathbf{x}_I) \equiv \prod_{i}\mathcal{O}_i(\mathbf{x}_i)\;,
\ee
where the index $i$ includes all of the light operator insertions in the preparation of the state. The state \eqref{eq:bdystateexc} has been appropriately normalized by the correlation function
\be\label{eq:norm1}
Z_I = \text{Tr}(\mathcal{T}\lbrace \Op(\tbeta_{\Ri}/2) \mathbb{O}_I(\mathbf{x}_I) \Op^\dagger(-\tbeta_{\Ri}/2)\mathbb{O}_I^\dagger(\mathbf{x}'_I)\rbrace)\;,
\ee 
where $\mathbf{x}'_I$ represents the time reflection-symmetric point $\tau_i \rightarrow - \tau_i$ of $\mathbf{x}_I$ on the Euclidean $\mathbf{S^1}\times \mathbf{S^{d-1}}$, formed by gluing the corresponding cylinders preparing the unnormalized bra and ket.

Following \cite{Chandra:2023dgq}, we now choose all of the operator insertions to be heavy enough to admit worldline saddle point approximations in the bulk, while still remaining within the probe approximation (cf. \cite{Chandra:2023dgq} for the black hole interior-to-boundary map). The corresponding primaries have conformal dimensions in the range $1\ll \Delta_i \ll N^2$. The restriction to this subset of operator insertions is sufficient for the purposes of the present paper and will greatly simplify our analysis because, with this choice, insertions of local operators in the CFT Euclidean path integral will correspond to insertions of approximately local bulk operators on the time reflection-symmetric slice $\Sigma$. 

In the Euclidean manifold $X$, the massive bulk fields $\phi_i$ dual to these primaries will follow geodesic worldlines with endpoints $\mathbf{x}_i, \mathbf{x}'_i \in \partial X$ at the conformal boundary, corresponding to the points where the dual operators are inserted in the path integral preparation, as illustrated in Fig. \ref{fig:cosmomapads}. For simplicity, we assume that all of the primary insertions are different so that all of these worldlines will intersect $\Sigma$ once. We classify the insertions in three families\footnote{Finding the full-fledged cosmology-to-boundary map $V_\psi$, valid for light bulk modes of the quantum fields and light operator insertions in the CFT, is beyond the scope of the present paper. The general expectation from intuition built upon the heat kernel is that, for local insertions of light CFT operators for which the geodesic approximation does not hold (i.e. with $\Delta_i \sim 1$), the three families are continuously related and the bulk wavefunctional always contains some support on both the cosmology and the AdS regions. Therefore, generic local operator insertions in the CFT would lead to a modification of the state both in the cosmology and in the AdS regions, and to a change in the entanglement between the two. In order to study the cosmology-to-boundary map for very light operators, one must thus consider local insertions of the bulk operators in the cosmology, which correspond to more complicated non-local insertions in the boundary CFT. Since restricting to operators for which the geodesic approximation applies is enough for our purposes, we will not investigate this more general case here.}, depending on the Euclidean time $\tau_i$ at which the CFT operator $\mathcal{O}_i(\mathbf{x}_i)$ is inserted:

\begin{figure}[h]
\centering
\includegraphics[width = .9\textwidth]{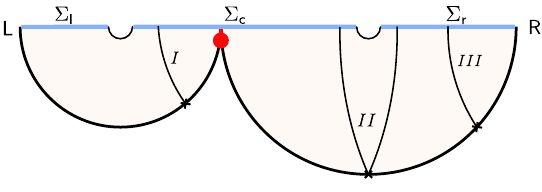}
\caption{The operator insertions with propagate into $X$ in the probe worldline approximation. There are three types of insertions, depending on the Euclidean time $\tau_i$ of the corresponding operator insertion: type $I$ creates a particle on $\Co$, type $II$ creates a linear superposition of a particle on $\Co$ and a particle on $\lef$ or $\ri$, and type $III$ creates a particle on $\lef$ or $\ri$.  }
\label{fig:cosmomapads}
\end{figure}

\begin{itemize}
    \item[$I:$] The operator is inserted at an Euclidean time $\tau_i \in [\frac{\tbeta_{\Le}}{4}+\epsilon_i, \frac{2\tbeta_{\Le} + \tbeta_{\Ri}}{4} - \epsilon_i]$, where $\epsilon_i \sim O(1/\Delta_i)$.  The dominant worldline intersects $\Sigma_\Co$ and the bulk state contains an additional particle in the cosmology.\\ 
    \item[$II:$] The operator is inserted at an Euclidean time $\tau_i \in (\frac{\tbeta_{\Le}}{4}-\epsilon_i,\frac{\tbeta_{\Le}}{4}+\epsilon_i) \cup (\frac{2\tbeta_{\Le} + \tbeta_{\Ri}}{4} - \epsilon_i, \frac{2\tbeta_{\Le} + \tbeta_{\Ri}}{4} + \epsilon_i)$. There are two degenerate worldlines, one intersecting $\Sigma_\Co$, and the other one intersection either $\Sigma_\lef$ or $\Sigma_\ri$. The bulk state contains a linear superposition of a particle in the cosmology and a particle in one of the AdS regions. \\
    \item[$III:$] The operator is inserted at an Euclidean time $\tau_i \in (0,\frac{\tbeta_{\Le}}{4}-\epsilon_i] \cup [\frac{2\tbeta_{\Le} + \tbeta_{\Ri}}{4} + \epsilon_i, \frac{\tbeta_{\Le} + \tbeta_{\Ri}}{2})$. The dominant geodesic intersects $\Sigma_\lef$ or $\Sigma_\ri$ and the bulk state contains an additional particle on the corresponding AdS region.
\end{itemize}

Case $II$ creates an approximate entangled pair of the form $\frac{1}{\sqrt{2}}(\ket{0_{m_i}}_{\Co}\ket{1_{m_i}}_{\lef,\ri} + \ket{1_{m_i}}_{\Co}\ket{0_{m_i}}_{\lef,\ri})$ between the modes in the cosmology and in the AdS regions, where $m_i$ is the mass of the particle. Therefore, adding type $II$ insertions allows one to increase the bulk entanglement with the cosmology by an $O(1)$ amount, within this overly restricted set of CFT states, as advocated more generally in Sec. \ref{sec:2.3}.

Microscopically, the initial increase of the holographic entanglement between the CFT$_\Le$ and CFT$_\Ri$ generated by the type $II$ insertions is more subtle. Originally, for the PETS, the entanglement between CFT$_\Le$ and CFT$_\Ri$ is small and the island formula is not sharp, as we have discussed. As we will see in the next subsection, the cosmological states are also expected to have $O(1)$ overlaps microscopically. In particular, this drastically reduces the amount of microscopic entanglement that type $II$ insertions can generate between the two holographic systems. In simple qubit models, it is possible to see that, under genericity assumptions, the regime of a sharp island formula is attained after the addition of $O(10)$ type $II$ insertions, even if the initial states overlap. After this, additional type $II$ insertions will increase the microscopic entanglement in accordance with the sharp island formula \eqref{eq:islandFLM2}, taking into account the $O(1)$ additional bulk entanglement generated by these insertions. 

\subsection*{Restricted cosmology-to-boundary map}

By fixing a given number of type $II$ operator insertions, we can prepare a bulk reference state $\ket{\psi}$ with a fixed number of entangled pairs between the cosmology and the AdS regions.\footnote{In general, we can also fix a given number of type $III$ insertions, corresponding to having additional particles in the AdS regions in the reference state $\ket{\psi}$.} On top of this state, we add type $I$ operator insertions which create additional particles in the cosmology. Note that these insertions only change the state of bulk fields in the cosmology, leaving the state in the two AdS regions and the entanglement between the cosmology and the AdS regions unchanged. The resulting bulk states will have the form
\be\label{eq:bulkstatec} 
\ket{\psi_I} = \dfrac{1}{\sqrt{z_I}}\prod_i \phi_i(y_i)\ket{\psi}\;,
\ee
where $z_I$ is a normalization constant and $y_i \in \Sigma_\Co$ is the intersection point of the particle's geodesic with the time reflection-symmetric slice $\Sigma_\Co$ (see Fig. \ref{fig:cosmomapads}). The operator $\phi_i(y_i)$ is smeared over the Compton wavelength of the particle $\lambda_i\sim 1/m_i$, and thus the state is normalizable.

In this way, our explicit construction defines the linear map
\be\label{eq:cosmotoboundaryrestrict} 
V_\psi\ket{\psi_I} = \ket{\Psi_I}\;,
\ee
restricted to states of the form \eqref{eq:bulkstatec}, where
\be
\ket{\Psi_I}=\ket{\dfrac{1}{\sqrt{Z_I}}\mathcal{T}\lbrace \Op(\tbeta_{\Ri}/2) \mathcal{O}_\Psi(\mathbf{x}_\Psi) \mathbb{O}_I(\mathbf{x}_I) \rbrace},
\label{eq:generalstate}
\ee
with $\Op(\tbeta_{\Ri}/2)$ shell operator, $\mathcal{O}_\Psi(\mathbf{x}_\Psi)$ the operator insertions necessary to prepare the reference state $\ket{\Psi}$, and $\mathbb{O}_I(\mathbf{x}_I)$ the type $I$ operator insertions associated with the bulk insertion of $\prod_i\phi_i(y_i)$. The restricted cosmology-to-boundary map \eqref{eq:cosmotoboundaryrestrict} is rich enough to contain many interesting features which are expected more generally for the full map \eqref{eq:cosmotobdyfull}. In the remainder of this section, we will study some of the properties of the restricted map $V_\psi$.

\subsection{Non-isometricity}
\label{sec:4.1}

\begin{figure}[h]
\centering
\includegraphics[width = .7\textwidth]{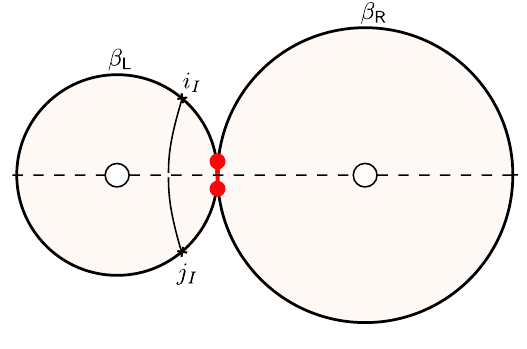}
\caption{The overlap $\overline{G_{ij}}=g_{ij}$, computed gravitationally by the overlap between two bulk states with two different particles, vanishes from the orthogonality of the bulk overlaps between the state of the particle $i_I$ and the particle $j_I$.}
\label{fig:naiveoverlap}
\end{figure}

At this point, we consider a collection of $K$ orthogonal cosmological states $S = \lbrace {\ket{\psi_{I_1}},...,\ket{\psi_{I_K}}}\rbrace $ in the bulk EFT, with each state consisting of one different particle in the cosmology added on top of the bulk reference state $\ket{\psi}$. This reference state is constructed from type $II$ insertions with the prescription described above, and it contains $S_{\lef}$ and $S_{\ri}$ entangled pairs between the cosmology and the AdS${}_\lef$ and AdS${}_\ri$ regions, respectively.\footnote{For the sake of simplifying the notation, we are omitting proportionality constants, and identifying the entanglement entropy of the reduced states $\rho_{\lef,\ri}$ directly with the number of entangled pairs $S_{\lef,\ri}$. The reason is that, since the thermal entropy in the original PETS is very small as we discussed in Sec. \ref{sec:2}, the quantities $S_{\lef,\ri}$ approximately reflect the total bulk entanglement entropy of the states on AdS${}_{\lef,\ri}$. These quantities arise from the number of type II insertions, which create approximate EPR pairs between the AdS regions and the cosmology.} The cosmology-to-boundary map generates a collection of CFT states, $\ket{\Psi_{I_i}} = V_\psi\ket{\psi_{I_i}}$ for $i=1,...,K$. In this setup, we define the respective Gram matrices of overlaps, 
\begin{gather} 
g_{ij} \equiv \bra{\psi_{I_i}}\ket{\psi_{I_j}} = \delta_{ij}\;,\label{eq:gram} \\[.4cm]
G_{ij} \equiv \bra{\Psi_{I_i}}\ket{\Psi_{I_j}} = \bra{\psi_{I_i}}V^\dagger_\psi V_\psi\ket{\psi_{I_j}}\;.\label{eq:gramCFT}
\end{gather}

The vanishing overlap between different bulk EFT states is represented in Fig. \ref{fig:naiveoverlap}. The non-vanishing of the off-diagonal elements of the CFT Gram matrix $G_{ij}$ provides a measure of the non-isometricity of the cosmology-to-boundary map $V_\psi$, i.e., it signals the amount by which $V_\psi^\dagger V_\psi \neq \mathbb{1}$.

The CFT gram matrix $G_{ij}$ is expected to be highly complex to determine microscopically, due to the presence of the operator which creates the thin shell, which is heavy and has very erratic matrix elements in the energy basis, even in the low-lying part of the spectrum of the CFT. We can assume that $G_{ij}$ inherits the pseudo-random structure from this operator,
\be\label{eq:randomgram}
G_{ij} = g_{ij} + R_{ij}\;,
\ee 
where $R_{ij}$ is some $K\times K$ random matrix.

From the construction of the states, the semiclassical connected computation of the CFT Gram matrix produces the EFT Gram matrix 
\be 
\overline{G_{ij}} = g_{ij} = \delta_{ij}\;.
\ee 
where the overline indicates that the quantity is evaluated using the bulk semiclassical description. In terms of \eqref{eq:randomgram}, this means that the EFT only produces the smooth part of the CFT overlaps, i.e. $\overline{R_{ij}} =0$ semiclassically.   

\subsection*{Universal moments of the CFT overlaps from wormholes}

\begin{figure}[h]
\centering
\includegraphics[width = .9\textwidth]{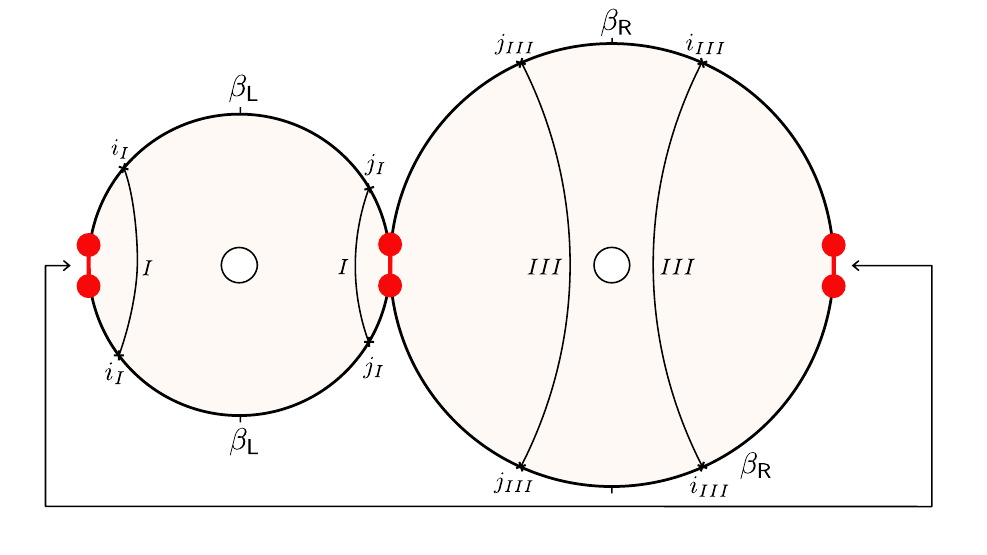}
\caption{Euclidean wormhole geometry $X_2$ contributing semiclassically to the second moment of the overlap $\overline{|G_{ij}|^2}$ in eq. \eqref{eq:overconnect}. It consists of two Euclidean AdS spaces, glued by the trajectories of two thin shells, where the shells on the left and right are identified. Each of the two states $\ket{\Psi_{I_i}}$, $\ket{\Psi_{I_j}}$ contains a different type $I$ and a type $III$ insertion (additional type $II$ insertions that prepare the reference state have been omitted). The type $I$ and type $III$ operator insertions correspond to different primaries, even if they are labeled by the same $i$ and $j$, which corresponds to the label of the bra and the ket in the overlap.}
\label{fig:overlapsq}
\end{figure}

\begin{figure}[h]
    \centering
    \includegraphics[width = .7\textwidth]{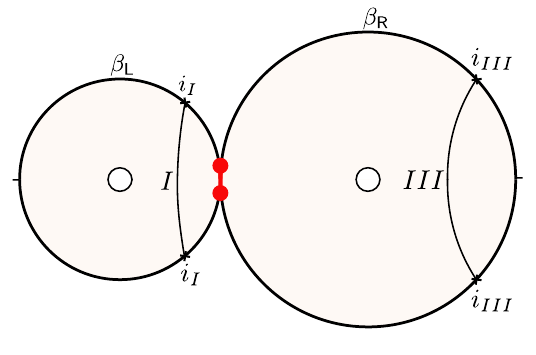}
    \caption{Saddle point geometry $X$ contributing to the semiclassical normalization of the CFT state $\ket{\Psi_{I_i}}$, $\overline{Z_{I_i}} = z_{I_i}$. The state contains a type $I$ insertion on the left, and a type $III$ insertion on the right (additional type $II$ insertions that prepare the reference state have been omitted). The Euclidean time elapsed by type $I$ geodesics is identical to that one elapsed in the wormhole geometry $X_2$ in Fig. \ref{fig:overlapsq}. Contrarily, the Euclidean time elapsed by type $III$ geodesics is smaller than that one elapsed in the wormhole geometry $X_2$ in Fig. \ref{fig:overlapsq}. In this case, both Euclidean times are related by $2\tau_*^{i,X_2}=2\left(\frac{\beta_\Ri}{2}-\tau_*^{i,X}\right)$ in the large shell's mass limit. This discrepancy causes the type $III$ overlaps to be suppressed.}
    \label{fig:normalizations}
\end{figure}

The semiclassical moments of the CFT overlaps, however, receive connected contributions (cf. \cite{Balasubramanian:2022gmo}) 
\be\label{eq:overconnect}
\overline{|G_{ij}|^2}  \sim  \dfrac{z^{ij}_2}{z_{I_i}z_{I_j}} \hspace{.8cm}(i\neq j)\;,
\ee 
coming from the Euclidean wormhole saddle point geometry $X_2$ constructed in Fig. \ref{fig:overlapsq}. The $z_{I_i},z_{I_j}$ factors appearing in the denominator correspond to the semiclassical normalizations of the CFT states $\ket{\Psi_{I_i}}$ and $\ket{\Psi_{I_j}}$, i.e. $\overline{Z_{I_i}} = z_{I_i}$. By construction, these quantities also correspond to the normalizations of the bulk states $\ket{\psi_{I_i}}$ and $\ket{\psi_{I_j}}$, as represented in Fig. \ref{fig:normalizations}. As shown in \cite{Sasieta:2022ksu}, the non-vanishing semiclassical variance given by eq. \eqref{eq:overconnect} corresponds to the microscopic variance $\overline{|R_{ij}|^2}$ over an ensemble of Gram matrices. This ensemble is induced from a coarse-grained version of the thin shell operator, which is characterized in terms of an ensemble of microscopic operators with identical semiclassical features, such as a their thermal two-point functions. Therefore, eq. \eqref{eq:overconnect} must be interpreted as a typical value of the erratic off-diagonal elements of the CFT Gram matrix, and therefore quantifies the non-isometricity of the cosmology-to-boundary map $V_\psi$.

To evaluate the connected overlap \eqref{eq:overconnect}, one needs to include the additional operator insertions propagating geodesically in this geometry and evaluate the quantity
\be\label{eq:overlapij}
z^{ij}_2 = Z[X_2,\phi] \exp(-m_iL_i[X_2] -m_jL_j[X_2])\;,
\ee
where $Z[X_2,\phi]$ is the gravitational action of $X_2$, with the addition of bulk operator insertions that prepare the bulk reference state $\ket{\psi}$ (in this case the type $II$ insertions). The last two factors correspond to the additional type $I$ worldline action of the particle that defines each cosmological state, where $m_i$ is the mass of this particle, and $L_i[X_2]$ is the length of the geodesic that it follows in the Euclidean wormhole $X_2$ (see Fig. \ref{fig:wh2}).

As explained in Sec. \ref{sec:3}, in the large mass limit for the thin shell, the manifold $X_2$ effectively factorizes, and the contributions become
\begin{gather}
Z[X_2,\phi] \approx Z_{\lef}(2\beta_{\Le})Z_{\ri}(2\beta_{\Ri}) (Z_0)^2\\
z_{I_i} \approx Z_{\lef}(\beta_{\Le})Z_{\ri}(\beta_{\Ri})Z_0 \exp(-m_iL_i[X])\;.
\end{gather}
where $Z_0= \exp(-2 m\ell \log (R_*/2\ell))$ is the intrinsic contribution from the shell and $L_i[X]$ is the length of the geodesic that the particle follows in the Euclidean geometry $X$ used to compute the norm of each state $\ket{\psi_{I_i}}$ semiclassically. 

Plugging these expressions back in \eqref{eq:overlapij}, the classical contributions from the shell cancels between numerator and denominator. For general insertions, the connected contribution to the squared of the overlap \eqref{eq:overconnect} is suppressed by a factor
\be\label{eq:suppression}
    \alpha=e^{-m_i\left(L_i[X_2]-L_i[X]\right)-m_j\left(L_j[X_2]-L_j[X]\right)}.
\ee

Since both $X$ and $X_2$ are locally AdS spaces, the geodesic lengths $L_i[X]$ and $L_i[X_2]$ only depend on the difference in Euclidean time between the endpoints of the trajectories on each manifold. In particular, the length of a generic radial geodesic connecting two points on the asymptotic boundary separated by an amount of global Euclidean time $2\tau_*$ is given by
\be
    L=2\ell \log\left[2\sinh\left(\frac{\tau_*}{\ell}\right)\right],
    \label{eq:geodesiclength}
\ee
where we neglected a universal divergent contribution coming from the near-boundary region, which cancels out in the combination between the different lengths in the suppression factor \eqref{eq:suppression}. In our time reflection-symmetric setup, $\tau_*$ represents the Euclidean time difference between the insertion point and the time reflection-symmetric slice $\tau=0,\beta/2$ for a thermal circle of length $\beta$. Therefore, for a given insertion point $\tau_i\in (0,\frac{\beta}{2})$, we have $\tau_*=\min \{\tau_i,\frac{\beta}{2}-\tau_i\}$.

For type $I$ insertions which create particles in the cosmology, the value of $\tau_*$ is the same in the geometries $X$ and $X_2$. In fact, restricting for simplicity to insertions in the left and to the large shell's mass limit for which $\tilde{\beta}_\Le\approx \beta_\Le$, type $I$ insertions in $X$ have $\tau_i\in [\frac{\beta_L}{4}+\epsilon_i,\frac{\beta_L}{2}-\epsilon_i]$, yielding $\tau_*^{i,X}=\frac{\beta_\Le}{2} -\tau_i\in (0,\frac{\beta_\Le}{4})$. In the wormhole geometry $X_2$, the inverse temperature of the left Euclidean AdS is $\beta_{\Le,2}=2\beta_{\Le}$, with the time reflection-symmetric slice given by $\tau=0,\beta_\Le$. In this case, type $I$ insertions have $\tau_i\in (0,\frac{\beta_\Le}{4}-\epsilon_i]\cup [\frac{3\beta_\Le}{4}+\epsilon_i,\beta_\Le)$. The value of $\tau_*^{i,X_2}\in (0,\frac{\beta_\Le}{4})$ for a given insertion is therefore the same as that one for the $X$ geometry (see Figs. \ref{fig:overlapsq} and \ref{fig:normalizations}). As a consequence, the geodesic lengths \eqref{eq:geodesiclength} for a given insertion satisfy $L_i[X]=L_i[X_2]$, and the suppression factor $\eqref{eq:suppression}$ is identically one.

This leads to the universal answer for the second moment of the overlap, independent of the nature of the type $I$ insertions,
\be\label{eq:overlapvariance}
\overline{|G_{ij}|^2}  =  \dfrac{Z_{\lef}(2\beta_{\Le})Z_{\ri}(2\beta_{\Ri})}{Z_{\lef}(\beta_{\Le})^2Z_{\ri}(\beta_{\Ri})^2} = \dfrac{1}{d_\lef d_\ri}\;.
\ee
The suppression factors are given by $\log d_{\lef,\ri}= S_2(\rho_{\lef,\ri})$, which correspond to the second R\'{e}nyi entropy of the bulk reduced states \eqref{eq:stateredl} and \eqref{eq:stateredr} to the AdS regions. 

If the second R\'{e}nyi entropies are sufficiently large, the square of the overlap is small, and states that are orthogonal in the bulk EFT due to the presence of different particles in the cosmology are mapped to nearly orthogonal states in the CFT. In this case, the cosmology-to-boundary map is approximately-isometric for general states (see Sec. \ref{sec:5.2} for a quantitatve analysis of the size of the entropies needed for the map to be approximately-isometric for \textit{all} simple states). On the other hand, if $d_\lef,d_\ri=O(1)$, states that are orthogonal in the bulk EFT are mapped to nearly parallel (but still linearly independent, and therefore in principle distinguishable) states in the CFT. In this case, the map is highly non-isometric. We want to remark that such non-isometricity does not obviously imply that the map is non-invertible, but simply that it does not preserve inner products. We will comment more on this point in Sec. \ref{sec:4.2}.

\begin{figure}[h]
    \centering
    \includegraphics[width=0.6\textwidth]{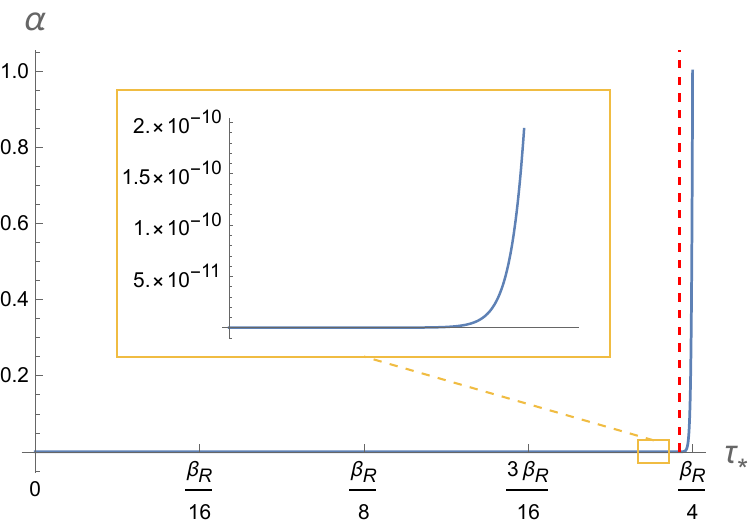}
    \caption{Suppression factor $\alpha$ given by eq. \eqref{eq:suppression} as a function of the Euclidean time difference between the insertion point of the operators and the time reflection-symmetric slice, where we set for simplicity $\tau_*^{i,X}=\tau_*^{j,X}\equiv \tau_*$. We chose $\ell=1$, $\beta_\Ri=1$, $m_i=50$, $m_j=40$ in eq. \eqref{eq:suppression}. Insertions to the right of the dashed red line at $\tau_*=\frac{\beta}{4}-\epsilon$ with $\epsilon=1/m_j\approx 1/\Delta_j$ are type $II$ and therefore are not relevant for our analysis. For all relevant values of the insertion point $\tau_*\in(0,\frac{\beta_\Ri}{4}-\epsilon)$, the suppression factor is very small and drops rapidly as we decrease $\tau_*$ (see the inset). This implies that the overlap computed by the gravitational Euclidean path integral for CFT states corresponding to (orthogonal) bulk EFT states with different particles in the AdS regions is very small. In this case, the bulk-to-boundary map is expected to be isometric.}
    \label{fig:suppression}
\end{figure}

It is interesting to evaluate $\overline{|G_{ij}|^2}$ also in the case in which the operator insertions are of type $III$, resulting in two orthogonal bulk EFT states $\ket{\psi_{I_i}}$ and $\ket{\psi_{I_j}}$ with two different particles in the right (or left) AdS region.\footnote{Note that these insertions modify the reference state $\ket{\psi}$ and therefore take us out of the Hilbert space $\mathcal{H}_\psi$ on which the cosmology-to-boundary map $V_\psi$ is defined.} In this case, most of the analysis carried out above still holds, with the difference that the suppression factor \eqref{eq:suppression} is non-vanishing. In fact, by a similar reasoning to the one used for type $I$ insertions, we can conclude that $\tau_*^{i,X}\in (0,\frac{\beta_\Ri}{4})$, and $\tau_*^{i,X_2}=\frac{\beta_\Ri}{2}-\tau_*^{i,X}\neq \tau_*^{i,X}$, see Figs. \ref{fig:overlapsq} and \ref{fig:normalizations}. It is then straightforward to show that the suppression factor takes the form
\be
\alpha=\left[\frac{\sinh\left(\frac{\beta_\Ri}{2\ell}-\frac{\tau_*^{i,X}}{\ell}\right)}{\sinh\left(\frac{\tau_*^{i,X}}{\ell}\right)}\right]^{-2m_i\ell}\left[\frac{\sinh\left(\frac{\beta_\Ri}{2\ell}-\frac{\tau_*^{j,X}}{\ell}\right)}{\sinh\left(\frac{\tau_*^{j,X}}{\ell}\right)}\right]^{-2m_j\ell}.
\ee
For $\tau_*^{i,X},\tau_*^{j,X}\to \frac{\beta_\Ri}{4}$ we obtain $\alpha\to 1$, whereas for $0<\tau_*^{i,X},\tau_*^{j,X}<\frac{\beta_\Ri}{4}$ the suppression factor decreases rapidly (we recall that $m_i\ell,m_j\ell\gg 1$ in order for the geodesic approximation to hold), see Fig. \ref{fig:suppression}.

We can therefore conclude that the CFT overlap computed by our replica calculation for states that are orthogonal in the bulk EFT due to the presence of different particles in AdS${}_{\lef,\ri}$ is extremely small. This result holds independently of the amount of entanglement between the cosmology and the AdS regions. We recall that the bulk replica calculation we carried out gives only an approximation to the true overlap between the CFT states. In fact, it relies on a saddle point approximation for the bulk geometry and on the geodesic approximation for the description of operator insertions. The fact that such an approximate overlap is very small and that the ordinary holographic dictionary applies to the two AdS components of our bulk spacetime suggests that the bulk-to-boundary map restricted to the Hilbert space of states generated by type $III$ insertions on top of some reference state is isometric, and the true CFT overlap vanishes.

The generalization to higher moments is also possible (with the AdS version of the Euclidean wormholes constructed in \cite{Sasieta:2022ksu,Balasubramanian:2022gmo} for the black hole case)
\be\label{eq:uniovern} 
\overline{G_{i_1i_2}G_{i_2i_3}...G_{i_ni_1}}|_{\text{conn.}}  \sim  \dfrac{Z_{\lef}(n\beta_{\Le})Z_{\ri}(n\beta_{\Ri})}{Z_{\lef}(\beta_{\Le})^n Z_{\ri}(\beta_{\Ri})^n} \;,
\ee
where the label conn. stands for the completely connected contribution between $n$ asymptotic boundaries. The particles travel through the Euclidean wormhole solution $X_n$ between the different components of the asymptotic boundary. The universal form of \eqref{eq:uniovern} arises in the large mass limit for the thin shell, for the same reason as for the second moment. Namely, the geodesic lengths for type $I$ insertions cancel out with the normalization factors, and the overlap becomes independent of the details of the states. \footnote{In fact, the situation is a little bit more subtle than in \cite{Balasubramanian:2022gmo}. In this case, since there is a single heavy shell $\Op$ for all states, there are in fact $n$ Euclidean wormholes, labelled by a cyclic permutation $g\in \text{Sym}(n)$, associated to the corresponding completely connected thin shell operator contractions through the bulk. It is easy to see that the dominant configuration consist of $g= \eta = (12...n)$, since in this configurations the additional light particles travel the smallest possible distance across the wormhole.}

\subsection{Cosmological bag of gold?}
\label{sec:4.2}

The rank of the $K\times K$ Gram matrices of overlaps \eqref{eq:gram} and \eqref{eq:gramCFT} provides information about the number of linearly independent states on $S$ and on $V_\psi S = \lbrace \ket{\Psi_{I_1}},...,\ket{\Psi_{I_K}}\rbrace $. For the former, there are $K$ independent states by assumption,
\be
\text{rank}(g_{ij}) = \text{dim} (\text{span}\lbrace S\rbrace) = K\;.
\ee

The question is to see how many independent states there are on $V_\psi S$,
\be
\text{rank}(G_{ij}) = \text{dim} (\text{span}\lbrace V_\psi S\rbrace) \leq K \;.
\ee

An important point to notice is that the question about $\text{rank}(G_{ij})$, formulated this way, has no relation with the non-isometricity of the map $V_\psi$, but rather with the existence of an inverse $V^{-1}_\psi$, such that $V^{-1}_\psi V_\psi S = S$. Any invertible map $V_\psi$, albeit non-isometric, will produce $\text{rank}(G_{ij}) = K$. Since the CFT Hilbert space is infinite dimensional, and all of the cosmological states contain thermal tails (i.e. when written as a superposition of energy eigenstates, they have exponentially suppressed but non-vanishing coefficients for all energy eigenstates), it is difficult to think how any finite collection of states built this way can conspire to be linearly dependent in the CFT. Therefore we expect all these states to be linearly independent, although not orthogonal to each other. An analogy of this situation is the set of coherent states of fixed average energy for the harmonic oscillator, which generate a basis of the Hilbert space precisely due to the tails in energy (see Appendix. \ref{app:C}).\footnote{The CFT states $\ket{\Psi_{I_i}}$ also have exponentially suppressed branches in their semiclassical wavefunction which corresponds to two-sided black hole microstates analogous to those constructed in \cite{Balasubramanian:2022gmo}. For this reason, it is reasonable that their tails can generate the high-energy density of states of the CFT as well.}

On the other hand, bag of gold paradoxes arise microcanonically in the CFT, where the number of states really remains finite. They have to do with the fact that different states of the cosmological EFT, with different features on the cosmology, have approximately the same energy, and thus must look linearly dependent when truncating the CFT Hilbert space by projecting onto the dominant microcanonical band. In this case, not all bulk EFT states are distinguishable from the CFT point of view, which means that the map $V_\psi$ has a non-trivial kernel when restricted to some microcanonical band in the CFT. We will now show that the universal overlaps \eqref{eq:uniovern} found in the previous subsection are consistent with the microcanonical density of states of the holographic CFT. In order do that, we consider the microcanonical version of the Gram matrix 
\be
G^E_{ij} =  \bra{\psi_{I_i}} V_\psi^\dagger \Pi_{E} V_\psi\ket{\psi_{I_j}} \;,
\ee
where $\Pi_{E}$ with $E = (E_\lef,E_\ri)$ is an orthogonal projection onto the corresponding microcanonical windows of energies $[E_{\lef,\ri},E_{\lef,\ri}+\Delta E_{\lef,\ri}]$, with $\Delta E_{\lef,\ri}\ll E_{\lef,\ri}$. 

We can expand the bulk partition functions as a sum over energies
\begin{gather}
Z_{\lef}(n\beta_\Le) = \int \dfrac{\text{d}E}{E} e^{S(E)} (\hat{\rho}_\lef^n)_{E,E}\\
Z_{\ri}(n\beta_\Ri) = \int \dfrac{\text{d}E}{E} e^{S(E)} (\hat{\rho}_\ri^n)_{E,E}\;,
\end{gather}
where $S(E)$ is the microcanonical entropy of the gas of particles in AdS. From standard AdS/CFT considerations, this same entropy corresponds to the microcanonical entropy of the CFT, in the microcanonical confined phase.

Given this, and assuming that the reduced bulk state is full rank and approximately constant on the microcanonical window of interest, the microcanonical contribution of the window at energy $E = (E_\lef,E_\ri)$ to the semiclassical overlaps \eqref{eq:uniovern} is given by
\be 
\overline{G^E_{i_1i_2}G^E_{i_2i_3}...G^E_{i_ni_1}}|_{\text{conn.}} \approx e^{-(n-1)\mathbf{S}} \;,
\ee 
where $\mathbf{S} = (S(E_\lef)+S(E_\ri))$ is the full microcanonical entropy of both CFTs.

\subsection*{Eigenvalue density of the microcanonical Gram matrix}

In linear algebra, a way to compute the rank of a matrix, in this case the microcanonical Gram matrix $G^E_{ij}$, is to consider its resolvent
\be \label{eq:resol}
R_{ij}(\lambda)\,\equiv\,\left( \frac{1}{\lambda \mathds{1} -G^E}\right)_{ij} \,=\,\frac{1}{\lambda}\,\delta_{ij}+\sum\limits_{n=1}^{\infty}\,\frac{1}{\lambda^{n+1}}\,[(G^E)^n]_{ij}\;.
\ee
The density of eigenvalues $\lambda$ of the Gram matrix $G^E_{ij}$, denoted by $D(\lambda)$, will follow from the discontinuity of the trace of its resolvent, $R(\lambda) = \sum\limits_{i=1}^K R_{ii}(\lambda)$, along the imaginary axis:
\be\label{eq:densitydisc}
D(\lambda)=\frac{1}{2\,\pi\,i}\left(\,R(\lambda-i\epsilon)-R(\lambda+i\epsilon)\,\right)\;.
\ee

The advantage of using this trick is that for our present purposes is that, in semiclassical gravity, each term in the series expansion of the trace of the resolvent can be computed directly from the semiclassical moments of the overlaps 
\be \label{resolav}
\lambda \overline{R(\lambda)}\,=\,\,K+\sum\limits_{n=1}^{\infty}\,\frac{1}{\lambda^{n}}\,\overline{\text{Tr}((G^E)^n)} \;.
\ee
Evaluating each term of the right hand side semiclassically produces a simple diagrammatic expansion, where \eqref{resolav} acquires the form of a Schwinger-Dyson equation (see \cite{Penington:2019npb,Balasubramanian:2022gmo} for details). The latter can be explicitly resumed 
\be 
\lambda \overline{R(\lambda)}\,=\,\,K+e^{\mathbf{S}}\sum\limits_{n=1}^{\infty}\,\,\left(\dfrac{\overline{R(\lambda)}}{e^{\mathbf{S}}}\right)^n = K + \frac{e^\textbf{S}\,\overline{R(\lambda)}}{e^\textbf{S}-\overline{R(\lambda)}}\;.
\ee

Thus, the trace of the semiclassical resolvent satisfies the following quadratic equation
\be 
\overline{R(\lambda)}^2+\left(\,\frac{e^\textbf{S}-K}{\lambda}-e^\textbf{S}\,\right)\,\overline{R(\lambda)}+\dfrac{K}{\lambda}\,e^\textbf{S}=0\;.
\ee
The density of states follows from the discontinuity across the real axis of the trace of the resolvent of the Gram matrix (eq.~\ref{eq:densitydisc}), which gives
\be \label{denG}
\overline{D(\lambda)}=\frac{e^\textbf{S}}{2\pi\lambda}\sqrt{\,\left[\lambda-\left(1-K^{1/2}\, e^{-\textbf{S}/2} \right)^2\,\right]\left[\,\left(1+K^{1/2} \,e^{-\textbf{S}/2} \right)^2-\lambda \right]}+\delta(\lambda)\left(K-e^{\mathbf{S}}\right)\theta(K-e^{\mathbf{S}})\;,
\ee

Such a density of states has a continuous part (first term) and a discontinuous part (second term). The continuous part accounts for the non-zero eigenvalues of the Gram matrix, which belong to the domain
\be 
\left(1-K^{1/2}\, e^{-\textbf{S}/2} \right)^2\,<\,\lambda\,<\,\left(1+K^{1/2} \,e^{-\textbf{S}/2} \right)^2\;.
\ee
The singular part counts the number of zero eigenvalues. The latter only contributes when $K>e^{\textbf{S}}$ due to the Heaviside factor, and gives a total of $K-e^{\textbf{S}}$ zero eigenvalues. Therefore, this shows that
\be\label{eq:maxstatescosmo}
\text{rank}(G^E_{ij}) = \text{min}\lbrace K,  e^{\textbf{S}}\rbrace\;. 
\ee 

Eq. \eqref{eq:maxstatescosmo} is the main result of this subsection. It shows that, when restricting to a CFT microcanonical band, the rank of the Gram matrix of overlaps, and therefore, the dimension of the Hilbert space generated by the cosmological microstates, is upper bounded by the microcanonical density of states of the CFTs, which is agreement with microscopic considerations. Therefore, by choosing $K>e^{\textbf{S}}$ orthogonal states in the cosmological EFT, not all of them will be mapped by $V_\psi$ to linearly independent states in the microcanonical window, and some of them will be linear combination of others. In particular, $e^{\textbf{S}}$ of such cosmological states will suffice to span the microcanonical window of the CFTs.\footnote{This statement is true with high probability, within the effective random matrix description that semiclassical gravity provides for the Gram matrix of overlaps. We always assume that the microscopic Gram matrix, which depends on details of the heavy operator, is typical in this ensemble of Gram matrices.}

\subsection{Reconstruction of operators in the cosmology}
\label{sec:4.3}

Given the explicit map $V_\psi$ in \eqref{eq:cosmotoboundaryrestrict}, we can ask how the operators that create particles in the cosmology look like in the CFT. In particular, consider the bulk EFT operator $\phi(y)$ defined in eq. \eqref{eq:bulkstatec}, which creates a particle of mass $m_i$ at $y\in \Sigma_{\Co}$. The reconstruction of this operator is formally given by
\be\label{eq:euclideanreconstruction}
\Phi = V_\psi\phi(y)V_\psi^{-1}\;.
\ee
By definition $\Phi\ket{\Psi_I}=\phi\ket{\psi_I}$, where $\ket{\Psi_I}=V_\psi\ket{\psi_I}$. In other words, the CFT operator $\Phi$ acts at the time reflection-symmetric slice on the state $\ket{\Psi_I}$ dual to a semiclassical state with no particle in the cosmology, and maps it to the state $\ket{\Psi'_I} = \Phi\ket{\Psi_I}$ dual to a semiclassical state with a particle in the cosmology.

The bulk weakly coupled field $\phi$ is dual to a CFT generalized free field $\mathcal{O}_{\phi}$, which can be identified in general using the extrapolate dictionary. In simple bulk setups such as a pure AdS spacetime, the insertion of the operator $\phi$ in the bulk corresponds to the insertion of the dual operator $\mathcal{O}_{\phi}$ smeared by a non-local kernel. This is the so-called HKLL dictionary \cite{Hamilton:2005ju,Hamilton:2006az}, which enables the reconstruction of operators within the causal wedge of a boundary subregion and is manifestly state-independent. This reconstruction can be applied in our case to insertions of the operator $\phi$ in one of the AdS regions, i.e. on $\Sigma_{\lef,\ri}$, leading to a state-independent reconstruction of bulk operators in AdS${}_{\lef,\ri}$.

On the other hand, the cosmology is causally disconnected from the dual boundary CFTs, implying that a bulk operator insertions on $\Sigma_\Co$ cannot be reconstructed using HKLL. However, as we have explained in Sec. \ref{sec:2} and \ref{sec:4}, we can associate the insertion of the bulk operator $\phi$ at some given point $y\in \Sigma_\Co$ to an insertion of the CFT operator $\mathcal{O}_\phi$ at a specific Euclidean time $\tau_\phi$ in the past of the Euclidean path integral preparing the state with a particle in the cosmology. The CFT operator $\Phi$ can thus be defined in terms of such an insertion in the Euclidean past. 

In this ``Euclidean reconstruction'', the Euclidean time-ordering of the operator insertions is clearly very important. A given state $\ket{\Psi_I}$ in the class of CFT states of our interest is built using insertions in the Euclidean past of several operators, see eq. \eqref{eq:generalstate}: the shell operator $\mathcal{O}$, type $II$ and type $III$ insertions (which we denoted by $\mathcal{O}_\Psi$ in \eqref{eq:generalstate}) necessary to prepare the reference state $\ket{\Psi}$, and type $I$ insertions (which we collectively indicate with $\mathbb{O}_I(\mathbf{x}_I)$) to prepare the final state. The operator $\mathcal{O}_\phi$ must be inserted at the specific Euclidean time $\tau_\phi$ in order for the resulting state prepared by the path integral to be given by $\Phi\ket{\Psi_I}$.

It is then clear that the Euclidean reconstruction we are considering is manifestly state-dependent and the correct boundary representation $V_\psi \phi(x) V_\psi^{-1}$ has to be reconstructed matrix element by matrix element, within the space of states representing the cosmological EFT in the CFT. Namely the matrix elements of the dual CFT operator need to be defined as
\be\label{eq:statedeprec}
\Phi_{IJ} = \bra{\Psi_I}V_\psi \phi(x) V_\psi^{-1}\ket{\Psi_J} = \dfrac{1}{\sqrt{Z_IZ_J}}\text{Tr}(\mathcal{T}\lbrace \Op(\tbeta_{\Ri}/2)\mathcal{O}_\Psi(\mathbf{x}_\Psi) \mathbb{O}_J(\mathbf{x}_J) \mathcal{O}_\phi(\mathbf{x}_\phi) \Op^\dagger(-\tbeta_{\Ri}/2)\mathcal{O}^\dagger_\Psi(\mathbf{x}_\Psi')\mathbb{O}_I^\dagger(\mathbf{x}'_I)\rbrace)\;,
\ee 
where $\mathbf{x}'_I,\mathbf{x}_\Psi'$ again indicate the time reflection-symmetric points associated with operator insertions at the point $\mathbf{x}_I,\mathbf{x}_\Psi$. The left hand side of \eqref{eq:statedeprec} is a Euclidean CFT path integral on $\mathbf{S}^1\times \mathbf{S}^{d-1}$, with the corresponding operator insertions, which defines the matrix element $\Phi_{IJ}$. 

By construction, this operator has the same matrix elements as the operator on the cosmology, when acting on cosmological microstates
\be 
\bra{\Psi_I} \Phi\ket{\Psi_J}  = \bra{\psi_I} \phi(y) \ket{\psi_J}\;.
\ee 

It is worth noticing that, although this reconstruction is state-dependent and we are not aware of a way to build a state-independent one, we did not prove explicitly that such a state-independent reconstruction does not exist. We leave this interesting endeavour to future work. Nonetheless, given the specific features of our setup (in particular, the fact that the cosmology is an island), it is reasonable to expect bulk reconstruction in the cosmology to be in general state-dependent. 

\vspace{1cm}

Before moving to the study of a TN model of our setup, which will allow us to study explicitly a number of features of the cosmology-to-boundary map, we would like to make two final remarks. First, if we are in the regime in which the island formula applies (see Sec. \ref{sec:3}) and the cosmology is therefore encoded either in the left or the right CFT, all the results we obtained in the present section (which apply in general to the two-sided cosmology-to-boundary map) are expected to also be valid for the single-sided quantum channel encoding the cosmology in one of the two CFTs. Second, in the highly non-isometric regime of the cosmology-to-boundary map, states that can be distinguished by a one-shot bulk EFT experiment in the cosmology (i.e. orthogonal states in the bulk) cannot be distinguished by a one-shot CFT experiment, because they are not mapped to orthogonal, nor even nearly orthogonal, CFT states. This in turn implies that well-defined experiments that can be performed by a bulk observer do not necessarily correspond to well-defined experiments for the CFT observer. In other words, the result of bulk experiments can only be predicted if one has control over the CFT, but such predictions require knowledge (to very good accuracy, as we have discussed) of the CFT state and of the cosmology-to-boundary map, which cannot simply be obtained by a one-shot boundary experiment. The detailed investigation of the consequences of this result is an interesting direction to be pursued in future work.

\section{Tensor network toy model}
\label{sec:5}

In this section, we present a simple TN toy model of the cosmological states, which qualitatively captures all of the gravitational features obtained in the previous sections, but which drastically simplifies the study of the system. Because we will be working with finite dimensional qubit models, we will need some care to interpret additional features of the model, such as the appearance of null states. We will use this tractable model to study the properties of the cosmology-to-boundary map.

Mimicking the Euclidean CFT preparation of the PETS, we consider a pair of thermofield-double states $\ket{\text{TFD}}_{\Le \aux}$ and $\ket{\text{TFD}}_{\aux' \Ri}$ at physical inverse temperatures $\beta_{\Le}$ and $\beta_{\Ri}$, respectively, defined on $\mathcal{H}_{\Le} \otimes \mathcal{H}_{\aux}$ and $\mathcal{H}_{\aux'} \otimes \mathcal{H}_{\Ri}$, for two auxiliary copies of the CFT, $A,A'$. Consider the maps $\mathcal{N}_{\aux}$ and $\mathcal{N}_{\aux'}$ acting on $\mathcal{H}_{\aux}$ and on $ \mathcal{H}_{\aux'}$ whose effect is to coarse-grain each auxiliary part of the TFD states, in the sense of the holographic RG flow, up to an energy scale set by the radius of the shell, $E_{s} \sim R_*/\ell^2 \gg \beta_{\Le,\Ri}^{-1}$. The unnormalized PETS \eqref{eq:PETS} will then arise from the projection 
\be\label{eq:stateproj}
\ket{\overline{\Psi}_{\Op}} = \bra{\Op}_{\aux \aux'}  (\mathcal{N}_{\aux}\ket{\text{TFD}}_{\Le \aux} \otimes \mathcal{N}_{\aux'}\ket{\text{TFD}}_{\aux'\Ri})\;,
\ee 
where projecting onto $\ket{\Op}_{\aux \aux'} = \Op^\dagger_{\aux} \ket{\text{MAX}}_{\aux \aux'} = \sum_{n,m} \Op^*_{nm} \ket{E_n}^*_{\aux}\otimes \ket{E_m}_{\aux'}$ has the effect of gluing the two AdS spaces, forming the cosmology. We indicated with $\ket{\text{MAX}}_{\aux \aux'}$ the maximally entangled state between $\aux$ and $\aux'$. To match \eqref{eq:PETS} one needs to impose that\footnote{The positive exponent in \eqref{eq:cg1} and \eqref{eq:cg2} is a manifestation of the fact that the natural time at the auxiliary systems, given by the Killing Hamiltonian, runs oppositely to the time at the CFT boundaries.}
\begin{gather}
\mathcal{N}_{\aux} = e^{\frac{\Delta \tau_{-}}{2} H_{\aux}} \;, \label{eq:cg1}\\
\mathcal{N}_{\aux'} = e^{\frac{\Delta \tau_{+}}{2} H_{\aux'}}\,.\label{eq:cg2}
\end{gather}

\begin{figure}[h]
\centering
\includegraphics[width = .7\textwidth]{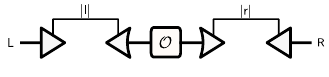}
\caption{TN for the unnormalized cosmological state $\ket{\overline{\Psi}_{\mathcal{O}}}$.}
\label{fig:TNPETS}
\end{figure}

The construction of the TN with \eqref{eq:stateproj} in mind is represented in Fig. \ref{fig:TNPETS}. Each TFD state at low temperatures is described by two spherical MERA TNs, represented by gray isometries $I_i$, for $i =\lef, \Co_{\lef}, \Co_{\ri}, \ri$ respectively. These are pairwise connected along the bulk dangling legs which represent the bulk entanglement, in the form of microcanonical EPR pairs. 

In total, there are two such TFDs, with bulk entanglement $S_{\lef}$ and $S_{\ri}$. The action of $\mathcal{N}_{\aux},\mathcal{N}_{\aux'}$ is to introduce a radial cutoff in the two auxiliary MERAs at a bulk radial distance $R_*$, which is depicted by curving the corresponding edge of the gray triangles. The effect of the projection onto $\bra{\Op}_{\aux \aux'}$ is to glue these two cutoff MERAs at $R_*$, adding the heavy matter operator $\mathcal{O}$ in between them.

The construction can naturally incorporate more general states with larger amounts of entanglement with the addition of EPR pairs between the AdS regions and the cosmology. This allows us to consider general values of $S_{\lef}$ and $S_{\ri}$ in the TN, at least within the microcanonical entropy bound $S_{\lef},S_{\ri} \lesssim N^2$. For convenience, we shall define $|\lef| = 2^{S_{\lef}}$ and $|\ri| = 2^{S_{\ri}}$ as the bond dimensions of the corresponding cuts in the network.

The unnormalized state represented by the TN has the norm 
\be\label{eq:normTN} 
Z_1 = \bra{\overline{\Psi}_{\mathcal{O}}}\ket{\overline{\Psi}_{\mathcal{O}}} = \text{Tr}(\mathcal{O}\,\Pi_\Co^{\lef} \mathcal{O}^\dagger \Pi^{\ri}_\Co)\,
\ee 
for the orthogonal projectors $\Pi_\Co^{\lef} = I_{\Co_\lef}I^\dagger_{\Co_\lef} $ and $\Pi_\Co^{\ri} =I_{\Co_\ri}I^\dagger_{\Co_\ri}$.

In the AdS/CFT construction of the cosmological microstates, the norm of the PETS given after eq. \eqref{eq:PETS} was positive-definite. This is in contrast with \eqref{eq:normTN}, which can vanish for certain choices of the $\mathcal{O}$ tensor, for which $\Pi_\Co^{\lef} \mathcal{O}^\dagger \Pi^{\ri}_\Co =0$.\footnote{Intuitively, the number of legs contracted with the tensor $\mathcal{O}$ can be larger than the number of legs connecting $\Co$ with $\Le$ and $\Ri$. Therefore, not all possible choices of $\mathcal{O}$ can be mapped to different states on the external legs of the TN.} This simply showcases the difference between finite dimensional qubit models and thermal states in an infinite-dimensional Hilbert space of a quantum field theory. Null states of this TN must be interpreted with some care: as we have discussed in Sec. \ref{sec:4.2}, they seem to arise from an operational point of view, namely due to the inability of the boundary observer to distinguish thermal tails of the wavefunction to arbitrary accuracy. This finite resolution is realized in the TN model by the restriction to a finite-dimensional Hilbert space.\footnote{A more faithful model, which we will not develop in this paper, would be to include the thermal tails of the bulk entanglement in an enlarged bulk Hilbert space, $\mathcal{H}_{\lef}\otimes \mathcal{H}_{\ri} \rightarrow \oplus_{i,j} (\mathcal{H}^i_{\lef}\otimes \mathcal{H}^j_{\ri})$, including different microcanonical bands.  The contribution of each band would need to be weighted by associated Boltzmann suppression factors.} Since the tensors $\mathcal{O}$ that make $Z_1$ vanish in this toy model are a measure-zero subset of all complex matrices, we shall ignore them for most of this paper, although their presence plays a role in the discussion of Sec. \ref{sec:5.2}.

Thus, in analogy with the gravitational preparation, given the TN state $\ket{\overline{\Psi}_{\mathcal{O}}}$ we will rescale it to get a normalized state  
\be 
\ket{\Psi_{\mathcal{O}}} = \dfrac{1}{\sqrt{Z_1}}\ket{\overline{\Psi}_{\mathcal{O}}}\in \mathcal{H}_\Le\otimes \mathcal{H}_\Ri\;.
\ee

Already at this level, any such TN state $\ket{\Psi_{\mathcal{O}}}$ captures by construction a weak version of the island formula
\be\label{eq:islandtn}
S(\rho_{\Le}), S(\rho_{\Ri}) \leq \text{min}\lbrace S_{\lef}, S_{\ri}\rbrace \;,
\ee 
whereas the strong version, corresponding to the approximate saturation of the inequality, will depend on the particular choice of tensors in the network. 

\subsection*{Coarse-grained shell tensor}

In semiclassical gravity, the heavy operator $\mathcal{O}$ is treated in a coarse-grained sense, as a gaussian random matrix in the energy basis, at the level of the two-point function (cf. \cite{Sasieta:2022ksu}). This suggests to follow the same logic here, and consider $\mathcal{O}_{ij}$ as a random matrix in the computational basis of the TN, at least for simple enough observables. In this vein, we pick $\bra{\Op}_{\aux\aux'}$ randomly from a complex gaussian probability distribution, with zero mean and unit variance, corresponding to the measure
\be\label{eq:measure}
\int \text{d}\mu[\mathcal{O}] =  \int \prod_{ij}\frac{\text{d}\Re{\mathcal{O}_{ij}}\text{d}\Im{\mathcal{O}_{ij}}}{2\pi} \,\exp(- \frac{1}{2}\text{Tr}(\mathcal{O}\mathcal{O}^\dagger))\;.
\ee

\subsection*{Island formula}

\begin{figure}[h]
\centering
\includegraphics[width = \textwidth]{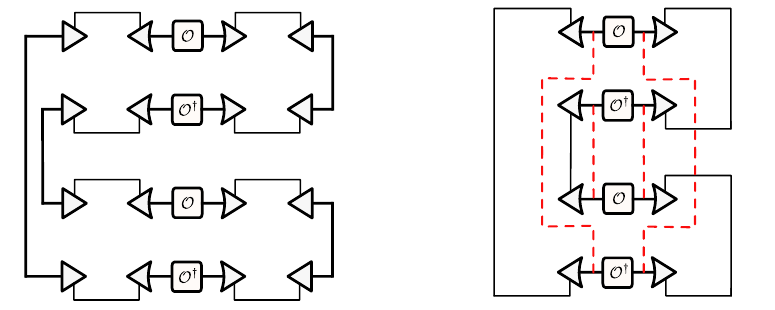}
\caption{Left: unnormalized purity $\text{Tr}(\hat{\rho}_{\Le}^2)$ for the TN state before gaussian average. Note that the contracted isometries on the left and right sides of the figure yield identities. Right: ``replica wormhole'' contribution to the average unnormalized purity, where the non-trivial Wick contraction of the random operator (depicted here by dashed red lines) amounts for the shell's propagation between different replicas. The outer isometries $I_{\lef}$ and $I_{\ri}$ do not play a role in this computation.}
\label{fig:tnpurity}
\end{figure}

For a typical choice of $\bra{\Op}_{\aux\aux'}$ from the gaussian ensemble, the TN state $\ket{\Psi_{\Op}}$ will approximately saturate the island formula \eqref{eq:islandtn}, which is generally expected in random TNs (cf. \cite{Hayden:2016cfa}). For the purity of the unnormalized state $\hat{\rho}_{\Le} = \text{Tr}_{\Ri}\ket{\overline{\Psi}_{\Op}}\bra{\overline{\Psi}_{\Op}}$, the gaussian average yields 
\be\label{eq:purityTN1}
\int \text{d}\mu[\Op]\, \text{Tr}(\hat{\rho}_{\Le}^2) = |\lef||\ri|^2 + |\lef|^2|\ri|\,.
\ee 
The trivial Wick contraction of the $\Op$ tensors produces the first term, while the second term comes from the non-trivial Wick contraction connecting different replicas. The latter has the same structure as the gravitational replica wormhole of Fig. \ref{fig:replica} (see Fig. \ref{fig:tnpurity}).

Likewise, the average normalization squared is
\be\label{eq:normTN2}
\int \text{d}\mu[\mathcal{O}]\, \text{Tr}(\hat{\rho}_{\Le})^2 = |\lef|^2|\ri|^2 + |\lef||\ri|\;,
\ee 
where, again, the first term arises from the trivial Wick contraction between copies of the norm, while the second arises from a connected contraction between the two copies, which has the interpretation of the Euclidean wormhole in Fig. \ref{fig:wh2}. 

Altogether, the average purity is approximately given by 
\be\label{eq:purityTN3}
\overline{\text{Tr}({\rho}_{\Le}^2)} \approx  \dfrac{|\lef| + |\ri|}{1 + |\lef| |\ri|}\;,
\ee 
in complete analogy with \eqref{eq:purity4}, which in this case is simply the average purity of a random state living in the cut $\mathcal{H}_{\lef}\otimes \mathcal{H}_{\ri}$ of the TN. Note that the approximation in \eqref{eq:purityTN3} consists in taking the gaussian average separately in the numerator \eqref{eq:purityTN1} and in the denominator \eqref{eq:normTN2}, for the unnormalized TN state $\hat{\rho}_{\Le} = \text{Tr}_{\Ri}\ket{\overline{\Psi}_{\mathcal{O}}}\bra{\overline{\Psi}_{\mathcal{O}}}$. This was also the case in the gravity calculation in Sec. \ref{sec:3}, where the semiclassical path integral computation gave the unnormalized R\'{e}nyi entropies and the average value $\overline{Z_1^n}$ separately, and then we took their ratio.\footnote{Dividing by $\overline{Z^n_1}$ instead of $(\overline{Z_1})^n$ accounts for the fact that \eqref{eq:purityTN3} gives an exactly pure state when $|\lef|=1$ or $|\ri|=1$, i.e. at the endpoints of the Page curve for the purity.} 

For the higher R\'{e}nyis, the formulas can be obtained from simple combinatorics (see e.g. \cite{Liu:2020jsv})
\be\label{eq:renyistn} 
\overline{\text{Tr}(\rho_{\Le}^n)} \approx \dfrac{\sum_{s=1}^n N(n,s) |\lef|^{n+1-s}|\ri|^s}{\sum_{s=1}^n N(n,s) |\lef|^{s}|\ri|^s} \;,
\ee
where $N(n,s) = \frac{1}{n}{n\choose s}{n\choose s-1}$ are the Narayana numbers. The analytic continuation of \eqref{eq:renyistn} is given by hypergeometric functions
\be\label{eq:renyistn2} 
\text{Tr}(\rho_{\Le}^n) \approx \dfrac{1}{|\ri|^{n-1}}\dfrac{_2F_1(1-n,-n;2;z)}{w\, _2F_1(1-n,-n;2;w)} = \dfrac{1}{|\lef|^{n-1}}\dfrac{_2F_1(1-n,-n;2;z^{-1})}{w\, _2F_1(1-n,-n;2;w)} \;,
\ee
for $z = |\ri|/|\lef|$ and $w = 1/|\lef||\ri|$.

The $n\rightarrow 1^+$ expansion of the hypergeometric functions can now be taken in either of the two expressions in \eqref{eq:renyistn2}, corresponding to the $|z|<1$ and $|z|>1$ branches. Inserting these in the expression \eqref{eq:replicatrick} for the von Neumann entropy leads to Page's result for a random state \cite{Page:1993df}
\be 
S(\rho_{\Le}) \approx \begin{cases}
    S_{\lef} - \dfrac{|\lef|}{2|\ri|} \hspace{1cm} \text{if } |\lef| < |\ri| \;,\\ 
    S_{\ri} - \dfrac{|\ri|}{2|\lef|} \hspace{1cm} \text{if } |\ri| < |\lef| \;,
\end{cases}   
\ee 
which is the island formula \eqref{eq:islandFLM2} in this toy model,
\be\label{eq:islandtn2}
S(\rho_{\Le}) \approx \text{min}\lbrace S_{\lef}, S_{\ri}\rbrace \;,
\ee 
with exponentially small corrections in $|S_{\lef}-S_{\ri}|$. Since the global state of the bulk fields is pure (it consists of a collection of EPR pairs in the toy model of the state), $S_{\ri}$ precisely corresponds to the entropy of the bulk state reduced to the AdS${}_\lef$ region and to the cosmology $\Co$, $S_{\ri} = S_{\lef \Co}$. Similarly, $S_{\lef} = S_{ \Co\ri}$, and naturally $S(\rho_\Le)=S(\rho_\Ri)$. Therefore, when $|S_\lef-S_\ri|\gg 1$, the cosmology is an island for the left boundary system if $S_\lef>S_\ri$ and it is an island for the right boundary system if $S_\lef<S_\ri$.

\subsection{Model of the cosmology-to-boundary map}

Our goal now is to upgrade the TN of the state $\ket{\Psi_{\Op}}$ to a map which can simultaneously account for many different bulk states of the cosmological EFT. In particular, as we have discussed in the previous sections, we are interested in studying a cosmology-to-boundary map, which is the bulk-to-boundary map restricted to a subset of bulk states with fixed entanglement between AdS regions and cosmology and a fixed state for bulk DOF on AdS${}_{\lef,\ri}$, whereas the state of bulk DOF in the cosmology can vary. In order to model this setup, we consider the bulk Hilbert space $\mathcal{H}_{\psi}$ formed by the collection of states obtained by acting with unitaries $U_{\Co}$ supported on the DOF of the cosmology on a given reference state $\ket{\psi}$ of the quantum fields on the cosmology and the two AdS spaces:
\be\label{eq:tncosmoeft}
\mathcal{H}_{\psi} = \lbrace U_{\Co}\ket{\psi} : U_{\Co}^\dagger U_{\Co} = \mathbb{1}_{\Co}\;,\, \text{supp}(U_{\Co}) = \Co \rbrace .
\ee
All of the bulk states in $\mathcal{H}_{\psi}$ have fixed entanglement $S_{\lef}$ and $S_{\ri}$ between the cosmology and the AdS regions and a fixed reduced density matrix $\rho_{\lef\ri}$ on the AdS components, because, by only acting with unitaries on the cosmology, we cannot change the entanglement between AdS regions and cosmology nor alter the bulk state in the AdS regions.

In a general setup, we can consider the presence of extra bulk dangling legs on the cosmological MERAs $I_{c_{\lef}}$ and $I_{c_{\ri}}$, corresponding to cosmological DOF which are not entangled with the AdS regions in the reference state of the bulk fields $\ket{\psi}$. Together with the entangled DOF, these legs form Hilbert spaces of dimension $|\Co_{\lef}|$ and $|\Co_{\ri}|$, respectively.

In the full-fledged AdS/CFT construction of the PETS, the entanglement of the quantum fields on the cosmology is thermal and thus all of the modes of the quantum field contain some entanglement with the AdS regions. We would call the extra bulk legs that we added \textit{weakly entangled} DOF with respect to $\ket{\psi}$, and identify them with high energy modes whose entanglement with modes in the AdS regions is suppressed due to Boltzmann factors controlling their weight in the wavefunction. In contrast, \textit{strongly entangled} modes would be those for which the contribution to the entanglement is large, such as the EPR pairs between the cosmology and the AdS region added by means of type $II$ insertions (see Sec. \ref{sec:4} for more details).

We can now define a cosmology-to-boundary map
\be
\tilde{V}_\psi: \tilde{\mathcal{H}}_\psi\rightarrow \mathcal{H}_\Le \otimes \mathcal{H}_\Ri
\ee
between the bulk EFT restricted to the states of our interest and the holographic CFTs, at fixed entanglement between the cosmology and the bulk AdS spaces. Here we have defined $\tilde{\mathcal{H}}_\psi=\mathcal{H}_{\psi} \setminus \text{Ker}(V_\psi)$ excluding the collection of null states.\footnote{As mentioned in the discussion after \eqref{eq:normTN}, null states of the toy model need to be interpreted with some care in AdS/CFT, where the number of entangled modes due to thermal tails is really infinite and, \textit{a priori}, states only look null from an operational point of view when the CFT observer has a finite resolution. Without such limitation, the cosmology-to-boundary map $V_\psi$ in AdS/CFT is \textit{a priori} invertible. In any case, null states form a measure zero subset of $\mathcal{H}_\psi$, so we will just ignore them in the present context, and comment on their role in Sec. \ref{sec:5.2}.}

\begin{figure}[h]
\centering
\includegraphics[width = .8\textwidth]{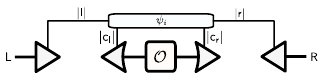}
\caption{TN for the cosmology-to-boundary map $V_\psi$. $\ket{\psi_i}\in \mathcal{H}_\psi$ is a state on all bulk DOF in the class of our interest. Here we only represented the DOF $|\lef|, |\ri|$ of the AdS regions which are entangled with the cosmology, omitting the unentangled ones.}
\label{fig:tnmap}
\end{figure}

The linear map $V_\psi$ defined by the TN and yielding unnormalized states in $\mathcal{H}_\Le \otimes \mathcal{H}_\Ri$ is given by (see Fig. \ref{fig:tnmap})
\be\label{eq:linearctbmap}
V_\psi\ket{\psi_i} = I_{\lef}I_{\ri}(\Pi_{\text{MAX}}^\lef \Pi_{\text{MAX}}^\ri\bra{\Pi_{\Op}}_{\Co})\ket{\psi_i}_{\lef \Co \ri}\;,
\ee
where $\ket{\psi_i}\in \mathcal{H}_{\psi}$ is a state on the cosmological EFT and the two AdS regions in the class of our interest, $\bra{\Pi_{\Op}}_{\Co} = \bra{\Op}I_{\Co_\lef}I_{\Co_\ri}$ is the projector exerted by the cosmology, and $\Pi^i_{\text{MAX}} = \ket{\text{MAX}_i}\bra{\text{MAX}_i}$ is the orthogonal projector to the maximally entangled state of $S_\lef$ and $S_\ri$ pairs, respectively. Note that the state $\ket{\psi}$ is a state on all bulk DOF, including those in the AdS regions which are unentangled with the cosmology. Because they are irrelevant in our discussion, we did not represent these DOF in Fig. \ref{fig:tnmap}.

The map \eqref{eq:linearctbmap} produces unnormalized states in $\mathcal{H}_\Le \otimes \mathcal{H}_\Ri$. The normalization can then be added \textit{a posteriori}, $\tilde{V}_\psi(\ket{\psi_i}) = V_\psi(\ket{\psi_i})/\sqrt{||V_\psi\ket{\psi_i}||}$, to produce unit vectors in $\mathcal{H}_\Le \otimes \mathcal{H}_\Ri$. This normalization makes sense only if we exclude null states from the bulk Hilbert space, hence the introduction of $\tilde{\mathcal{H}}_\psi$. Normalizing states this way is reminiscent of the Euclidean CFT preparation of cosmological states, where the resulting CFT state is generally unnormalized, and the normalization is added separately.\footnote{In fact, ways to assign a TN directly to the Euclidean preparation to model $V$ more precisely have been proposed in \cite{Chandra:2023dgq}. We will not pursue this direction in this paper.}

In this form, the linear cosmology-to-boundary map $V_\psi$ defined by the TN in Fig. \ref{fig:tnmap} can be directly regarded as a static version of the toy model of \cite{Akers:2022qdl} for the interior-to-boundary map of an evaporating black hole, with the appropriate identifications. The main difference between the two is that for the cosmology-to-boundary map $V_\psi$ the entanglement to the AdS regions is non-geometrical and appears through the state of the bulk fields. In our case the dimension of the cosmological EFT $|\Co|$ is fixed, and it is independent of what $|\lef|$ and $|\ri|$ are. Moreover, the map $V_\psi$ is gaussian random due to the presence of the operator $\Op$ on the cosmology rather than Haar random, which affects the precise way in which the measure concentrates at large dimension (see also \cite{Kar:2022qkf} for a gaussian random model in JT gravity). 

Although we will not focus on entanglement wedge reconstruction in what follows, we remark that the same TN can be also used to model the single-sided quantum channel encoding the cosmology on one of the two boundary systems, as well as the Petz map that allows to recover this information \cite{Cotler:2017erl,Chen:2019gbt}. Such a construction should only be reliable in a regime in which the sharp island formula holds (see Sec. \ref{sec:3}).

\subsection*{Entanglement dressing}

Given that we are dealing with closed spaces and gravity, one might find disturbing that we are allowing to generate new states by acting with local unitaries on the cosmology, $U_{\Co}$, in order to build the EFT Hilbert space $\mathcal{H}_{\psi}$. Perturbatively, when $G\neq 0$, the gravitational Gauss' law needs to be implemented to get physically sensible states, and no local operator on $\Co$ should be gauge invariant. The same issue could be raised in the full AdS/CFT setup of the previous subsections, where we generated new states in the bulk EFT by acting with local unitary operators in the cosmology.

The subterfuge in this case is that these unitaries are not gauge invariant in a conventional (state-independent) sense. There is nevertheless a natural sense in which entanglement serves to dress them to an asymptotic boundary, thus making them physical. Consider the simplest case consisting of an EPR pair for some mode $\omega$ in the cosmology with the respective mode in the AdS${}_\lef$ region
\be 
\ket{\text{MAX}} =\dfrac{1}{\sqrt{2}}( \ket{0_{\omega}}_{\lef}\ket{0_{\omega}}_{\Co}+\ket{1_{\omega}}_{\lef}\ket{1_{\omega}}_{\Co} ) \;.
\ee 

The action of any unitary $U^\omega_{\Co}$ on the mode in the cosmology can likewise be implemented by a unitary $U^\omega_{\lef}$ acting on the entangled mode $\omega$ on the AdS${}_\lef$ region, 
\be 
U^\omega_{\Co}\ket{\text{MAX}} = U^\omega_{\lef}\ket{\text{MAX}}\;.
\ee
The unitary $U^\omega_{\lef}$ is a \textit{state-specific} reconstruction of $U^\omega_{\Co}$ from $\lef$ for this particular state, in the sense of \cite{Akers:2021fut}. Now, $U^\omega_{\lef}$ acts on the AdS${}_\lef$ space, and can therefore be conventionally dressed to the asymptotic boundary by a suitable gravitational Wilson line, making it gauge invariant.\footnote{We assume that there is always a (perhaps state-specific) Wilson line which does not affect the strongly entangled modes so that the action of the dressed operator on the state is still that of $U^\omega_{\Co}$, at least within some approximate error.} It is always possible in this way to create any state of the mode in the cosmology by acting with a proper gauge invariant operator on its EPR partner.

In the general case where the entanglement structure between all of the modes of the cosmology and the AdS regions is more involved, one needs other less physical but still mathematically valid ways to implement a general unitary $U_\Co$, for example by means of associated non-unitary operators acting on the AdS regions.\footnote{Heuristically, this is the mechanism behind the Reeh-Schlieder theorem in local quantum field theory, where the vacuum contains an infinite number of ``EPR pairs'' at all scales. In our full AdS/CFT case, the spectrum is discrete, the associated bulk von Neumann algebras are of type I, and the entanglement is finite. However, the reduced states $\rho_{\lef,\ri}$ are themselves of full rank in the bulk EFT Hilbert space, since the number of weakly entangled pairs involved in the bulk state is really infinite. These are the conditions which need to be met for the corresponding version of the ``Reeh-Schlieder theorem'' to apply for type I algebras.} Particular choices of $U_\Co$ might still be approximately reconstructible in a state-specific way using other unitaries (see \cite{Akers:2021fut,Akers:2022qdl}), while others might not. 

In any case, given that the bulk state $\ket{\psi}$ in AdS/CFT contains thermal entanglement between all of the modes of the EFT, i.e. the reduced states $\rho_{\lef,\ri}$ are full rank, we will assume that there is always a mathematical way to dress the unitaries $U_\Co$ (or the local operators inserted in the full AdS/CFT setup) in this way, and define $\mathcal{H}_{\psi}$ in a gauge invariant way when $G\neq 0$. This way islands---and cosmological islands in this particular case---can exist in Einstein gravity, evading conventional dressing arguments.

\subsection{Non-isometric to approximately-isometric transition}
\label{sec:5.2}

At large spatial volume, the dimension of the cosmological EFT $\mathcal{H}_{\psi}$, given by $|\Co|$ in this model, can be taken to be fixed. Mainly, it corresponds to the microcanonical dimension of two gases of particles in AdS, with a Planck scale cutoff for the modes, $\omega \lesssim M_{P}$, and total energy $E_{\text{max}}\ell  \lesssim N^2$ so that the states can be considered ``thermal gas''-like (as opposed to black hole microstates).\footnote{$N$ here regulates as usual the (logarithm of the) bond dimension of the geometric part of the TN as well as the number of external boundary legs.} A rough estimate of this dimension is,
\be\label{eq:cosmodimension}
|\Co| =|\Co_{\lef}||\Co_{\ri}| \sim 4^{N^\frac{2d}{d+1}}\;,
\ee 
which arises from the microcanonical density of states of a local QFT in $d+1$ dimensions, in a box of size $\ell$, with total energy $E_{\text{max}}\ell \sim N^2$. The factor of $4$ accounts for the two sides of the cosmology.

Clearly, the map $V_\psi$ defined by the TN in Fig. \ref{fig:tnmap} can at most have $|\lef||\ri|$ independent states as an outcome, 
\be\label{eq:imagetn}
\text{dim}\lbrace \text{Im}(V_\psi)\rbrace \leq |\lef||\ri|\;,
\ee 
where $\text{Im}(V_\psi)$ is the image of $V_\psi$ on $\mathcal{H}_\Le \otimes \mathcal{H}_\Ri$. To see that the inequality is approximately saturated, we run the resolvent calculation for TN states in Appendix \ref{app:E}, by computing the average eigenvalue density of the Gram matrix in the gaussian ensemble \eqref{eq:measure} of $\Op$ tensors. This qualitatively reproduces the gravitational features of Sec. \ref{sec:4} on each microcanonical band of the CFT, and provides the saturation of \eqref{eq:imagetn} for general choices of the $\Op$ tensor.

When $|\lef| |\ri| < |\Co|$, there are null states\footnote{These are null states associated with the bulk dangling legs, in addition to the ``geometrical'' null states arising for specific choices of the shell operator $\mathcal{O}$.} in the cosmological EFT, which span a non-trivial kernel for the map $V_\psi$, $\text{dim}\lbrace\text{Ker}(V_\psi)\rbrace = |\Co| - |\lef| |\ri| > 0$. For the linear map $V_\psi$ this implies that this map cannot preserve the inner product structure, i.e. the cosmology-to-boundary map must be \textit{non-isometric}, $V_\psi^\dagger V_\psi \neq \mathbb{1}_{\Co}$.\footnote{Note that the existence of a non-trivial kernel implies non-isometricity of the map, but the converse is not true. A non-isometric map can in general have a trivial kernel and therefore be invertible. This is expected to be the case for the full AdS/CFT map discussed in Sec. \ref{sec:4}.} A way to quantify the non-isometricity of the map is to take two states $\ket{\psi_1},\ket{\psi_2}\in \mathcal{H}_{\psi}$ and compute how the overlap $\bra{\psi_1}V_\psi^\dagger V_\psi \ket{\psi_2}$ differs from the EFT overlap $\bra{\psi_1} \ket{\psi_2}$ on average,
\begin{gather}
\int \text{d}\mu[\Op] \left(\bra{\psi_1}V_\psi^\dagger V_\psi \ket{\psi_2} - \bra{\psi_1} \ket{\psi_2}\right) = 0\;,\\
\int \text{d}\mu[\Op]\, \left| \bra{\psi_1}V_\psi^\dagger V_\psi \ket{\psi_2} - \bra{\psi_1} \ket{\psi_2}\right|^2 = \text{Tr}(\rho^{\psi_1}_{\lef \ri}\rho^{\psi_2}_{\lef\ri}) = \dfrac{1}{|\lef||\ri|}\;, \label{eq:varianceisom}
\end{gather}
where $\rho^{\psi_1}_{\lef \ri}$, $\rho^{\psi_2}_{\lef \ri}$ are the bulk reduced states to $\lef \ri$, which are equal by the definition of the cosmological EFT \eqref{eq:tncosmoeft}, $\rho^{\psi_1}_{\lef \ri} = \rho^{\psi_2}_{\lef \ri} = \rho_{\lef \ri}$, where $\rho_{\lef \ri}$ is the reduced state in the global reference state $\ket{\psi}$. Given our construction, the reference state $\ket{\psi}$ is chosen to be maximally entangled between the $\lef\ri$ DOF and the cosmology.

The non-vanishing variance \eqref{eq:varianceisom} shows that the map $V_\psi$ is indeed non-isometric. The normalized map $\tilde{V}_\psi$ will likewise be non-isometric, and on average it will be the same as the map $V_\psi$, since $\overline{||V_\psi\ket{\psi_i}||} = 1$. The average overlap \eqref{eq:varianceisom} is consistent with the microcanonical version of the gravitational overlap \eqref{eq:overlapvariance} computed in Sec. \ref{sec:4.1}. 

The goal of the discussion in the rest of the present subsection will be to determine, under the assumed genericity of the map $V_\psi$, for what values of $|\lef||\ri|$ we can consider that \textit{all} operationally ``simple'' states that a cosmological observer can prepare are approximately isometrically encoded in $\mathcal{H}_{\Le}\otimes \mathcal{H}_{\Ri}$, with high probability.

In particular, the expression \eqref{eq:varianceisom} only holds on average, and does not contain information of what the dimension of the cosmological EFT $|\Co|$ is. Following \cite{Akers:2022qdl}, we note that it is also possible to make stronger statements than those obtained on average by taking into account the way the gaussian measure for the $\Op$ tensor \eqref{eq:measure} concentrates around the perimeter of the half-space in the limit of large bond dimension for this tensor (in analogy with how the Haar measure concentrates around the equator of the higher dimensional sphere). In fact, the gaussian random version of the black hole model was already worked out in \cite{Kar:2022qkf}. We will translate their central results in what follows.

Consider a set of $K$ states in the cosmological EFT $S = \lbrace \ket{\psi_i}\in \mathcal{H}_{\psi}\rbrace$, for $i=1,...,K$. Then \cite{Kar:2022qkf} 
\be\label{eq:gaussianconcentration} 
\text{Pr}\left[\underset{\ket{\psi_1},\ket{\psi_2} \in S}{\text{max}}\left|\bra{\psi_1}V_\psi^\dagger V_\psi \ket{\psi_2} - \bra{\psi_1} \ket{\psi_2}\right|\geq \sqrt{18}(|\lef||\ri|)^{-\gamma}\right] \leq 12 {K \choose 2} \exp\left(-\dfrac{(|\lef||\ri|)^{1-2\gamma}}{2}\right)\;.
\ee 
for any fixed $0<\gamma < \frac{1}{2}$ and $|\lef||\ri|>4$.\footnote{Note that the gaussian measure concentrates exponentially stronger than the Haar measure, since the coefficient in the exponent in \eqref{eq:gaussianconcentration} is $1/2$ instead of $1/24$ for the Haar measure (cf. \cite{Akers:2022qdl}).}

In order to understand for what values of $|\lef||\ri|$ all ``simple'' cosmological states are isometrically encoded in $\mathcal{H}_{\Le}\otimes \mathcal{H}_{\Ri}$ with high probability, we need to quantify how many ``simple'' cosmological states we have. A generic state of $n$ qubits prepared using only $k$-local gates with $k\geq 2$ has complexity $\mathcal{C}=\frac{n}{k}T$, where $T$ is the number of layers in the circuit preparing the state and we assumed for simplicity that $n$ is a multiple of $k$. A rough estimate of the number of states with complexity $\mathcal{C}$ is given by\footnote{For simplicity, we are omitting the logarithmic dependence on the size of the regulator $\varepsilon$, given a tessellation of the Hilbert space in $\varepsilon$-balls.}
\be
    K_{\mathcal{C}} \sim \left[\frac{ {n \choose k}{n-k \choose k} ... {k\choose k}}{(n/k)!}\right]^{T}=\left[\frac{n!}{(k!)^{n/k}(n/k)!}\right]^T\approx n^{\left(1-\frac{1}{k}\right)nT}=\exp[(k-1)\mathcal{C}\log n]
\ee
where in the approximation we used $n\gg k$ and dropped multiplicative coefficients that would give subleading contributions to the entropy bound. Since $K_{\mathcal{C}}$ scales exponentially with the complexity and, for our purposes, $\mathcal{C}$ is parametrically large, $K_{\mathcal{C}}$ gives a good approximation also to the number of states with complexity smaller than or equal to $\mathcal{C}$. In our cosmological setup, we can set $n=\log_2 |\Co|$.

Now we shall apply eq. \eqref{eq:gaussianconcentration} to the collection of states $S_{\mathcal{C}}$ in the cosmological EFT with complexity upper bounded by $\mathcal{C}$, i.e.
\be\label{eq:gaussianconcentration2} 
\text{Pr}\left[\underset{\ket{\psi_1},\ket{\psi_2} \in S_{\mathcal{C}}}{\text{max}}\left|\bra{\psi_1}V_\psi^\dagger V_\psi \ket{\psi_2} - \bra{\psi_1} \ket{\psi_2}\right|\geq \sqrt{18}(|\lef||\ri|)^{-\gamma}\right] \lesssim 6  \exp\left(2(k-1)\,\mathcal{C}\log_2 \log_2 |\Co|-\dfrac{(|\lef||\ri|)^{1-2\gamma}}{2}\right)\;.
\ee 
Therefore, we can conclude that the encoding of these states will be nearly-isometric with high probability if
\be 
|\lef||\ri| \gtrsim \left(4(k-1)\,\mathcal{C}\log_2 \log_2 |\Co|\right)^\frac{1}{1-2\gamma}\;.
\ee 

What should a good estimate of $\mathcal{C}$ be for ``simple'' cosmological states? First of all, the proper time $t_{\text{max}}$ that any cosmological observer can live is bounded by the presence of the big bang singularity in the past and the big crunch singularity in the future, and it is given by $t_{\text{max}} \sim \ell \sim N^{\frac{2}{d-1}} \ell_P$. Assume that the observer is powerful enough to have access to all of the microcanonical entropy in the cosmology during that time, with the restriction of only implementing $k$-local gates, where $k$ is some $O(1)$ number. The maximum complexity of any such state is naturally
\be\label{eq:complexityobserver} 
\mathcal{C} \sim \dfrac{t_{\text{max}}}{\tau_{c}} \frac{\log_2 |\Co|}{k} \lesssim (\log_2 |\Co|)^{\frac{d^2+1}{d(d-1)}}\;,
\ee 
where the circuit time $\tau_c$ has been taken to be Planckian to get the upper bound, we dropped irrelevant multiplicative constants, and we used \eqref{eq:cosmodimension} for $|\Co|$. That is, the complexity of any such state is polynomial in $\log_2 |\Co|$, with the particular exponent given by \eqref{eq:complexityobserver}.

Given this assumption, then the encoding of \textit{all} ``simple'' states that the cosmological observer can prepare is approximately-isometric with extremely high probability if
\be 
|\lef||\ri| \gtrsim  (\log_2 |\Co|)^{\frac{d^2+1}{d(d-1)}\frac{1}{1-2\gamma}} \sim N^{\frac{2(d^2+1)}{d^2-1}\frac{1}{1-2\gamma}} \;,
\ee 
where we have used \eqref{eq:cosmodimension} and neglected multiplicative constants and the subleading logarithmic dependence. We note that this occurs at 
\be\label{eq:boundentropy}
S_{\lef}+S_{\ri}  \gtrsim  \dfrac{2(d^2+1)}{d^2-1}\dfrac{1}{1-2\gamma}\,\log_2 N \;,
\ee 
which is well below the microcanonical entropy bound $S_{\lef},S_{\ri}\lesssim N^2$ for the AdS gases to collapse into a small black hole. We remark that a similar conclusion would hold if we replaced the complexity \eqref{eq:complexityobserver} by any polinomial in $\log |\Co|$ so that more complex states arising from the past singularity can be included, as well as the internal states of the particles that form the shell.

\begin{figure}[h]
    \centering
    \includegraphics[width=0.7\textwidth]{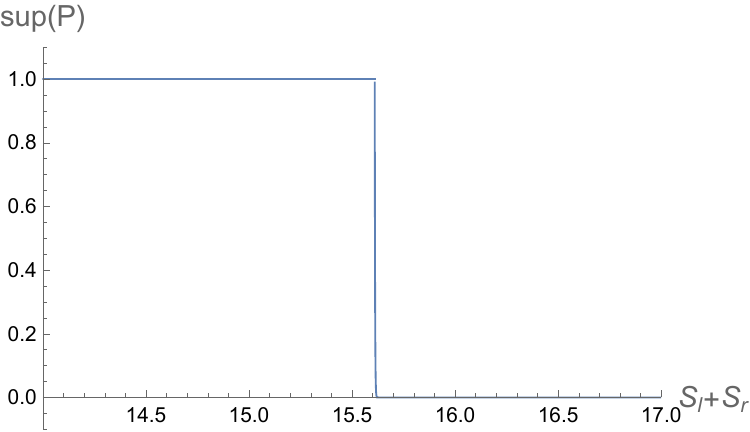}
    \caption{Upper bound for the probability of non-isometric encoding---given by the minimum between 1 and the right-hand side of eq. \eqref{eq:gaussianconcentration2} with $\mathcal{C}$ given by eq. \eqref{eq:complexityobserver}---as a function of the entanglement $S_\lef+S_\ri$ between cosmology and AdS regions. We chose here $d=3$, $k=5$, $\gamma=0.25$, $|\Co|=1000$. For $S_\lef+S_\ri=O(\log_2\log_2|\Co|)$ a sharp non-isometric to approximately-isometric transition takes place. In particular, with the parameters of our choice the right hand side of eq. \eqref{eq:boundentropy} is $\sim 16.6$.}
    \label{fig:transition}
\end{figure}

The conclusion is that, by entangling more and more modes between the cosmology and the AdS regions microcanonically and therefore increasing the value of $S_{\lef}+S_{\ri}$ to the point where condition \eqref{eq:boundentropy} is satisfied, the cosmology-to-boundary map $V_\psi$ (and consequently $\tilde{V}_\psi$) will undergo a non-isometric to approximately-isometric transition for basically all of the states that a cosmological low-energy observer can prepare (see Fig. \ref{fig:transition}). The transition occurs at a relatively low value for the microcanonical entropy, given by \eqref{eq:boundentropy}. Although this result was derived using a TN toy model in which an explicit calculation is possible, we expect the non-isometric to approximately-isometric transition to also occur in our full AdS/CFT setup.

\section{Discussion}

\label{sec:6}
 
The construction presented in this paper provides a specific example of a holographic model of closed cosmology within the standard framework of AdS/CFT. The inhomogeneous big bang-big crunch cosmology is supported by a shell of heavy matter and is described microscopically as a low-temperature partially entangled thermal state of two holographic CFTs. In the bulk EFT, the bulk quantum fields in the cosmology purify the state of the quantum fields in two additional disconnected AdS spacetimes.

If the entanglement between the cosmology and the AdS spacetimes is sufficiently large, the cosmology arises as an entanglement island of either the right or the left CFT. We have seen that the cosmology-to-boundary map, encoding the cosmological physics into the boundary theory, is non-isometric, and provided a state-dependent reconstruction of bulk operators acting on the cosmology. Using a TN toy model of our setup, we have argued that the cosmology-to-boundary map generically undergoes a non-isometric to approximately-isometric transition for all operationally ``simple'' states of the cosmology, as a function of the entanglement between the AdS regions and the cosmology.

A number of open questions related to our results indicate the direction for future research on these topics. 

A central conceptual challenge raised by our construction is whether lowering the bulk entanglement between the AdS regions and the cosmology, and in particular reaching a regime in which the inequality \eqref{eq:boundentropy} is violated, implies any physical breakdown of the EFT on the cosmology. 
From the bulk point of view, there is \textit{a priori} no reason to expect such an intrinsic quantum gravitational regime, given that the cosmology is macroscopic and under semiclassical control. Another piece of evidence favoring the same conclusion is that the gravitational replica trick used to derive the island formula \eqref{eq:islandFLM2} in this paper heavily relies on the validity of the bulk EFT. The island formula \eqref{eq:islandFLM2} should still hold at parametrically smaller entanglement than that given by the bound \eqref{eq:boundentropy}. 

Given these considerations, the most conservative possibility is simply that a version of ``cosmological complementarity'' holds for this system \cite{Susskind:1993if}. The cosmology is always mathematically encoded in the CFT through the invertible cosmology-to-boundary map $V_\psi$, but the decoding of cosmological physics, given by the inverse of this map, $V_\psi^{-1}$, becomes highly non-trivial to obtain for any CFT observer with finite resolution and limited access to the states. This is particularly true when the inequality \eqref{eq:boundentropy} is not satisfied, and therefore when $V_\psi^{-1}$ and $V_\psi^\dagger$ start to differ even for simple states. Thus, any physically reasonable CFT observer might not be able to resolve the cosmology if the entanglement is too low.

A closely related issue that would be interesting to investigate in more detail is the one-shot distinguishability of cosmological states. Since the cosmology-to-boundary map $V_\psi$ is non-isometric, states that are orthogonal in the bulk EFT are mapped to non-orthogonal states in the CFT (and even to almost-parallel states if the entanglement between the AdS regions and the cosmology is very small). This suggests that, while a bulk EFT observer can naturally distinguish two states by performing a one-shot experiment, a CFT observer might not be able to predict the outcome of the experiment easily, given that this would require precise knowledge of the CFT states and of the cosmology-to-boundary map. A deeper understanding of the relationship between experiments in the bulk EFT and experiments in the microscopic boundary theory could shed new light on the features of holographic duality at the operational level, as well as determine the viability of the implementation of different classes of semiclassical bulk experiments on quantum devices.

Another open problem which we leave to future work is the exact mechanism by which observables in the closed cosmological spacetime can be made gauge-invariant when $G\neq 0$. In our TN model we explained how this dressing works in extremely simple cases, and suggested that the presence of an infinite number of entangled modes in the full AdS/CFT setup can be used to dress all bulk observables to the boundaries of the two AdS spaces. However, a precise quantitative analysis of this issue would be useful to understand in more detail the relationship between bulk EFT observables in the cosmology and associated dual observables in the microscopic boundary theory.

A different topic that also requires further investigation is that one regarding the state-dependence of the reconstruction of local operators of the cosmological EFT. In Sec. \ref{sec:4.2} we were able to give an explicit state-dependent prescription to reconstruct bulk operators on the cosmology from the boundary theory. However, we did not prove that a state-independent reconstruction does not exist. This problem is related to a more general question regarding the exact relationship between state-dependence and non-isometric encoding in AdS/CFT. Understanding whether and under which conditions these two features are equivalent is an interesting direction that we plan to explore in the near future.

Finally, the exact zero temperature limit of our construction, with $\tbeta_{\Le},\tbeta_{\Ri}\rightarrow \infty$, requires a separate treatment. In fact, in this case the (infinite) Euclidean evolution acts as a sharp projector onto the unique supersymmetric ground state of the CFTs, $\ket{0}_{\Le}\otimes \ket{0}_{\Ri}$, regardless of the operator insertions performed in the Euclidean path integral. This in turn implies that there is a single bulk state and somehow all different cosmological states that we discussed in the present paper are equivalent at the fundamental level. In this case, the inverse map $V_\psi^{-1}$ clearly does not exist. This presents some similarities with the puzzle raised in \cite{Lin:2022rzw,Lin:2022zxd} for the degeneracy of states of supersymmetric extremal black holes. In our setup, however, the lack of spatial connectivity and, in the zero temperature limit, of bulk entanglement between the cosmology and the AdS regions might come to the rescue. In fact, under these conditions it would be impossible to dress operators in the cosmology to make them gauge-invariant. Consequently, it is reasonable to expect that, upon a careful definition of the bulk EFT which takes dressing into account, there is only one state of the closed universe, even from a bulk semiclassical gravity point of view. This is in accordance with the claim, common in the literature, that the intrinsic Hilbert space of closed universes is one-dimensional (see e.g. \cite{McNamara:2020uza} and references therein).

\section*{Acknowledgments}

We thank Vijay Balasubramanian, Ning Bao, Jos\'{e} Barb\'{o}n, Charles Cao, Wissam Chemissany, Roberto Emparan, Albion Lawrence, Javier Mag\'{a}n, Geoff Penington, Simon Ross, Petar Simidzija, Mark Van Raamsdonk, and Chris Waddell for stimulating conversations. We acknowledge support from the U.S. Department of Energy through DE-SC0009986 (M.S. and B.S.) and QuantISED DE-SC0020360 (M.S.), and from the U.S. Department of Energy, Office of Science, Office of Advanced Scientific Computing Research, Accelerated Research for Quantum Computing program “FAR-QC” (S.A.). This preprint is assigned the code BRX-TH-6714.

\appendix

\section{Subdominant saddles and a Hawking-Page transition}
\label{app:hawkingpage}

In this appendix we show that the Euclidean saddle point geometry $X$ described in Sec. \ref{sec:2} provides the dominant contribution to the semiclassical wavefunction of the bulk state dual to the PETS $\ket{\Psi_\Op}$ \eqref{eq:PETS}, at low preparation temperatures $\tbeta_{\Le}, \tbeta_{\Ri} \gg \ell$. 

For this matter, we consider the normalization of the PETS $\ket{\Psi_\Op}$, defined in \eqref{eq:PETS} as an Euclidean two-point correlation function of the shell operator
\be\label{eq:normPETS}
Z_1 = \text{Tr}(e^{-\tbeta_{\Le} H}\mathcal{O}^\dagger e^{\tbeta_{\Ri} H}\mathcal{O})\;.
\ee 
The preparation of the PETS amounts to cutting the Euclidean CFT path integral computing \eqref{eq:normPETS} open in half (see Fig. \ref{fig:PETS}). For this reason, the dominant semiclassical saddle point of \eqref{eq:normPETS} will also correspond to the dominant semiclassical state in the bulk wavefunction.

Including $X$, there are in total four obvious saddle points contributing to $Z_1$. The other three consist of either Euclidean black hole solutions glued to Euclidean AdS spaces, or of a pair of Euclidean black holes glued together. We shall denote them collectively by $X_{i|j} = X_i \cup \mathcal{W}\cup X_j$, where $X_{\text{AdS}}$ represents an Euclidean AdS component and $X_{\text{BH}}$ represents an Euclidean Schwarzschild-AdS black hole component. In this notation, $X \equiv X_{\text{AdS}|\text{AdS}}$, and $X_{\text{BH}|\text{BH}}$ corresponds to the Euclidean solution represented in Fig. \ref{fig:bh}. The last two, $X_{\text{BH}|\text{AdS}}$ and $X_{\text{AdS}|\text{BH}}$, contain mixed components, and provide initial data of one-sided black hole microstates entangled to AdS regions, similar to those described in \cite{Chandra:2022fwi}.  Note that the black hole component $X_{\text{BH}}$ only exists above the critical temperature for AdS black hole solutions. Moreover, we only consider the thermally stable branch of large AdS black holes solutions. 

With this in mind, the semiclassical contribution to \eqref{eq:normPETS} becomes
\be\label{eq:twoptfn}
Z_1 \sim \sum_{i,j}e^{-I[X_{i|j}]} Z_{\text{bulk}}[X_{i|j}]\;,
\ee
interpreted in the asymptotic $G\rightarrow 0$ expansion, where $I[X_{i|j}] = O(G^{-1})$ is the classical gravitational action \eqref{eq:action} and $Z_{\text{bulk}}[X_{i|j}] = O(G^0)$ is the one-loop determinant of the quantum fields, for each corresponding saddle point $X_{i|j}$. 

In the saddle point $X_{i|j}$, the shell will again propagate according to $\dot{R}^2 +V_{\text{eff}}(R) =0$, for the effective potential
\be 
V^{i,j}_{\text{eff}}(R) = -f_j(R) + \left(\dfrac{M_j-M_i}{m} - \dfrac{4\pi G m}{(d-1)V_\Omega R^{d-2}}\right)^2\;,
\ee 
where $M_i$ is the ADM mass associated with the component $X_i$. The solutions will have physical temperatures $\beta_{\Le}^{i,j} = \tbeta_{\Le} + \Delta \tau^{i,j}_-$ and $\beta_{\Ri}^{i,j} = \tbeta_{\Ri} + \Delta \tau^{i,j}_+$, where the Euclidean time elapsed by the shell $\Delta \tau^{i,j}_\pm$ is determined by the analog of \eqref{eq:shifttime} in the corresponding background (we refer to the reader to \cite{Balasubramanian:2022gmo} for details in the case of $X_{\text{BH}|\text{BH}}$). 

The classical action $I[X_{i|j}]$ can be evaluated to be
\be \label{eq:action}
I[X_{i|j}] = \tbeta_{\Le} F_i(\beta_{\Le}) + \tbeta_{\Ri} F_j(\beta_{\Ri}) + I_{s}[X_{i|j}]\;,
\ee 
where $F_i(\beta)$ is the Gibbons-Hawking free energy associated to the saddle $X_i$, and $I_{s}[X_{i|j}]$ is the intrinsic contribution from the shell. After the addition of suitable counterterms the free energy becomes (cf. \cite{Emparan:1999pm})
\be
F_{i}(\beta) = \dfrac{V_\Omega}{8\pi G}\left(-r_i^{d}+ r_i^{d-2} + c_d\right)
\ee
where $r_i$ represents the radius of the horizon associated to $X_i$, at inverse temperature $\beta$ (with $r_{\text{AdS}}=0$). The constant $c_d$ gives the Casimir energy of the CFT in even dimensions \cite{Balasubramanian:1999re} ($c_d = -\frac{1}{2},\frac{3}{8}, -\frac{5}{16},\ldots$ in $d=2,4,6,\ldots$). 

The intrinsic contribution coming from the shell in \eqref{eq:action} can be expressed as (cf. \cite{Balasubramanian:2022gmo})
\be\label{eq:Ishell2}
	I_{s}[X_{i|j}] = \dfrac{d}{8\pi G}\,\left(\text{Vol}(X^i_{s}) + \text{Vol}(X^j_{s})\right)\,+\,m\dfrac{d-2}{d-1}  L[\gamma_{\mathcal{W}}]  \;,
\ee	
for
\begin{gather}
	L[\gamma_{\mathcal{W}}] = 2\int_{R_*}^{r_\infty} \dfrac{\text{d}R}{\sqrt{-V^{i,j}_{\text{eff}}(R)}}\;,\label{eq:length}\\[.4cm]
	\text{Vol}(X^i_{s}) = \dfrac{2V_\Omega}{d} \int_{R_*}^{r_\infty}\dfrac{\text{d}R}{f_i(R)}\, \sqrt{\dfrac{f_i(R) + V^{i,j}_{\text{eff}}(R)}{- V^{i,j}_{\text{eff}}(R)}}\,(R^d-r_i^d)\;.\label{eq:vol}
\end{gather}

The phase diagram of \eqref{eq:twoptfn} will thus depend on the presence of the shell operator $\mathcal{O}$, since the operator is heavy and modifies the microcanonical saddle point of \eqref{eq:normPETS} in the form of bulk backreaction in the classical limit, namely through \eqref{eq:Ishell2}. However, the effects of the shell will be small in the phase transition of \eqref{eq:twoptfn}. 

To illustrate this point, we shall take the large mass limit of the shell, $m\ell \rightarrow \infty$. In this limit, the manifolds $X_{i|j}$ become effectively disconnected, and the intrinsic contribution from the shell $I_{s}[X_{i|j}]$ no longer depends on the solution that it propagates in (cf. \cite{Balasubramanian:2022gmo}). The Euclidean time elapsed by the shell vanishes $\Delta \tau^{i,j}_\pm \rightarrow 0$, and the preparation temperatures become the physical temperatures of all of the solutions, $\beta_{\Le}^{i,j} = \tbeta_{\Le}$ and $\beta_{\Ri}^{i,j} = \tbeta_{\Ri}$. The classical contribution from each of the components of $X_{i|j}$ factorizes in this limit,
\be\label{eq:partitionfunction2}
Z_1 \sim Z_0\sum_{i,j}Z_{i}(\beta_{\Le})Z_{j}(\beta_{\Ri})\;\;,
\ee
where $Z_0 = \exp(-2 m\ell \log R_*/2\ell)$, with $R_*$ given by \eqref{eq:rstar}, is the constant part coming from the intrinsic contribution of the shell (cf. \cite{Balasubramanian:2022gmo}). Each factor of $Z_{i}(\beta) = \exp(-\beta F_i(\beta)) Z_{\text{bulk}}[X_{i}]$ corresponds to the Gibbons-Hawking partition function of $X_i$ at inverse temperature $\beta$, including the contribution of the quantum fields on each background geometry at one-loop.

Each partition function $Z_{i}(\beta)$ undergoes a well-known first order phase transition at $G\rightarrow 0$, dual to the confinement/deconfinement phase transition of the holographic CFT, at the Hawking-Page temperature $\beta^{-1}_{\text{HP}}$ \cite{HawkingPage}. Thus, in this case, the different semiclassical branches of the wavefunction of the PETS will exchange dominance at the preparation temperatures
\be 
\tbeta_{\Le}, \tbeta_{\Ri} = \beta_{\text{HP}}\;.
\ee 
At low temperatures $\tbeta_{\Le}, \tbeta_{\Ri}\geq\beta_{\text{HP}}$, the Euclidean AdS saddle point $X_{\text{AdS}|\text{AdS}}$ will provide the dominant contribution to \eqref{eq:partitionfunction2}. This shows that in this regime, the semiclassical bulk dual to the PETS \eqref{eq:PETS} prepares the closed cosmology described in the previous subsection.

\begin{figure}[h]
    \centering
    \includegraphics[width=0.85\textwidth]{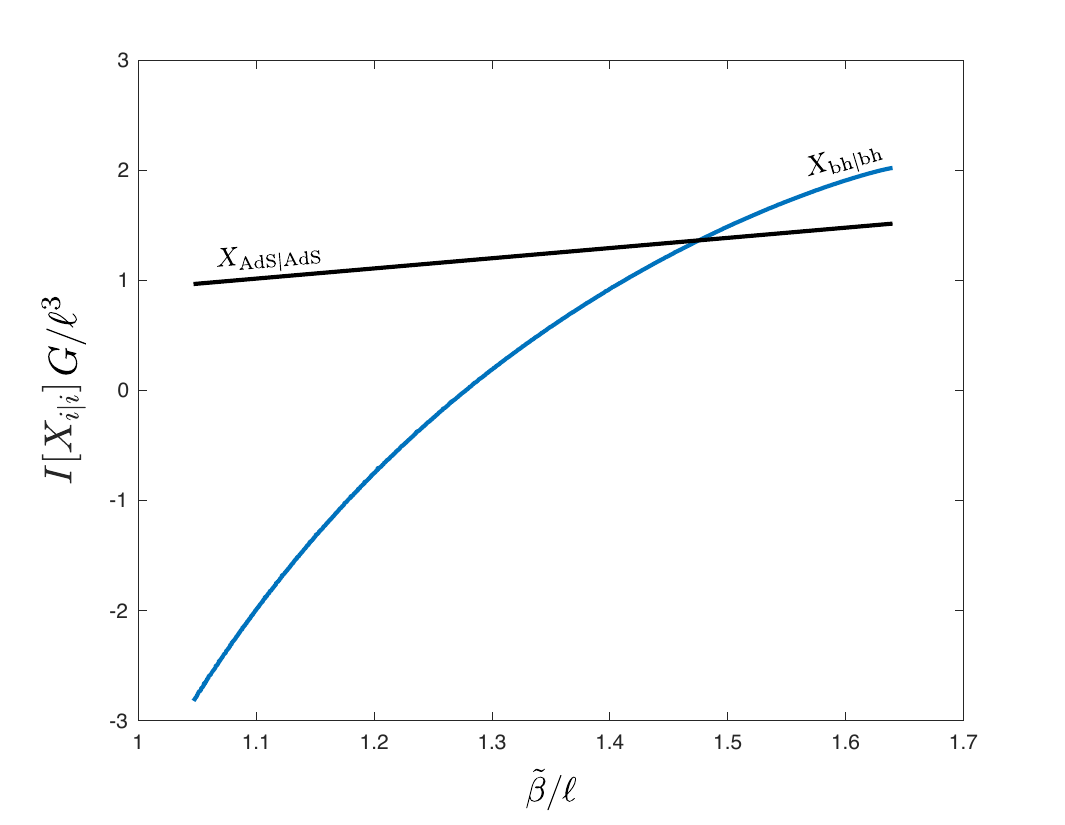}
    \caption{In the symmetric case $\tbeta_{\Le} = \tbeta_{\Ri} = \tbeta$ in AdS$_{5}$ ($d=4$), the exchange in dominance between the Euclidean black hole saddle $X_{\text{BH}|\text{BH}}$ and the Euclidean AdS saddle $X_{\text{AdS}|\text{AdS}}$ occurs at the inverse preparation temperature $\tilde{\beta} \approx 0.70 \,\beta_{\text{HP}} \approx 1.47 \ell$ for a shell of mass $m = \ell^2/G$, where $\beta_{\text{HP}} \approx 2\pi\ell/3$.}
    \label{fig:ptrans}
\end{figure}

At finite $m\ell$, the effects of the operator $\mathcal{O}$ can be incorporated numerically (see Fig. \ref{fig:ptrans}). These effects do not have a substantial effect in the exchange in dominance between the different semiclassical saddles in the wavefunction of the PETS. In the opposite limit, taking the shell to be a probe particle, $m\ell \rightarrow0$, the operator $\mathcal{O}$ becomes light, and the PETS \eqref{eq:PETS} becomes the TFD state with the insertion of a light operator, provided that $\tbeta_{\Le} = \tbeta_{\Ri}$. In the bulk, the semiclassical state at high temperatures is that of an eternal black hole of physical temperature $\tbeta_{\Le} = \tbeta_{\Ri} = \beta/2$, with a probe particle sitting at the horizon.\footnote{The mixed solutions $X_{\text{BH}|\text{AdS}}$ and $X_{\text{AdS}|\text{BH}}$ do not exist in this limit, since they need a heavy enough shell to support them.} Since the operator does not backreact, the phase transition in this case occurs when the physical temperature of the black hole equals the Hawking-Page temperature, that is, at a preparation temperature $\tbeta_{\Le}, \tbeta_{\Ri} = \beta_{\text{HP}}/2$. In the intermediate regime, i.e. at finite $m\ell$, the overall effect of the operator is to decrease the critical temperature to some value in the range $\beta_{\text{HP}}/2 \leq \tbeta_{\Le}, \tbeta_{\Ri} \leq \beta_{\text{HP}}$. 

\subsection{Other possible contributions}

All the Euclidean saddle point geometries discussed above are constructed assuming the existence of a single, coherent spherical shell of matter whose trajectory in the Euclidean geometry connects the operator insertions in the Euclidean past ($\Op$) and future ($\Op^\dagger$). In other words, each one of the massive particles forming the shell follows an Euclidean trajectory which connects the operator insertion creating that particle in the bra to the respective operator insertion creating the same particle in the ket. Clearly, this is not the only possibility in general settings.\footnote{We thank Mark Van Raamsdonk for pointing this out to us.}

In fact, if all the particles composing the shell are identical, i.e. they are created by the insertion of the same conformal primary $\mathcal{O}_\Delta$, there are alternative Euclidean configurations where the particles' trajectories connect operator insertions within the $\Op$ and the $\Op^\dagger$ insertions. In this disconnected case, there are two separate ``shells'', one in the Euclidean past and one in the Euclidean future, and no trajectory intersecting the time reflection-symmetric slice. Thus these configurations are expected to represent branches of the semiclassical wavefunction of the PETS corresponding to two AdS spaces and no cosmology.

Since, as we have assumed throughout the paper, the number of local operator insertions in $\Op$ scales with $N^2$, we expect that the Euclidean geometry associated to these contributions is non-perturbatively modified by the presence of the two disconnected shells we just described. Unlike for the single shell case, there is no obvious way to build the backreacted geometries in this case, because the matter contractions explicitly break spherical symmetry.  As a consequence, we are not able to evaluate the on-shell gravitational action for such solutions, assuming that they exist, and compare it to the on-shell action for the saddle of our interest to determine which saddle is dominant in the gravitational Euclidean path integral.

From the CFT side, the assumption that the connected spherical shell solution dominates the correlation function $Z_1$ translates into the technical assumption of ref. \cite{Anous:2016kss}. Namely, the correlation functions of the thin shell operator $\Op$ are assumed to be dominated by intermediate identity conformal blocks, in the global OPE channel in which each one of the local operator insertions in $\Op$ contracts with the corresponding operator insertion in $\Op^\dagger$ at the same point on the sphere.\footnote{Actually, this does not determine the OPE channel completely. One needs to maximize over paths in the complex plane for fixed contractions (see \cite{Anous:2016kss}).} This assumption is motivated by the behavior of heavy Virasoro conformal blocks in the large-$c$ expansion of two-dimensional CFTs.

Alternatively, a way to suppress the additional configurations further is to consider that each particle that forms the shell carries some definite internal global charge, for a global symmetry which can be broken in the bulk by the $1/N$-suppressed interactions. In this case, we choose $\Op$ as the operator which creates a large number of classical particles with definite global charge, which together form a macroscopic classical object. The operator $\Op^\dagger$ annihilates the classical shell, while other annihilation channels are highly suppressed, in the same manner as for any classical object comprised of a large number of particles. Under this assumption, the operator $\Op$ creates global charge, and therefore its one-point function vanishes to leading order. For this reason any disconnected contribution to $Z_1$ vanishes in this approximation, and the connected saddle is clearly dominant. Another way to achieve a similar effect is to consider that different local insertions that form the shell consist of different primaries, with different conformal dimensions $\Delta_i$, in such a way that the mass density is approximately homogeneous along the sphere. 

For these reasons, we will restrict our attention to the cases described above in which the connected shell saddle point contributions of interest are dominant. Studying disconnected shell configurations in more detail is an interesting and non-trivial endeavour that we leave for future work.

\section{Estimate of thermal entropy in type IIB supergravity}
\label{app:B}

In this appendix, we estimate the thermal partition function of type IIB supergravity in AdS$_5 \times \mathbf{S}^5$ below the Hawking-Page temperature. The full computation would require to compute the heat kernell of the field content in AdS, in zeta function regularization \cite{Hawking:1977}. To simplify things, we shall approximate the space by a $9$-dimensional box of size $\ell$. Recall that the thermal partition function of a massless scalar/fermion in a $D$-dimensional spatial box of size $\ell$ is 
\begin{gather}
\log Z_s(\beta) = \sum_{\nvec} - \dfrac{\beta\omega_{\nvec}}{2} - \log (1-e^{-\beta \omega_{\nvec}}) \\ \,
\log Z_f(\beta) = \sum_{\nvec}  \dfrac{\beta\omega_{\nvec}}{2}  + \log (1+e^{-\beta \omega_{\nvec}})\;,
\end{gather}
where $\mathbf{n} \in \mathbf{N}^{D-1}$ and $\omega_{\nvec} = |\nvec|/\ell$.

The ground state free energy will vanish due to the $\mathcal{N} =(2,0)$ supersymmetry in $D+1=10$ dimensions. The free energy of each kind of particle is then
\begin{gather}
\log Z_s(\beta) \approx -\text{Vol}(\mathbf{S}^{D-1}) \int \dfrac{\text{d}|\nvec|}{(2\pi)^D} |\nvec|^{D-1} \log (1-e^{-\beta |\nvec|/\ell}) = c_D\dfrac{\ell^D}{\beta^D}\;,\\
\log Z_f(\beta) \approx \text{Vol}(\mathbf{S}^{D-1}) \int \dfrac{\text{d}|\nvec|}{(2\pi)^D} |\nvec|^{D-1}\log (1+e^{-\beta |\nvec|/\ell}) = c'_D\dfrac{\ell^D}{\beta^D}\;,
\end{gather} 
where 
\begin{gather}
c_D = \dfrac{\Gamma(D) \text{Li}_{D+1}(1)}{\Gamma\left(\frac{D}{2}\right)2^{D-1}\pi^{\frac{D}{2}}}\;, \\
c_D' = \dfrac{(1 - 2^{-D}) \Gamma(D) \zeta(D+1) \text{Li}_{D+1}(1)}{\Gamma\left(\frac{D}{2}\right)2^{D-1}\pi^{\frac{D}{2}}}\;,
\end{gather}  
with $\text{Li}_s(z)$ the polylogarithm function.

The free energy of the theory is just
\be 
F(\beta) \approx -\log Z(\beta) = -(N_{\text{NS-NS}}+N_{\text{R-R}})\left(c_D + c_D'\right) \dfrac{\ell^D}{\beta^D}
\ee 
where $N_{\text{NS-NS}} = N_{\text{R-R}} =(D-1)^2 = 56$ corresponds to the DOF of each bosonic sector. In total, there are $128$ bosonic and $128$ fermionic local DOF. The thermal entropy is then
\be 
S_{\text{gas}}(\beta) \approx (1-\beta \partial_\beta )\log Z(\beta) \approx  221 \dfrac{\ell^9}{\beta^9}\;.
\ee 

At the Hawking-Page temperature in AdS$_{5}$, $\beta_{\text{HP}} \approx \frac{2\pi}{3}\ell$, the entropy of the supergravity gas,
\be 
S(\beta_{\text{HP}})_{\text{IIB}} \approx 0.29\;,
\ee 
is very small. The reason is that the temperature is of the scale of the gap of the system, and that the number of species is not extremelly large. For the considerations in Secs. \ref{sec:3}, \ref{sec:4} and \ref{sec:5} to apply sharply to this system, we will need to include more entanglement between $\Co$ and $\lef,\ri$,by adding entangled modes microcanonically on top of the low-temperature thermal state.

\section{Low energy coherent states in the harmonic oscillator}
\label{app:C}

The bulk DOF in each asymptotically AdS region are weakly coupled fields which can be treated as collections of bosonic or fermionic oscillators in the first approximation. In this appendix, we review a simplified version of this setting to illustrate some points in the main text. In particular, we will show that the Hilbert space of a harmonic oscillator can be spanned by coherent states with a fixed average number of quanta thanks to the presence of energy tails. This should be regarded as a simple explicit realization of the discussion of Sec. \ref{sec:4.2}.

Consider a single harmonic oscillator with annihilation and creation operators $a$ and $a^\dagger$, respectively. Coherent states $\ket{\alpha}$ obey
\be
    a |\alpha\rangle = \alpha | \alpha \rangle.
\ee
It is well known that the set of all coherent states form an overcomplete basis for the oscillator Hilbert space. It is also possible to obtain an overcomplete basis just from coherent states with a fixed average number of quanta.

The average number of quanta in a coherent state is 
\be
    \langle \alpha | a^\dagger a | \alpha \rangle = |\alpha|^2,
\ee
so we need only consider states with a fixed magnitude $|\alpha| = r$. The uniform mixture over such coherent states is
\be
    \int \frac{d\theta}{2\pi} | r e^{i \theta} \rangle \langle r e^{i \theta} | =   \int \frac{d\theta}{2\pi} \sum_{n,m} \frac{r^{n+m} e^{i (n-m)\theta}}{\sqrt{n! m!}} |n \rangle \langle m| = \sum_n \frac{r^{2n}}{n!} |n\rangle \langle n | .
\ee
Because this uniform mixture is full rank, the fixed magnitude coherent states also form an overcomplete basis. For example, a given Fock basis state can be obtained as
\be
    | n\rangle = \frac{\sqrt{n!}}{r^n} \int \frac{d \theta}{2 \pi} e^{- i n \theta} | r e^{i \theta} \rangle.
\ee

A resolution of the identity can be formally obtained via
\be
    I = \int \frac{d\theta}{2\pi} \frac{d \theta'}{2\pi} | r e^{i \theta} \rangle \langle r e^{i \theta'} | g(\theta-\theta')
\ee
where 
\be
    g(\theta - \theta') = \sum_{\ell = 0}^\infty \frac{ \ell!}{r^{2\ell}} e^{- i \ell (\theta - \theta')}.
\ee
This is formal because the series for $g$ does not converge. However, defining
\be
    g_\Lambda(\theta -\theta') = \sum_{\ell=0}^\Lambda  \frac{ \ell!}{r^{2\ell}} e^{- i \ell (\theta - \theta')},
\ee
we see that
\be
    I_\Lambda = \int \frac{d\theta}{2\pi} \frac{d \theta'}{2\pi} | r e^{i \theta} \rangle \langle r e^{i \theta'} | g_\Lambda(\theta-\theta')
\ee
where
\be
    I_\Lambda = \sum_{n=0}^\Lambda | n \rangle \langle n |
\ee
is a partial resolution in the Fock basis up to level $\Lambda$.

\section{Cosmology as a random purification}
\label{app:D}

The entanglement spectrum \eqref{eq:renyis} of the microscopic CFT state is characteristic of a \textit{random purification} of the bulk reduced density matrices
\be\label{eq:randtherm}
\ket{\Psi_{\Op}} \rightarrow \ket{ \sqrt{\rho_\lef} \, R \,\sqrt{\rho_\ri}} = \dfrac{1}{Z_1}\sum_{n,p,q,m} (\sqrt{\rho_\lef})_{np} R_{pq} (\sqrt{\rho_\ri})_{qm} \ket{E_n}^*_\Le\otimes \ket{E_m}_\Ri \;,
\ee
where $R$ is a complex gaussian random matrix of zero mean and unit variance. The density matrices correspond to \eqref{eq:stateredl} and \eqref{eq:stateredr} in the low-lying spectrum of the CFT.

Such a random state originates from the microscopic description of the heavy operator $\mathcal{O}$, assuming that it satisfies a generalized version of the eigenstate thermalization hypothesis \cite{Deutsch,Srednicki1994},\footnote{The \textit{ansatz} \eqref{eq:genETH} for the heavy operator $\mathcal{O}$ is justified since it can still be considered ``simple'' in the large-$N$ internal space of the holographic CFT. In fact, it consists of roughly the product of $\sim N^2$ local conformal primaries of low conformal dimension at different points of the sphere, as opposed to complex operators consisting of random polynomials of these primaries. Moreover, even if the low-lying spectrum of the CFT becomes integrable in the large-$N$ limit, the operator $\mathcal{O}$ does not become sparse in the energy basis, becuase it is composed of $\sim N^2$ light operator insertions and thus its structure is still expected to resemble the one for a chaotic system even at low energies.}
\be\label{eq:genETH}
\mathcal{O}_{nm} \,= e^{-f(E_n,E_m)/2} \,R_{nm}\;,
\ee
for a smooth envelope function $f(E_n,E_m)$ which determines the semiclassical two-point function of the operator (cf. \cite{Sasieta:2022ksu}). In the relevant low-energy spectrum, the envelope function reads
\be\label{eq:envelope}
f(E_n,E_m) \approx \Delta\tau \,(E_n + E_m) + 2m\ell\log R_*/2\ell \,.
\ee 

Under these considerations, the Euclidean saddle point geometries $X_g$ for $g\in \text{Sym}(n)$ in the gravitational replica trick of Sec. \ref{sec:3} can be equivalently derived directly from the microscopic CFT description in the large-$N$ asymptotic expansion of the corresponding quantities, by assuming an effective gaussian random matrix description for the $R_{nm}$ coefficients of the shell operator in \eqref{eq:genETH}. 

\subsection*{The replica saddles from the CFT}

To see this, consider the unnormalized density matrix $\rhohat_{\Le} = \sqrt{\rho_\lef} R \rho_\ri R^\dagger \sqrt{\rho_\lef}$ obtained from \eqref{eq:PETS} under the partial trace of the $\Ri$ subsystem. The unnormalized purity corresponds to
\be
\text{Tr}\rhohat^2_{\Le} = \text{Tr}(\rho_\lef R \rho_\ri R^\dagger \rho_\lef R \rho_\ri R^\dagger) \;.
\ee 

Assuming gaussian statistics for the $R_{nm}$ coefficients, 
\be\label{eq:dgaussianr}
\overline{R_{nm}R^*_{pq}R_{n'm'}R^*_{p'q'}} = \delta_{np}\delta_{mq}\delta_{n'p'}\delta_{m'q'} + \delta_{np'}\delta_{mq'}\delta_{n'p}\delta_{m'q}\;,
\ee 
the unnormalized purity gives
\be\label{eq:dpurityun} 
\overline{\text{Tr}\rhohat_{\Le}^2}  =  Z_\lef(2\beta_\Le)Z_\ri(\beta_\Ri)^2 +  Z_\lef(\beta_\Le)^2Z_\ri(2\beta_\Ri)   \;,
\ee 
where $Z_\lef(n\beta_\Le) = \text{Tr}(\rho_\lef^n)$ and $Z_\ri(n\beta_\Ri) = \text{Tr}(\rho_\ri^n)$. The first term is associated to the replica-disconnected contractions of the $R_{nm}$ coefficients, while the second term is associated to the replica-connected contractions, in analogy with eqs. \eqref{eq:gpur1} and \eqref{eq:gpur2} in the Euclidean gravity calculation.

Similarly, the Euclidean saddle point geometries computing the normalization can also be derived in this way. One gets that
\be\label{eq:dnorm}
\overline{(\text{Tr}\rhohat_{\Le})^2} = Z_\lef(\beta_\Le)^2 Z_\ri(\beta_\Ri)^2 + Z_\lef(2\beta_\Le)Z_\ri(2\beta_\Ri)\;,
\ee 
where the first term comes from the disconnected contraction of the $R_{nm}$ coefficients, while the second term comes from the connected contraction, in complete analogy with eqs. \eqref{eq:gnorm1} and \eqref{eq:gnorm2} in the Euclidean gravity calculation.

The ratio between \eqref{eq:dpurityun} and \eqref{eq:dnorm} gives the desired result
\be 
\overline{\text{Tr}\rho_{\Le}^2} \approx \dfrac{d_{\ri} + d_{\lef}}{1 + d_{\lef}d_{\ri}}
\ee 
for $d^{-1}_{\lef,\ri} = \text{Tr}(\rho_{\lef,\ri}^2)$. This formula yields the same result as the semiclassical purity \eqref{eq:purity4}.

In a completely analogous way, it is possible to derive the higher semiclassical R\'{e}nyi entropies \eqref{eq:renyis} from the gaussian statistics of the $R_{nm}$ coefficients \eqref{eq:dgaussianr}. The universality of the entanglement spectrum \eqref{eq:renyis} arises naturally in the large mass limit from the CFT perspective, by thinking of $m\ell$ as a label which parametrizes a trajectory of the cosmological states at fixed entanglement in the Hilbert space of the CFTs. As $m\ell$ increases, the state vector $\ket{\Psi_{\mathcal{O}}}$ ergodically explores the Hilbert space generated by the cosmological states. At large values of $m\ell$, the entanglement spectrum equilibrates to the value for a random purification.\footnote{For the PETS, this reproduces the entanglement spectrum of a thermal random state (cf. \cite{Liu:2020jsv}). An analogous spectrum is generated by the equilibration in time under a chaotic Hamiltonian, which is relevant for the study of the evaporating black hole (cf. \cite{Sasieta:2021pzj}).}

\section{Tensor network resolvent}
\label{app:E}

In this appendix, we carry out a resolvent calculation similar to the one reported in Sec. \ref{sec:4.2}, but in the random TN toy model introduced in Sec. \ref{sec:5}. Let us consider a collection of $K$ orthogonal bulk states $S = \lbrace {\ket{\psi_{I_1}},...,\ket{\psi_{I_K}}}\rbrace \subset \mathcal{H}_{\psi} $ of the cosmological EFT, and the corresponding TN states $V_{\psi}S = \lbrace {\ket{\Psi_{I_1}},...,\ket{\Psi_{I_K}}}\rbrace \subset \mathcal{H}_{\Le}\otimes  \mathcal{H}_{\Ri}$. We can then build the TN $K\times K$ Gram matrix of overlaps
\be\label{eq:tngm}
G_{ij} \equiv \bra{\Psi_{I_i}}\ket{\Psi_{I_j}} = \bra{\psi_{I_i}}V^\dagger_\psi V_\psi\ket{\psi_{I_j}}\;.
\ee 

The moments of the TN Gram matrix can be easily computed from the TN diagramatics upon the gaussian average over the $\Op$ tensor. For $i\neq j$, the second moment is given by eq. \eqref{eq:varianceisom}, which we import here
\be 
\overline{|G_{ij}|^2} = \dfrac{1}{|\lef| |\ri|}\;.
\ee 

Similarly, it is straightforward to see that the completely connected contribution to the $n$-th moment reads
\be 
\overline{G_{i_{1}i_{2}}G_{i_{2}i_{3}}...G_{i_{n}i_{1}}}|_{\text{conn.}} \approx \dfrac{1}{|\lef|^{n-1} |\ri|^{n-1}}\;.
\ee 
In this case there are $n-1$ different contractions of the $\Op$ tensor, which lead to completely connected contributions. The dominant one comes from the $\eta = (12...n)$ permutation, in analogy with eq. \eqref{eq:uniovern} for the dominant Euclidean wormhole configuration to the gravitational moments of the Gram matrix.

In order to find the dimension of the Hilbert space that the cosmological TN states span, we want to compute the average eigenvalue density of the Gram matrix \eqref{eq:tngm}. To do that, we again invoke the resolvent
\be \label{eq:resolapp}
R_{ij}(\lambda)\,\equiv\,\left( \frac{1}{\lambda \mathds{1} -G}\right)_{ij} \,=\,\frac{1}{\lambda}\,\delta_{ij}+\sum\limits_{n=1}^{\infty}\,\frac{1}{\lambda^{n+1}}\,(G)^n_{ij}\;,
\ee
and write down the trace of the TN resolvent in Schwinger-Dyson form, which allows for the resummation
\be 
\lambda \overline{R(\lambda)}\,= \,\,K+2^{\mathbf{S}}\sum\limits_{n=1}^{\infty}\,\,\left(\dfrac{\overline{R(\lambda)}}{2^{\mathbf{S}}}\right)^n = K + \frac{2^\textbf{S}\,\overline{R(\lambda)}}{2^\textbf{S}-\overline{R(\lambda)}}\;
\ee
with $\mathbf{S} = \log_2 |\lef||\ri| = S_\lef + S_\ri$.

This leads to the quadratic equation for the trace of the resolvent
\be 
\overline{R(\lambda)}^2+\left(\,\frac{2^\textbf{S}-K}{\lambda}-2^\textbf{S}\,\right)\,\overline{R(\lambda)}+\dfrac{K}{\lambda}\,2^\textbf{S}= 0\;.
\ee

The density of eigenvalues of the Gram matrix follows from the discontinuity of the trace of the resolvent along the imaginary axis \eqref{eq:densitydisc},
\be \label{denGtn}
\overline{D(\lambda)}=\frac{2^\textbf{S}}{2\pi\lambda}\sqrt{\,\left[\lambda-\left(1-K^{1/2}\, 2^{-\textbf{S}/2} \right)^2\,\right]\left[\,\left(1+K^{1/2} \,2^{-\textbf{S}/2} \right)^2-\lambda \right]}+\delta(\lambda)\left(K-2^{\mathbf{S}}\right)\theta(K-2^{\mathbf{S}})\;.
\ee

Therefore, when $K> 2^{\mathbf{S}}$, the rank of the Gram matrix, which is equal to the dimension of the image of the map $V_\psi$, saturates at a value $|\lef||\ri|$, according to the sharp curve
\be\label{eq:imagetnapp}
\text{dim}\lbrace \text{Im}(V_\psi)\rbrace = \text{rank}(G_{ij}) = \text{min}\lbrace K, |\lef||\ri|\rbrace\;.
\ee

\bibliographystyle{utphys}
\bibliography{bibliography}

\providecommand{\href}[2]{#2}\begingroup\raggedright\begin{thebibliography}{100}

\bibitem{tHooft:1993dmi}
G.~'t~Hooft, ``{Dimensional reduction in quantum gravity}'', {\em Conf. Proc.
  C} {\bfseries 930308} (1993) 284--296,
  \href{http://arxiv.org/abs/gr-qc/9310026}{{\ttfamily arXiv:gr-qc/9310026}}.

\bibitem{Susskind:1994vu}
L.~Susskind, ``{The World as a hologram}'',
  \href{http://dx.doi.org/10.1063/1.531249}{{\em J. Math. Phys.} {\bfseries 36}
  (1995) 6377--6396}, \href{http://arxiv.org/abs/hep-th/9409089}{{\ttfamily
  arXiv:hep-th/9409089}}.

\bibitem{Maldacena:1997re}
J.~M. Maldacena, ``{The Large N limit of superconformal field theories and
  supergravity}'', \href{http://dx.doi.org/10.1023/A:1026654312961}{{\em Adv.
  Theor. Math. Phys.} {\bfseries 2} (1998) 231--252},
  \href{http://arxiv.org/abs/hep-th/9711200}{{\ttfamily arXiv:hep-th/9711200}}.

\bibitem{Ryu2006a}
S.~Ryu and T.~Takayanagi, ``{Aspects of Holographic Entanglement Entropy}'',
  \href{http://dx.doi.org/10.1088/1126-6708/2006/08/045}{{\em JHEP} {\bfseries
  08} (2006) 045},
\href{http://arxiv.org/abs/hep-th/0605073}{{\ttfamily arXiv:hep-th/0605073
  [hep-th]}}.

\bibitem{Swingle:2009bg}
B.~Swingle, ``{Entanglement Renormalization and Holography}'',
  \href{http://dx.doi.org/10.1103/PhysRevD.86.065007}{{\em Phys. Rev. D}
  {\bfseries 86} (2012) 065007},
  \href{http://arxiv.org/abs/0905.1317}{{\ttfamily arXiv:0905.1317
  [cond-mat.str-el]}}.

\bibitem{VanRaamsdonk:2010pw}
M.~Van~Raamsdonk, ``{Building up spacetime with quantum entanglement}'',
  \href{http://dx.doi.org/10.1142/S0218271810018529}{{\em Gen. Rel. Grav.}
  {\bfseries 42} (2010) 2323--2329},
  \href{http://arxiv.org/abs/1005.3035}{{\ttfamily arXiv:1005.3035 [hep-th]}}.

\bibitem{Lewkowycz:2013nqa}
A.~Lewkowycz and J.~Maldacena, ``{Generalized gravitational entropy}'',
  \href{http://dx.doi.org/10.1007/JHEP08(2013)090}{{\em JHEP} {\bfseries 08}
  (2013) 090}, \href{http://arxiv.org/abs/1304.4926}{{\ttfamily arXiv:1304.4926
  [hep-th]}}.

\bibitem{Maldacena:2013xja}
J.~Maldacena and L.~Susskind, ``{Cool horizons for entangled black holes}'',
  \href{http://dx.doi.org/10.1002/prop.201300020}{{\em Fortsch. Phys.}
  {\bfseries 61} (2013) 781--811},
  \href{http://arxiv.org/abs/1306.0533}{{\ttfamily arXiv:1306.0533 [hep-th]}}.

\bibitem{Faulkner:2013ana}
T.~Faulkner, A.~Lewkowycz, and J.~Maldacena, ``{Quantum corrections to
  holographic entanglement entropy}'',
  \href{http://dx.doi.org/10.1007/JHEP11(2013)074}{{\em JHEP} {\bfseries 11}
  (2013) 074}, \href{http://arxiv.org/abs/1307.2892}{{\ttfamily arXiv:1307.2892
  [hep-th]}}.

\bibitem{Stanford:2014jda}
D.~Stanford and L.~Susskind, ``{Complexity and Shock Wave Geometries}'',
  \href{http://dx.doi.org/10.1103/PhysRevD.90.126007}{{\em Phys. Rev. D}
  {\bfseries 90} no.~12, (2014) 126007},
  \href{http://arxiv.org/abs/1406.2678}{{\ttfamily arXiv:1406.2678 [hep-th]}}.

\bibitem{Almheiri:2014lwa}
A.~Almheiri, X.~Dong, and D.~Harlow, ``{Bulk Locality and Quantum Error
  Correction in AdS/CFT}'',
  \href{http://dx.doi.org/10.1007/JHEP04(2015)163}{{\em JHEP} {\bfseries 04}
  (2015) 163}, \href{http://arxiv.org/abs/1411.7041}{{\ttfamily arXiv:1411.7041
  [hep-th]}}.

\bibitem{Maldacena:2015waa}
J.~Maldacena, S.~H. Shenker, and D.~Stanford, ``{A bound on chaos}'',
  \href{http://dx.doi.org/10.1007/JHEP08(2016)106}{{\em JHEP} {\bfseries 08}
  (2016) 106}, \href{http://arxiv.org/abs/1503.01409}{{\ttfamily
  arXiv:1503.01409 [hep-th]}}.

\bibitem{Pastawski:2015qua}
F.~Pastawski, B.~Yoshida, D.~Harlow, and J.~Preskill, ``{Holographic quantum
  error-correcting codes: Toy models for the bulk/boundary correspondence}'',
  \href{http://dx.doi.org/10.1007/JHEP06(2015)149}{{\em JHEP} {\bfseries 06}
  (2015) 149}, \href{http://arxiv.org/abs/1503.06237}{{\ttfamily
  arXiv:1503.06237 [hep-th]}}.

\bibitem{Brown:2015bva}
A.~R. Brown, D.~A. Roberts, L.~Susskind, B.~Swingle, and Y.~Zhao,
  ``{Holographic Complexity Equals Bulk Action?}'',
  \href{http://dx.doi.org/10.1103/PhysRevLett.116.191301}{{\em Phys. Rev.
  Lett.} {\bfseries 116} no.~19, (2016) 191301},
  \href{http://arxiv.org/abs/1509.07876}{{\ttfamily arXiv:1509.07876
  [hep-th]}}.

\bibitem{Hayden:2016cfa}
P.~Hayden, S.~Nezami, X.-L. Qi, N.~Thomas, M.~Walter, and Z.~Yang,
  ``{Holographic duality from random tensor networks}'',
  \href{http://dx.doi.org/10.1007/JHEP11(2016)009}{{\em JHEP} {\bfseries 11}
  (2016) 009}, \href{http://arxiv.org/abs/1601.01694}{{\ttfamily
  arXiv:1601.01694 [hep-th]}}.

\bibitem{Penington:2019npb}
G.~Penington, ``{Entanglement Wedge Reconstruction and the Information
  Paradox}'', \href{http://dx.doi.org/10.1007/JHEP09(2020)002}{{\em JHEP}
  {\bfseries 09} (2020) 002}, \href{http://arxiv.org/abs/1905.08255}{{\ttfamily
  arXiv:1905.08255 [hep-th]}}.

\bibitem{Almheiri:2019psf}
A.~Almheiri, N.~Engelhardt, D.~Marolf, and H.~Maxfield, ``{The entropy of bulk
  quantum fields and the entanglement wedge of an evaporating black hole}'',
  \href{http://dx.doi.org/10.1007/JHEP12(2019)063}{{\em JHEP} {\bfseries 12}
  (2019) 063}, \href{http://arxiv.org/abs/1905.08762}{{\ttfamily
  arXiv:1905.08762 [hep-th]}}.

\bibitem{Penington:2019kki}
G.~Penington, S.~H. Shenker, D.~Stanford, and Z.~Yang, ``{Replica wormholes and
  the black hole interior}'',
  \href{http://dx.doi.org/10.1007/JHEP03(2022)205}{{\em JHEP} {\bfseries 03}
  (2022) 205}, \href{http://arxiv.org/abs/1911.11977}{{\ttfamily
  arXiv:1911.11977 [hep-th]}}.

\bibitem{Almheiri:2019qdq}
A.~Almheiri, T.~Hartman, J.~Maldacena, E.~Shaghoulian, and A.~Tajdini,
  ``{Replica Wormholes and the Entropy of Hawking Radiation}'',
  \href{http://dx.doi.org/10.1007/JHEP05(2020)013}{{\em JHEP} {\bfseries 05}
  (2020) 013}, \href{http://arxiv.org/abs/1911.12333}{{\ttfamily
  arXiv:1911.12333 [hep-th]}}.

\bibitem{Hartle:2012qb}
J.~B. Hartle, S.~W. Hawking, and T.~Hertog, ``{Accelerated Expansion from
  Negative $\Lambda$}'', \href{http://arxiv.org/abs/1205.3807}{{\ttfamily
  arXiv:1205.3807 [hep-th]}}.

\bibitem{Peebles:1987ek}
P.~J.~E. Peebles and B.~Ratra, ``{Cosmology with a Time Variable Cosmological
  Constant}'', \href{http://dx.doi.org/10.1086/185100}{{\em Astrophys. J.
  Lett.} {\bfseries 325} (1988) L17}.

\bibitem{Ratra:1987rm}
B.~Ratra and P.~J.~E. Peebles, ``{Cosmological Consequences of a Rolling
  Homogeneous Scalar Field}'',
  \href{http://dx.doi.org/10.1103/PhysRevD.37.3406}{{\em Phys. Rev. D}
  {\bfseries 37} (1988) 3406}.

\bibitem{Tsujikawa:2013fta}
S.~Tsujikawa, ``{Quintessence: A Review}'',
  \href{http://dx.doi.org/10.1088/0264-9381/30/21/214003}{{\em Class. Quant.
  Grav.} {\bfseries 30} (2013) 214003},
  \href{http://arxiv.org/abs/1304.1961}{{\ttfamily arXiv:1304.1961 [gr-qc]}}.

\bibitem{Antonini:2022ptt}
S.~Antonini, P.~Simidzija, B.~Swingle, and M.~Van~Raamsdonk, ``{Accelerating
  Cosmology from a Holographic Wormhole}'',
  \href{http://dx.doi.org/10.1103/PhysRevLett.130.221601}{{\em Phys. Rev.
  Lett.} {\bfseries 130} no.~22, (2023) 221601},
  \href{http://arxiv.org/abs/2206.14821}{{\ttfamily arXiv:2206.14821
  [hep-th]}}.

\bibitem{Antonini:2022fna}
S.~Antonini, P.~Simidzija, B.~Swingle, M.~Van~Raamsdonk, and C.~Waddell,
  ``{Accelerating cosmology from \ensuremath{\Lambda} \ensuremath{<} 0
  gravitational effective field theory}'',
  \href{http://dx.doi.org/10.1007/JHEP05(2023)203}{{\em JHEP} {\bfseries 05}
  (2023) 203}, \href{http://arxiv.org/abs/2212.00050}{{\ttfamily
  arXiv:2212.00050 [hep-th]}}.

\bibitem{VanRaamsdonk:2023ion}
M.~Van~Raamsdonk and C.~Waddell, ``{Possible hints of decreasing dark energy
  from supernova data}'', \href{http://arxiv.org/abs/2305.04946}{{\ttfamily
  arXiv:2305.04946 [astro-ph.CO]}}.

\bibitem{Strominger:2001pn}
A.~Strominger, ``{The dS / CFT correspondence}'',
  \href{http://dx.doi.org/10.1088/1126-6708/2001/10/034}{{\em JHEP} {\bfseries
  10} (2001) 034}, \href{http://arxiv.org/abs/hep-th/0106113}{{\ttfamily
  arXiv:hep-th/0106113}}.

\bibitem{Alishahiha:2004md}
M.~Alishahiha, A.~Karch, E.~Silverstein, and D.~Tong, ``{The dS/dS
  correspondence}'', \href{http://dx.doi.org/10.1063/1.1848341}{{\em AIP Conf.
  Proc.} {\bfseries 743} no.~1, (2004) 393--409},
  \href{http://arxiv.org/abs/hep-th/0407125}{{\ttfamily arXiv:hep-th/0407125}}.

\bibitem{dsds}
X.~Dong, E.~Silverstein, and G.~Torroba, ``{De Sitter Holography and
  Entanglement Entropy}'',
  \href{http://dx.doi.org/10.1007/JHEP07(2018)050}{{\em JHEP} {\bfseries 07}
  (2018) 050},
\href{http://arxiv.org/abs/1804.08623}{{\ttfamily arXiv:1804.08623 [hep-th]}}.

\bibitem{Gorbenko:2018oov}
V.~Gorbenko, E.~Silverstein, and G.~Torroba, ``{dS/dS and $ T\overline{T} $}'',
  \href{http://dx.doi.org/10.1007/JHEP03(2019)085}{{\em JHEP} {\bfseries 03}
  (2019) 085}, \href{http://arxiv.org/abs/1811.07965}{{\ttfamily
  arXiv:1811.07965 [hep-th]}}.

\bibitem{Coleman:2021nor}
E.~Coleman, E.~A. Mazenc, V.~Shyam, E.~Silverstein, R.~M. Soni, G.~Torroba, and
  S.~Yang, ``{De Sitter microstates from T$ \overline{T} $ +
  \ensuremath{\Lambda}$_{2}$ and the Hawking-Page transition}'',
  \href{http://dx.doi.org/10.1007/JHEP07(2022)140}{{\em JHEP} {\bfseries 07}
  (2022) 140}, \href{http://arxiv.org/abs/2110.14670}{{\ttfamily
  arXiv:2110.14670 [hep-th]}}.

\bibitem{Susskind:2021dfc}
L.~Susskind, ``{Black Holes Hint Towards De Sitter-Matrix Theory}'',
  \href{http://arxiv.org/abs/2109.01322}{{\ttfamily arXiv:2109.01322
  [hep-th]}}.

\bibitem{Susskind:2021omt}
L.~Susskind, ``{De Sitter Holography: Fluctuations, Anomalous Symmetry, and
  Wormholes}'', \href{http://dx.doi.org/10.3390/universe7120464}{{\em Universe}
  {\bfseries 7} no.~12, (2021) 464},
  \href{http://arxiv.org/abs/2106.03964}{{\ttfamily arXiv:2106.03964
  [hep-th]}}.

\bibitem{Susskind:2021esx}
L.~Susskind, ``{Entanglement and Chaos in De Sitter Space Holography: An SYK
  Example}'', \href{http://dx.doi.org/10.22128/jhap.2021.455.1005}{{\em JHAP}
  {\bfseries 1} no.~1, (2021) 1--22},
  \href{http://arxiv.org/abs/2109.14104}{{\ttfamily arXiv:2109.14104
  [hep-th]}}.

\bibitem{Shaghoulian:2022fop}
E.~Shaghoulian and L.~Susskind, ``{Entanglement in De Sitter space}'',
  \href{http://dx.doi.org/10.1007/JHEP08(2022)198}{{\em JHEP} {\bfseries 08}
  (2022) 198}, \href{http://arxiv.org/abs/2201.03603}{{\ttfamily
  arXiv:2201.03603 [hep-th]}}.

\bibitem{McFadden:2009fg}
P.~McFadden and K.~Skenderis, ``{Holography for Cosmology}'',
  \href{http://dx.doi.org/10.1103/PhysRevD.81.021301}{{\em Phys. Rev. D}
  {\bfseries 81} (2010) 021301},
  \href{http://arxiv.org/abs/0907.5542}{{\ttfamily arXiv:0907.5542 [hep-th]}}.

\bibitem{Afshordi:2017ihr}
N.~Afshordi, E.~Gould, and K.~Skenderis, ``{Constraining holographic cosmology
  using Planck data}'',
  \href{http://dx.doi.org/10.1103/PhysRevD.95.123505}{{\em Phys. Rev. D}
  {\bfseries 95} no.~12, (2017) 123505},
  \href{http://arxiv.org/abs/1703.05385}{{\ttfamily arXiv:1703.05385
  [astro-ph.CO]}}.

\bibitem{Nastase:2018cbf}
H.~Nastase, ``{Solution of the cosmological constant problem within holographic
  cosmology}'', \href{http://dx.doi.org/10.1016/j.physletb.2019.135168}{{\em
  Phys. Lett. B} {\bfseries 801} (2020) 135168},
  \href{http://arxiv.org/abs/1812.07597}{{\ttfamily arXiv:1812.07597
  [hep-th]}}.

\bibitem{Nastase:2019rsn}
H.~Nastase and K.~Skenderis, ``{Holography for the very early Universe and the
  classic puzzles of Hot Big Bang cosmology}'',
  \href{http://dx.doi.org/10.1103/PhysRevD.101.021901}{{\em Phys. Rev. D}
  {\bfseries 101} no.~2, (2020) 021901},
  \href{http://arxiv.org/abs/1904.05821}{{\ttfamily arXiv:1904.05821
  [hep-th]}}.

\bibitem{Maldacena:2004rf}
J.~M. Maldacena and L.~Maoz, ``{Wormholes in AdS}'',
  \href{http://dx.doi.org/10.1088/1126-6708/2004/02/053}{{\em JHEP} {\bfseries
  02} (2004) 053}, \href{http://arxiv.org/abs/hep-th/0401024}{{\ttfamily
  arXiv:hep-th/0401024}}.

\bibitem{McInnes:2004nx}
B.~McInnes, ``{Answering a basic objection to bang / crunch holography}'',
  \href{http://dx.doi.org/10.1088/1126-6708/2004/10/018}{{\em JHEP} {\bfseries
  10} (2004) 018}, \href{http://arxiv.org/abs/hep-th/0407189}{{\ttfamily
  arXiv:hep-th/0407189}}.

\bibitem{McInnes:2005sa}
B.~McInnes, ``{Pre-inflationary spacetime in string cosmology}'',
  \href{http://dx.doi.org/10.1016/j.nuclphysb.2006.04.035}{{\em Nucl. Phys. B}
  {\bfseries 748} (2006) 309--332},
  \href{http://arxiv.org/abs/hep-th/0511227}{{\ttfamily arXiv:hep-th/0511227}}.

\bibitem{Freivogel:2005qh}
B.~Freivogel, V.~E. Hubeny, A.~Maloney, R.~C. Myers, M.~Rangamani, and
  S.~Shenker, ``{Inflation in AdS/CFT}'',
  \href{http://dx.doi.org/10.1088/1126-6708/2006/03/007}{{\em JHEP} {\bfseries
  03} (2006) 007}, \href{http://arxiv.org/abs/hep-th/0510046}{{\ttfamily
  arXiv:hep-th/0510046}}.

\bibitem{Engelhardt:2014mea}
N.~Engelhardt, T.~Hertog, and G.~T. Horowitz, ``{Holographic Signatures of
  Cosmological Singularities}'',
  \href{http://dx.doi.org/10.1103/PhysRevLett.113.121602}{{\em Phys. Rev.
  Lett.} {\bfseries 113} (2014) 121602},
  \href{http://arxiv.org/abs/1404.2309}{{\ttfamily arXiv:1404.2309 [hep-th]}}.

\bibitem{Cooper:2018cmb}
S.~Cooper, M.~Rozali, B.~Swingle, M.~Van~Raamsdonk, C.~Waddell, and D.~Wakeham,
  ``{Black hole microstate cosmology}'',
  \href{http://dx.doi.org/10.1007/JHEP07(2019)065}{{\em JHEP} {\bfseries 07}
  (2019) 065}, \href{http://arxiv.org/abs/1810.10601}{{\ttfamily
  arXiv:1810.10601 [hep-th]}}.

\bibitem{Antonini:2019qkt}
S.~Antonini and B.~Swingle, ``{Cosmology at the end of the world}'',
  \href{http://dx.doi.org/10.1038/s41567-020-0909-6}{{\em Nature Phys.}
  {\bfseries 16} no.~8, (2020) 881--886},
  \href{http://arxiv.org/abs/1907.06667}{{\ttfamily arXiv:1907.06667
  [hep-th]}}.

\bibitem{Chen:2020tes}
Y.~Chen, V.~Gorbenko, and J.~Maldacena, ``{Bra-ket wormholes in gravitationally
  prepared states}'', \href{http://dx.doi.org/10.1007/JHEP02(2021)009}{{\em
  JHEP} {\bfseries 02} (2021) 009},
  \href{http://arxiv.org/abs/2007.16091}{{\ttfamily arXiv:2007.16091
  [hep-th]}}.

\bibitem{VanRaamsdonk:2020tlr}
M.~Van~Raamsdonk, ``{Comments on wormholes, ensembles, and cosmology}'',
  \href{http://dx.doi.org/10.1007/JHEP12(2021)156}{{\em JHEP} {\bfseries 12}
  (2021) 156}, \href{http://arxiv.org/abs/2008.02259}{{\ttfamily
  arXiv:2008.02259 [hep-th]}}.

\bibitem{VanRaamsdonk:2021qgv}
M.~Van~Raamsdonk, ``{Cosmology from confinement?}'',
  \href{http://dx.doi.org/10.1007/JHEP03(2022)039}{{\em JHEP} {\bfseries 03}
  (2022) 039}, \href{http://arxiv.org/abs/2102.05057}{{\ttfamily
  arXiv:2102.05057 [hep-th]}}.

\bibitem{Antonini:2022blk}
S.~Antonini, P.~Simidzija, B.~Swingle, and M.~Van~Raamsdonk, ``{Cosmology from
  the vacuum}'', \href{http://arxiv.org/abs/2203.11220}{{\ttfamily
  arXiv:2203.11220 [hep-th]}}.

\bibitem{Fallows:2022ioc}
S.~Fallows and S.~F. Ross, ``{Constraints on cosmologies inside black holes}'',
  \href{http://dx.doi.org/10.1007/JHEP05(2022)094}{{\em JHEP} {\bfseries 05}
  (2022) 094}, \href{http://arxiv.org/abs/2203.02523}{{\ttfamily
  arXiv:2203.02523 [hep-th]}}.

\bibitem{Ross:2022pde}
S.~F. Ross, ``{Cosmologies inside hyperbolic black holes}'',
  \href{http://dx.doi.org/10.1007/JHEP11(2022)168}{{\em JHEP} {\bfseries 11}
  (2022) 168}, \href{http://arxiv.org/abs/2209.11620}{{\ttfamily
  arXiv:2209.11620 [hep-th]}}.

\bibitem{Sahu:2023fbx}
A.~Sahu, P.~Simidzija, and M.~Van~Raamsdonk, ``{Bubbles of cosmology in
  AdS/CFT}'', \href{http://arxiv.org/abs/2306.13143}{{\ttfamily
  arXiv:2306.13143 [hep-th]}}.

\bibitem{Akers:2022max}
C.~Akers, T.~Faulkner, S.~Lin, and P.~Rath, ``{The Page curve for reflected
  entropy}'', \href{http://dx.doi.org/10.1007/JHEP06(2022)089}{{\em JHEP}
  {\bfseries 06} (2022) 089}, \href{http://arxiv.org/abs/2201.11730}{{\ttfamily
  arXiv:2201.11730 [hep-th]}}.

\bibitem{Hartman:2020khs}
T.~Hartman, Y.~Jiang, and E.~Shaghoulian, ``{Islands in cosmology}'',
  \href{http://dx.doi.org/10.1007/JHEP11(2020)111}{{\em JHEP} {\bfseries 11}
  (2020) 111}, \href{http://arxiv.org/abs/2008.01022}{{\ttfamily
  arXiv:2008.01022 [hep-th]}}.

\bibitem{Balasubramanian:2020xqf}
V.~Balasubramanian, A.~Kar, and T.~Ugajin, ``{Islands in de Sitter space}'',
  \href{http://dx.doi.org/10.1007/JHEP02(2021)072}{{\em JHEP} {\bfseries 02}
  (2021) 072}, \href{http://arxiv.org/abs/2008.05275}{{\ttfamily
  arXiv:2008.05275 [hep-th]}}.

\bibitem{Balasubramanian:2020coy}
V.~Balasubramanian, A.~Kar, and T.~Ugajin, ``{Entanglement between two disjoint
  universes}'', \href{http://dx.doi.org/10.1007/JHEP02(2021)136}{{\em JHEP}
  {\bfseries 02} (2021) 136}, \href{http://arxiv.org/abs/2008.05274}{{\ttfamily
  arXiv:2008.05274 [hep-th]}}.

\bibitem{Balasubramanian:2021wgd}
V.~Balasubramanian, A.~Kar, and T.~Ugajin, ``{Entanglement between two
  gravitating universes}'',
  \href{http://dx.doi.org/10.1088/1361-6382/ac3c8b}{{\em Class. Quant. Grav.}
  {\bfseries 39} no.~17, (2022) 174001},
  \href{http://arxiv.org/abs/2104.13383}{{\ttfamily arXiv:2104.13383
  [hep-th]}}.

\bibitem{Bousso:2022gth}
R.~Bousso and E.~Wildenhain, ``{Islands in closed and open universes}'',
  \href{http://dx.doi.org/10.1103/PhysRevD.105.086012}{{\em Phys. Rev. D}
  {\bfseries 105} no.~8, (2022) 086012},
  \href{http://arxiv.org/abs/2202.05278}{{\ttfamily arXiv:2202.05278
  [hep-th]}}.

\bibitem{Papadodimas:2012aq}
K.~Papadodimas and S.~Raju, ``{An Infalling Observer in AdS/CFT}'',
  \href{http://dx.doi.org/10.1007/JHEP10(2013)212}{{\em JHEP} {\bfseries 10}
  (2013) 212}, \href{http://arxiv.org/abs/1211.6767}{{\ttfamily arXiv:1211.6767
  [hep-th]}}.

\bibitem{Brown:2019rox}
A.~R. Brown, H.~Gharibyan, G.~Penington, and L.~Susskind, ``{The
  Python\textquoteright{}s Lunch: geometric obstructions to decoding Hawking
  radiation}'', \href{http://dx.doi.org/10.1007/JHEP08(2020)121}{{\em JHEP}
  {\bfseries 08} (2020) 121}, \href{http://arxiv.org/abs/1912.00228}{{\ttfamily
  arXiv:1912.00228 [hep-th]}}.

\bibitem{Engelhardt:2021mue}
N.~Engelhardt, G.~Penington, and A.~Shahbazi-Moghaddam, ``{A world without
  pythons would be so simple}'',
  \href{http://dx.doi.org/10.1088/1361-6382/ac2de5}{{\em Class. Quant. Grav.}
  {\bfseries 38} no.~23, (2021) 234001},
  \href{http://arxiv.org/abs/2102.07774}{{\ttfamily arXiv:2102.07774
  [hep-th]}}.

\bibitem{Engelhardt:2021qjs}
N.~Engelhardt, G.~Penington, and A.~Shahbazi-Moghaddam, ``{Finding pythons in
  unexpected places}'', \href{http://dx.doi.org/10.1088/1361-6382/ac3e75}{{\em
  Class. Quant. Grav.} {\bfseries 39} no.~9, (2022) 094002},
  \href{http://arxiv.org/abs/2105.09316}{{\ttfamily arXiv:2105.09316
  [hep-th]}}.

\bibitem{Akers:2022qdl}
C.~Akers, N.~Engelhardt, D.~Harlow, G.~Penington, and S.~Vardhan, ``{The black
  hole interior from non-isometric codes and complexity}'',
  \href{http://arxiv.org/abs/2207.06536}{{\ttfamily arXiv:2207.06536
  [hep-th]}}.

\bibitem{Cao:2023gkw}
C.~Cao, W.~Chemissany, A.~Jahn, and Z.~Zimbor\'as, ``{Approximate observables
  from non-isometric maps: de Sitter tensor networks with overlapping
  qubits}'', \href{http://arxiv.org/abs/2304.02673}{{\ttfamily arXiv:2304.02673
  [hep-th]}}.

\bibitem{Balasubramanian:2022gmo}
V.~Balasubramanian, A.~Lawrence, J.~M. Magan, and M.~Sasieta, ``{Microscopic
  origin of the entropy of black holes in general relativity}'',
  \href{http://arxiv.org/abs/2212.02447}{{\ttfamily arXiv:2212.02447
  [hep-th]}}.

\bibitem{Sasieta:2022ksu}
M.~Sasieta, ``{Wormholes from heavy operator statistics in AdS/CFT}'',
  \href{http://dx.doi.org/10.1007/JHEP03(2023)158}{{\em JHEP} {\bfseries 03}
  (2023) 158}, \href{http://arxiv.org/abs/2211.11794}{{\ttfamily
  arXiv:2211.11794 [hep-th]}}.

\bibitem{Wheeler:1964qna}
J.~A. Wheeler, ``{Geometrodynamics and the issue of final state}'', in {\em
  {Les Houches Summer Shcool of Theoretical Physics}: {Relativity, Groups and
  Topology}}, pp.~317--522.
\newblock 1964.

\bibitem{Fu:2019oyc}
Z.~Fu and D.~Marolf, ``{Bag-of-gold spacetimes, Euclidean wormholes, and
  inflation from domain walls in AdS/CFT}'',
  \href{http://dx.doi.org/10.1007/JHEP11(2019)040}{{\em JHEP} {\bfseries 11}
  (2019) 040}, \href{http://arxiv.org/abs/1909.02505}{{\ttfamily
  arXiv:1909.02505 [hep-th]}}.

\bibitem{Hamilton:2005ju}
A.~Hamilton, D.~N. Kabat, G.~Lifschytz, and D.~A. Lowe, ``{Local bulk operators
  in AdS/CFT: A Boundary view of horizons and locality}'',
  \href{http://dx.doi.org/10.1103/PhysRevD.73.086003}{{\em Phys. Rev. D}
  {\bfseries 73} (2006) 086003},
  \href{http://arxiv.org/abs/hep-th/0506118}{{\ttfamily arXiv:hep-th/0506118}}.

\bibitem{Hamilton:2006az}
A.~Hamilton, D.~N. Kabat, G.~Lifschytz, and D.~A. Lowe, ``{Holographic
  representation of local bulk operators}'',
  \href{http://dx.doi.org/10.1103/PhysRevD.74.066009}{{\em Phys. Rev. D}
  {\bfseries 74} (2006) 066009},
  \href{http://arxiv.org/abs/hep-th/0606141}{{\ttfamily arXiv:hep-th/0606141}}.

\bibitem{Goel:2018ubv}
A.~Goel, H.~T. Lam, G.~J. Turiaci, and H.~Verlinde, ``{Expanding the Black Hole
  Interior: Partially Entangled Thermal States in SYK}'',
  \href{http://dx.doi.org/10.1007/JHEP02(2019)156}{{\em JHEP} {\bfseries 02}
  (2019) 156}, \href{http://arxiv.org/abs/1807.03916}{{\ttfamily
  arXiv:1807.03916 [hep-th]}}.

\bibitem{Lin:2022rzw}
H.~W. Lin, J.~Maldacena, L.~Rozenberg, and J.~Shan, ``{Holography for people
  with no time}'', \href{http://arxiv.org/abs/2207.00407}{{\ttfamily
  arXiv:2207.00407 [hep-th]}}.

\bibitem{Lin:2022zxd}
H.~W. Lin, J.~Maldacena, L.~Rozenberg, and J.~Shan, ``{Looking at
  supersymmetric black holes for a very long time}'',
  \href{http://dx.doi.org/10.21468/SciPostPhys.14.5.128}{{\em SciPost Phys.}
  {\bfseries 14} (2023) 128}, \href{http://arxiv.org/abs/2207.00408}{{\ttfamily
  arXiv:2207.00408 [hep-th]}}.

\bibitem{Chandra:2023rhx}
J.~Chandra, ``{Euclidean wormholes for individual 2d CFTs}'',
  \href{http://arxiv.org/abs/2305.07183}{{\ttfamily arXiv:2305.07183
  [hep-th]}}.

\bibitem{Kourkoulou:2017zaj}
I.~Kourkoulou and J.~Maldacena, ``{Pure states in the SYK model and
  nearly-$AdS_2$ gravity}'', \href{http://arxiv.org/abs/1707.02325}{{\ttfamily
  arXiv:1707.02325 [hep-th]}}.

\bibitem{Chandra:2022fwi}
J.~Chandra and T.~Hartman, ``{Coarse graining pure states in AdS/CFT}'',
  \href{http://arxiv.org/abs/2206.03414}{{\ttfamily arXiv:2206.03414
  [hep-th]}}.

\bibitem{Balasubramanian:2022lnw}
V.~Balasubramanian, A.~Lawrence, J.~M. Magan, and M.~Sasieta, ``{Microscopic
  origin of the entropy of astrophysical black holes}'',
  \href{http://arxiv.org/abs/2212.08623}{{\ttfamily arXiv:2212.08623
  [hep-th]}}.

\bibitem{Anous:2016kss}
T.~Anous, T.~Hartman, A.~Rovai, and J.~Sonner, ``{Black Hole Collapse in the
  1/c Expansion}'', \href{http://dx.doi.org/10.1007/JHEP07(2016)123}{{\em JHEP}
  {\bfseries 07} (2016) 123}, \href{http://arxiv.org/abs/1603.04856}{{\ttfamily
  arXiv:1603.04856 [hep-th]}}.

\bibitem{deHaro:2000vlm}
S.~de~Haro, S.~N. Solodukhin, and K.~Skenderis, ``{Holographic reconstruction
  of space-time and renormalization in the AdS / CFT correspondence}'',
  \href{http://dx.doi.org/10.1007/s002200100381}{{\em Commun. Math. Phys.}
  {\bfseries 217} (2001) 595--622},
  \href{http://arxiv.org/abs/hep-th/0002230}{{\ttfamily arXiv:hep-th/0002230}}.

\bibitem{Skenderis:2002wp}
K.~Skenderis, ``{Lecture notes on holographic renormalization}'',
  \href{http://dx.doi.org/10.1088/0264-9381/19/22/306}{{\em Class. Quant.
  Grav.} {\bfseries 19} (2002) 5849--5876},
  \href{http://arxiv.org/abs/hep-th/0209067}{{\ttfamily arXiv:hep-th/0209067}}.

\bibitem{Darmois:1927}
G.~Darmois, {\em Les \'equations de la gravitation einsteinienne}.
\newblock No.~25 in M\'emorial des sciences math\'ematiques. Gauthier-Villars,
  1927.
\newblock \url{http://www.numdam.org/item/MSM_1927__25__1_0/}.

\bibitem{Israel}
W.~Israel, ``{Singular hypersurfaces and thin shells in general relativity}'',
  \href{http://dx.doi.org/10.1007/BF02710419}{{\em Il Nuovo Cimento B
  (1965-1970)} {\bfseries 44} no.~1, (1966) 1--14}.

\bibitem{Sasieta:2021pzj}
M.~Sasieta, ``{Ergodic equilibration of R\'enyi entropies and replica
  wormholes}'', \href{http://dx.doi.org/10.1007/JHEP08(2021)014}{{\em JHEP}
  {\bfseries 08} (2021) 014}, \href{http://arxiv.org/abs/2103.09880}{{\ttfamily
  arXiv:2103.09880 [hep-th]}}.

\bibitem{Takayanagi:2011zk}
T.~Takayanagi, ``{Holographic Dual of BCFT}'',
  \href{http://dx.doi.org/10.1103/PhysRevLett.107.101602}{{\em Phys. Rev.
  Lett.} {\bfseries 107} (2011) 101602},
  \href{http://arxiv.org/abs/1105.5165}{{\ttfamily arXiv:1105.5165 [hep-th]}}.

\bibitem{Fujita:2011fp}
M.~Fujita, T.~Takayanagi, and E.~Tonni, ``{Aspects of AdS/BCFT}'',
  \href{http://dx.doi.org/10.1007/JHEP11(2011)043}{{\em JHEP} {\bfseries 11}
  (2011) 043}, \href{http://arxiv.org/abs/1108.5152}{{\ttfamily arXiv:1108.5152
  [hep-th]}}.

\bibitem{Climent:2023}
A.~Climent, R.~Emparan, J.~Mag\'{a}n, M.~Sasieta, and A.~Vilar-L\'{o}pez, ``to
  appear'',.

\bibitem{Page:1993df}
D.~N. Page, ``{Average entropy of a subsystem}'',
  \href{http://dx.doi.org/10.1103/PhysRevLett.71.1291}{{\em Phys. Rev. Lett.}
  {\bfseries 71} (1993) 1291--1294},
  \href{http://arxiv.org/abs/gr-qc/9305007}{{\ttfamily arXiv:gr-qc/9305007}}.

\bibitem{HawkingPage}
S.~W. Hawking and D.~N. Page, ``{Thermodynamics of black holes in anti-de
  Sitter space}'', {\em Communications in Mathematical Physics} {\bfseries 87}
  no.~4, (1982) 577 -- 588.

\bibitem{Witten:2021nzp}
E.~Witten, ``{A Note On Complex Spacetime Metrics}'',
  \href{http://arxiv.org/abs/2111.06514}{{\ttfamily arXiv:2111.06514
  [hep-th]}}.

\bibitem{Kontsevich:2021dmb}
M.~Kontsevich and G.~Segal, ``{Wick Rotation and the Positivity of Energy in
  Quantum Field Theory}'', \href{http://dx.doi.org/10.1093/qmath/haab027}{{\em
  Quart. J. Math. Oxford Ser.} {\bfseries 72} no.~1-2, (2021) 673--699},
  \href{http://arxiv.org/abs/2105.10161}{{\ttfamily arXiv:2105.10161
  [hep-th]}}.

\bibitem{Bousso:2021sji}
R.~Bousso and A.~Shahbazi-Moghaddam, ``{Island Finder and Entropy Bound}'',
  \href{http://dx.doi.org/10.1103/PhysRevD.103.106005}{{\em Phys. Rev. D}
  {\bfseries 103} no.~10, (2021) 106005},
  \href{http://arxiv.org/abs/2101.11648}{{\ttfamily arXiv:2101.11648
  [hep-th]}}.

\bibitem{Geng:2021hlu}
H.~Geng, A.~Karch, C.~Perez-Pardavila, S.~Raju, L.~Randall, M.~Riojas, and
  S.~Shashi, ``{Inconsistency of islands in theories with long-range
  gravity}'', \href{http://dx.doi.org/10.1007/JHEP01(2022)182}{{\em JHEP}
  {\bfseries 01} (2022) 182}, \href{http://arxiv.org/abs/2107.03390}{{\ttfamily
  arXiv:2107.03390 [hep-th]}}.

\bibitem{Geng:2020qvw}
H.~Geng and A.~Karch, ``{Massive islands}'',
  \href{http://dx.doi.org/10.1007/JHEP09(2020)121}{{\em JHEP} {\bfseries 09}
  (2020) 121}, \href{http://arxiv.org/abs/2006.02438}{{\ttfamily
  arXiv:2006.02438 [hep-th]}}.

\bibitem{Chandra:2023dgq}
J.~Chandra and T.~Hartman, ``{Toward random tensor networks and holographic
  codes in CFT}'', \href{http://dx.doi.org/10.1007/JHEP05(2023)109}{{\em JHEP}
  {\bfseries 05} (2023) 109}, \href{http://arxiv.org/abs/2302.02446}{{\ttfamily
  arXiv:2302.02446 [hep-th]}}.

\bibitem{Liu:2020jsv}
H.~Liu and S.~Vardhan, ``{Entanglement entropies of equilibrated pure states in
  quantum many-body systems and gravity}'',
  \href{http://dx.doi.org/10.1103/PRXQuantum.2.010344}{{\em PRX Quantum}
  {\bfseries 2} (2021) 010344},
  \href{http://arxiv.org/abs/2008.01089}{{\ttfamily arXiv:2008.01089
  [hep-th]}}.

\bibitem{Kar:2022qkf}
A.~Kar, ``{Non-isometric quantum error correction in gravity}'',
  \href{http://dx.doi.org/10.1007/JHEP02(2023)195}{{\em JHEP} {\bfseries 02}
  (2023) 195}, \href{http://arxiv.org/abs/2210.13476}{{\ttfamily
  arXiv:2210.13476 [hep-th]}}.

\bibitem{Cotler:2017erl}
J.~Cotler, P.~Hayden, G.~Penington, G.~Salton, B.~Swingle, and M.~Walter,
  ``{Entanglement Wedge Reconstruction via Universal Recovery Channels}'',
  \href{http://dx.doi.org/10.1103/PhysRevX.9.031011}{{\em Phys. Rev. X}
  {\bfseries 9} no.~3, (2019) 031011},
  \href{http://arxiv.org/abs/1704.05839}{{\ttfamily arXiv:1704.05839
  [hep-th]}}.

\bibitem{Chen:2019gbt}
C.-F. Chen, G.~Penington, and G.~Salton, ``{Entanglement Wedge Reconstruction
  using the Petz Map}'', \href{http://dx.doi.org/10.1007/JHEP01(2020)168}{{\em
  JHEP} {\bfseries 01} (2020) 168},
  \href{http://arxiv.org/abs/1902.02844}{{\ttfamily arXiv:1902.02844
  [hep-th]}}.

\bibitem{Akers:2021fut}
C.~Akers and G.~Penington, ``{Quantum minimal surfaces from quantum error
  correction}'', \href{http://dx.doi.org/10.21468/SciPostPhys.12.5.157}{{\em
  SciPost Phys.} {\bfseries 12} no.~5, (2022) 157},
  \href{http://arxiv.org/abs/2109.14618}{{\ttfamily arXiv:2109.14618
  [hep-th]}}.

\bibitem{Susskind:1993if}
L.~Susskind, L.~Thorlacius, and J.~Uglum, ``{The Stretched horizon and black
  hole complementarity}'',
  \href{http://dx.doi.org/10.1103/PhysRevD.48.3743}{{\em Phys. Rev. D}
  {\bfseries 48} (1993) 3743--3761},
  \href{http://arxiv.org/abs/hep-th/9306069}{{\ttfamily arXiv:hep-th/9306069}}.

\bibitem{McNamara:2020uza}
J.~McNamara and C.~Vafa, ``{Baby Universes, Holography, and the Swampland}'',
  \href{http://arxiv.org/abs/2004.06738}{{\ttfamily arXiv:2004.06738
  [hep-th]}}.

\bibitem{Emparan:1999pm}
R.~Emparan, C.~V. Johnson, and R.~C. Myers, ``{Surface terms as counterterms in
  the AdS / CFT correspondence}'',
  \href{http://dx.doi.org/10.1103/PhysRevD.60.104001}{{\em Phys. Rev. D}
  {\bfseries 60} (1999) 104001},
  \href{http://arxiv.org/abs/hep-th/9903238}{{\ttfamily arXiv:hep-th/9903238}}.

\bibitem{Balasubramanian:1999re}
V.~Balasubramanian and P.~Kraus, ``{A Stress tensor for Anti-de Sitter
  gravity}'', \href{http://dx.doi.org/10.1007/s002200050764}{{\em Commun. Math.
  Phys.} {\bfseries 208} (1999) 413--428},
  \href{http://arxiv.org/abs/hep-th/9902121}{{\ttfamily arXiv:hep-th/9902121}}.

\bibitem{Hawking:1977}
S.~W. Hawking, ``{Zeta function regularization of path integrals in curved
  spacetime}'', \href{http://dx.doi.org/10.1007/BF01626516}{{\em Communications
  in Mathematical Physics} {\bfseries 55} no.~2, (1977) 133 -- 148}.

\bibitem{Deutsch}
J.~M. Deutsch, ``Quantum statistical mechanics in a closed system'',
  \href{http://dx.doi.org/10.1103/PhysRevA.43.2046}{{\em Phys. Rev. A}
  {\bfseries 43} (Feb, 1991) 2046--2049}.
  \url{https://link.aps.org/doi/10.1103/PhysRevA.43.2046}.

\bibitem{Srednicki1994}
M.~Srednicki, ``Chaos and quantum thermalization'',
  \href{http://dx.doi.org/10.1103/PhysRevE.50.888}{{\em Phys. Rev. E}
  {\bfseries 50} (Aug, 1994) 888--901}.
  \url{https://link.aps.org/doi/10.1103/PhysRevE.50.888}.

\end{thebibliography}\endgroup




\end{document}